\AtBeginDocument{
	\paperwidth=\dimexpr
	1in + \oddsidemargin
	+ \textwidth
	+ 1in + \oddsidemargin
	\relax
	\paperheight=\dimexpr
	1in + \topmargin
	+ \headheight + \headsep
	+ \textheight
	+ 1in + \topmargin
	\relax
	\usepackage[pass]{geometry}\relax
}

\documentclass[smallextended]{svjour3}
\DeclareMathAlphabet{\mathbbcal} {OMS}{cmsy}{m}{n}
\usepackage{amsmath}
\usepackage{times}
\usepackage{newtxmath}
\usepackage{multicol}        
\usepackage[bottom]{footmisc}
\usepackage{bm}
\usepackage{overpic}
\usepackage{algorithm}
\usepackage{algpseudocode}
\usepackage{color}
\usepackage{enumitem}   
\usepackage{mathtools}
\usepackage[colorlinks=true,allcolors=blue]{hyperref}
\usepackage{graphbox}

\newcommand{\R}{\mathbb R}

\newcommand{\N}{\mathbbcal{N}}

\newcommand{\F}{{F}}

\def\E{{\mathbb E}}
\newcommand{\Cov}{\mathbb{C}\textmd{ov}}

\newcommand{\norm}[1]{\left\lVert#1\right\rVert}

\ifx \undefined \varvec    \def \varvec    #1{\text{\boldmath$#1$}} \fi

\usepackage[authoryear]{natbib}
\bibpunct{(}{)}{}{a}{}{;}
\newcommand{\subsubsubsection}[1]{\paragraph{#1}\mbox{}\\}
\begin{document}
	
\title{Multivariate Simulation Using A Locally Varying Coregionalization Model}

\titlerunning{Multivariate Simulation Using A Locally Varying Coregionalization Model}
\author{\'Alvaro I. Riquelme         \and
	Julian M. Ortiz}

\institute{\'Alvaro I. Riquelme \at
	Robert M. Buchan Department of Mining, Queen’s University, Kingston, Canada \\
	\email{alvaro.riquelme@queensu.ca}
	\and
	Julian M. Ortiz \at
	Robert M. Buchan Department of Mining, Queen’s University, Kingston, Canada \\
	\email{julian.ortiz@queensu.ca}  
}

\maketitle

\begin{abstract}
Multivariate spatial modeling is key to understanding the behavior of materials downstream in a mining operation. The ore recovery depends on the mineralogical composition, which needs to be properly captured by the model to allow for good predictions. Multivariate modeling must also capture the behavior of tailings and waste materials to understand the environmental risks involved in their disposal. However, multivariate spatial modeling is challenging when the variables show complex relationships, such as non-linear correlation, heteroscedastic behavior, or spatial trends.  
This contribution proposes a novel methodology for general multivariate contexts, with the idea of disaggregating the global non-linear behavior among variables into the spatial domain in a piece-wise linear fashion. We demonstrate that the complex multivariate behavior can be reproduced by looking at local correlations between Gaussianized variables at sample locations, inferred from a local neighborhood, and interpolating these local linear dependencies by using a non-stationary version of the Linear Model of Coregionalization. This mixture of locally varying linear correlations is combined to reproduce the global complex behavior seen in the multivariate distribution.
The main challenge is to solve appropriately the interpolation of the known correlation matrices over the domain, as these local correlations defined at sample locations can be endowed with a manifold structure, on which the Euclidean distance is not a suitable metric for interpolation of such correlations. This is addressed by using tools from Riemannian geometry: correlation matrices are interpolated using a weighted Fr\'echet mean of the correlations inferred at sample locations. An application of the procedure is shown in a real case study with good results in terms of accuracy and reproduction of the reference multivariate 
distributions and semi-variograms.

\end{abstract}

\section{Introduction}
\label{intro}

Geostatistical applications often consider multivariate data samples as a starting point for  interpolation, simulation, or statistical modeling. Typically, when dealing with individual properties, these are treated as random variables (RV), and they are interpolated with kriging relying on the Random Function (RF) model  (\citeauthor{metheron1971theory} \citeyear{metheron1971theory}). In the case of multiple attributes, interpolation generally improves when addressing it as full multivariate problem via co-kriging (\citeauthor{Chiles} \citeyear{Chiles}; \citeauthor{goovaerts1997geostatistics} \citeyear{goovaerts1997geostatistics}; \citeauthor{wackernagel2013multivariate} \citeyear{wackernagel2013multivariate}). However, these results are constrained to the case of linear dependence between the variables, by use of a linear model of coregionalization.

In the case of simulation of multivariate spatial data, sequential gaussian co-simulation can be used for uncertainty assessment purposes. It sequentially simulates a spatially correlated Gaussian random vector conditioned to the previously simulated vectors, based on cokriging (\citeauthor{almeida1994joint} \citeyear{almeida1994joint}; \citeauthor{gomez1993joint} \citeyear{gomez1993joint}; \citeauthor{verly1993sequential} \citeyear{verly1993sequential}). The conventional procedure is to transform each variable to a Gaussian distribution one at a time. Therefore, an implicit assumption in the process is that the multivariate distribution is also Gaussian. Alternative workflows consider transforming the data into independent variables through linear transformations such as principal component analysis (PCA) (\citealt{pearson1896vii}; \citealt{david1988handbook}) or minimum/maximum auto-correlation factors \citep{switzer1984min}. A stronger premise is related to the stationarity assumption on direct- and cross-variograms, leaving them constant over the domain. Both assumptions are justified when the practitioner does not find a reason to falsify them, that is, if the data do not blatantly contradict these assumptions.

Due to the mineralogical and physical complexity of geological phenomena, geologic data rarely conform to such well-behaved distributions. A particular problem arises when these relationships are non-stationary over the geological domain, meaning that attributes showing a correlated behavior at a given location of the domain may show a different relationship at another location. On top of this non-stationary behavior, relationships among data attributes may show heteroscedasticity, non-linear relations or some kinds of compositional constraints. In these cases, common geostatistical tools are unsuccessful in capturing these features both globally and locally. One of the challenges in multivariate geostatistical modeling is, therefore, to reproduce complex relationships between the variables in space.

In order to incorporate higher levels of complexity in spatial modeling, different tools have been developed over the years to handle the non-stationarity. Among these techniques, the spatial deformation (\citeauthor{monestiez1991semiparametric} \citeyear{monestiez1991semiparametric}; \citeauthor{almendral2008multidimensional} \citeyear{almendral2008multidimensional}; \citeauthor{boisvert2009kriging} \citeyear{boisvert2009kriging}, \citeauthor{fouedjio2015estimation} \citeyear{fouedjio2015estimation}), introduced by \citeauthor{sampson1992nonparametric} (\citeyear{sampson1992nonparametric}), consists in mapping the current spatial domain into a higher dimensional space, where the spatial process can be modeled as stationary. A second approach derives from the convolution method (\citeauthor{yaglom1957some} \citeyear{yaglom1957some}; \citeauthor{metheron1971theory} \citeyear{metheron1971theory}; \citeauthor{journel1978mining} \citeyear{journel1978mining}; \citeauthor{oliver1995moving} \citeyear{oliver1995moving}), where an array of random normal deviates is convolved with a kernel, obtained from a decomposition of the covariance, to obtain the outcome of the RV at a given location. By varying the characteristics of the kernel from point to point, one obtains a non-stationary RF (\citeauthor{higdon1998process} \citeyear{higdon1998process}, \citeyear{higdon1999non}; \citeauthor{paciorek2004nonstationary} \citeyear{paciorek2004nonstationary}; \citeauthor{paciorek2006spatial} \citeyear{paciorek2006spatial}).
A complete review of non-stationary techniques is provided by \citeauthor{fouedjio2017second} (\citeyear{fouedjio2017second}). These techniques have proven to be satisfactory for the univariate case where available data is abundant enough.

In the realm of multivariate modeling, two rather separate paths have been followed to incorporate higher levels of complexity. The first one is related with multivariate transformation techniques, mapping the available data to a Gaussian space, thus making it compatible with Gaussian simulation techniques. Multivariate transformation techniques are required when standard methods, such as the normal score transform (\citeauthor{deutsch1998gslib} \citeyear{deutsch1998gslib}), fail in generating a multivariate normal distribution when applied independently on each variable. Among the tools in this category we find Stepwise Conditional Transformation (SCT) (\citeauthor{rosenblatt1952remarks} \citeyear{rosenblatt1952remarks};  \citeauthor{leuangthong2003stepwise} \citeyear{leuangthong2003stepwise}), which applies normal-score transformation to the first variable, and then hierarchically transforms subsequent variables, conditioned to classes of the previous transforms. Recently, \citeauthor{de2021direct} (\citeyear{de2021direct}) implemented a direct-sampling algorithm based on SCT. Projection Pursuit Multivariate Transform (\citeauthor{barnett2014projection} \citeyear{barnett2014projection}) is another algorithm on the same category that iteratively transforms variables individually to normal scores via a quantile matching, followed by the iterative gaussianization along the direction that maximizes the projection index. The back-transformation to raw values is based on a nearest neighbors search, making the procedure prone to generating values that escape from the original probability distribution function (PDF), as the nearest neighbors may stand close on the attributes space but far on the spatial domain (\citeauthor{barnett2014projection} \citeyear{barnett2014projection}; \citeauthor{madani2019application} \citeyear{madani2019application}). \citeauthor{mueller2017truly} (\citeyear{mueller2017truly}) and \citeauthor{van2017affine} (\citeyear{van2017affine}) seek for a transformation algorithm independent of the logratio transformation applied to the multivariate probability distribution, a property known as affine equivariance, implementing for this purpose the Flow Anamorphosis, shifting the multivariate data towards the centre of a multivariate normal distribution via equations of movement applied to the probability mass.

On the other hand, we find methods dealing with the non-stationarity of the linear dependency among attributes. The obvious approach consists in relaxing the stationarity assumption on the Linear Model of Coregionalization (LMC) (\citeauthor{wackernagel2013multivariate} \citeyear{wackernagel2013multivariate}), allowing the linear correlation at distance zero among two variables to vary over space. This model was introduced first by \citeauthor{gelfand2003spatial} (\citeyear{gelfand2003spatial}) as a method for generating spatial non-stationary RVs from stationary independent factors. An extended version of this model, considering non-stationary independent factors, is treated by \citeauthor{fouedjio2018fully} (\citeyear{fouedjio2018fully}) with the purpose of including locally varying anisotropy in the univariate RVs. A similar approach has been taken by \citeauthor{menafoglio2021kriging} (\citeyear{menafoglio2021kriging}) differing from the previous author in the inference of the semi-variogram used for interpolation of the linear correlation at unsampled locations, based on a kernel estimator using non-Euclidean distances of covariance matrices.

Currently, the two paths to finding a solution to this multivariate problem do not overlap, as the first one treats multivariate complexity globally by gaussianizing the full multivariate probability, missing the spatial nature of the geological phenomena, while the second treats it locally but applies only for the case of Gaussian RVs.

In this paper, we face the problem of simulating a multiple set of $ p $ regionalized variables, $ Z_1,\dots,Z_p $, under a global non-linear and non-stationary framework. We demonstrate that global non-linear features can be reproduced by means of a non-stationary LMC model. The method looks at the local linear correlations between variables at sample locations, inferred after applying a Gaussian transformation in a local neighborhood. Then, these local correlations defined at sample locations can be interpolated on the spatial domain by mapping them into the space of correlation matrices, which form a Riemannian manifold. As Euclidean distances are no longer a suitable metric on this Riemannian space, the main challenge is to find an appropriate metric to measure closeness among correlation matrices, with the purpose of interpolating between known correlations at specific sample locations. This task is addressed by using tools from Riemannian Geometry. An application of the procedure is shown in a real case study, focusing on the essential steps of the methodology. This example demonstrates how the proposed methodology honors the multivariate configuration of data in the attributes space, as well as agreeing with spatial experimental features such as cross-semi-variograms.

The paper is structured as follows: Sect. \ref{Extending} introduces the LMC, and explains in detail how to obtain a non-stationary model from it; the methodology is summarized within the section. Sect. \ref{Syn Study} implements a synthetic case study that demonstrate the capability for retrieving the independent underlying factors and inference of local correlation. In Sect. \ref{Case Study} we implement the methodology on a simulation study considering six cross-correlated variables from a blast hole campaign belonging to a Nickel-Laterite deposit. Section \ref{Conclusions} provides the discussion of the results and conclusions of the work. In Appendix A we provide the reader with the required notions of Riemannian geometry used in this work, and also discusses the interpolation of correlation matrices.

\section{The Locally Varying Linear Model of Coregionalization (LVLMC)}
\label{Extending}

The interpretation provided in this work for the occurrence of global non-linear multivariate properties in the attribute space
is to see the geological process as a linear mixture of independent RVs defined on a spatial domain $  D $, with local properties that change smoothly throughout the different positions $ \textbf{u} \in D $. In particular, we consider these local features to be captured by the correlation matrix at location $ \textbf{u} $, which leads to the global reproduction of the complex non-linear features among the variables. 

\subsection{The Gaussian Setting}

Let $ {{\textbf{Y}}}= [{Y}_1(\textbf{u}),\dots, {Y}_p(\textbf{u})]^T $ be the vector-valued RF considering $p$ simultaneous zero mean and unit variance Gaussian RFs  ${Y}_i=\{{Y}_i(\mathbf{u}): \mathbf{u} \in D \subseteq \R^3\}$, indexed by $ i $ ranging in the set  $I=\{1,\dots, p\}$. The collection of sampling data is given by the multivariate vectors $ {{{\textbf{y}}}}_\alpha = [{y}_1(\textbf{u}_\alpha),\dots,{y}_p(\textbf{u}_\alpha)]^T $, $ \alpha \in \{1,\dots,n\} $, with $ n $ the number of samples available.

Let us assume a given theoretical linear correlation among the variables at step zero for all pair of variables $ \{Y_i(\textbf{u}_\alpha),Y_j(\textbf{u}_\alpha)\} $, given by $\rho_{Y_iY_j}(\textbf{u}_\alpha)$ (or simply $\rho_{ij}(\textbf{u}_\alpha)$) with $i,j \in I$, noticing that the correlation may vary according to the location $ \textbf{u}_\alpha $. We can represented it by the matrix
 
\begin{equation}
	\textbf{C}(\textbf{u}_\alpha)=\begin{pmatrix}
		1&\quad\rho_{12}(\textbf{u}_\alpha)&\quad\cdots&\quad\rho_{1p}(\textbf{u}_\alpha) \\
		\rho_{21}(\textbf{u}_\alpha)&\quad1&\quad\cdots&\quad\rho_{2p}(\textbf{u}_\alpha) \\
		\vdots    &\quad\vdots      &\quad\ddots&\quad\vdots \\
		\rho_{p1}(\textbf{u}_\alpha)&\quad\rho_{p2}(\textbf{u}_\alpha)&\quad\cdots&\quad1
	\end{pmatrix},\nonumber
\end{equation}
or just $ \textbf{C}_\alpha $.

A simple way of building a spatially coherent Gaussian model, with varying step-zero correlation, is to consider the LMC, where the vector of correlated variables $ \textbf{Y} $ is the result of applying an affine transformation \textbf{A} to a vector of independent Gaussian RFs $ \tilde{\textbf{Y}} $

\begin{equation}
\textbf{Y} =  \textbf{A}\tilde{\textbf{Y}},
\end{equation}
or more explicitly, each variable $ Y_i $ consisting of a weighted sum of $ p $ independent factors, $ \tilde{Y}_j $:
\begin{equation}
	Y_i(\mathbf{u}) = \sum_{j=1}^{p}a_{ij}\tilde{Y}_j(\mathbf{u}),
\end{equation}
with $ a_{ij} $ the $ ij$ entry of \textbf{A}. For simplicity, we consider the number of factors equal to the number of attributes to avoid the ill-definition of the linear system (the problem of working in the stationary case with a number of factors different than the number of attributes has been tackled first by \cite{bourgault1991multivariable}, and recently by \cite{pinto2021decomposition}).

Then, the direct and cross covariance structure between variables at different locations are given by (\citeauthor{wackernagel2013multivariate} \citeyear{wackernagel2013multivariate};  \citeauthor{gelfand2003spatial} \citeyear{gelfand2003spatial}; \citeauthor{fouedjio2018fully} \citeyear{fouedjio2018fully}):

\begin{eqnarray*}
	\Cov\big(Y_l(\textbf{u}),Y_m(\textbf{u}')\big)&=&\E\bigg(\sum_{j=1}^{p}a_{lj}\tilde{Y}_j(\mathbf{u})\cdot\sum_{k=1}^{p}a_{mk}\tilde{Y}_k(\mathbf{u})\bigg)\nonumber\\
	&=&\sum_{j=1}^{p}\sum_{k=1}^{p}a_{lj}a_{mk}\E\big(\tilde{Y}_j(\mathbf{u})\tilde{Y}_k(\mathbf{u})\big)\nonumber\\ &=&\sum_{j=1}^{p}a_{lj}a_{mj}C_j(\mathbf{u}-\mathbf{u'}),\nonumber
\end{eqnarray*}
implying that the covariance at step-zero among $ \{Y_l(\textbf{u}),Y_m(\textbf{u})\} $ pairs is given by
\begin{eqnarray*}
	\Cov\big(Y_l(\textbf{u}),Y_m(\textbf{u})\big)=\sum_{j=1}^{p}a_{lj}a_{mj},\nonumber
\end{eqnarray*}
or, in matrix notation, as
\begin{eqnarray*}
	\textbf{C}(\textbf{u})=\textbf{A}\textbf{A}^T,\nonumber
\end{eqnarray*}
with $ (\textbf{C})_{lm}=\Cov\big(Y_l(\textbf{u}),Y_m(\textbf{u})\big) $ the entries of $ \textbf{C} $. This relation allows us to find the required affine transformation \textbf{A}, needed to compute the vector of independent Gaussian RFs $ \tilde{\textbf{Y}} $, just by inferring the step-zero correlation among Gaussian variables ${\textbf{Y}} $.

Therefore, to obtain a set of Gaussian RFs with varying step-zero correlation on the Gaussian variables of vector ${\textbf{Y}} $ as a function of \textbf{u}, it is required  to set an affine transformation varying smoothly on the domain, \textbf{A}(\textbf{u}) (\citeauthor{gelfand2003spatial} \citeyear{gelfand2003spatial}).

\subsection{Decomposing the Correlation Matrix}

We take a closer look into the decomposition of $ \textbf{C}(\textbf{u})=\textbf{A}(\textbf{u})\textbf{A}^T(\textbf{u}) $, with the purpose of decoupling $ \textbf{Y} $ and work with independent variables $\tilde{\textbf{Y}}(\textbf{u})  =  \textbf{A}(\textbf{u})^{-1}\textbf{Y}(\textbf{u}) \sim \N(\textbf{0},\textbf{I}_p)$, with $ \textbf{I}_p $ the identity matrix of $ p \times p $ size.

One may suggest the use of eigen-decomposition $ \textbf{C} = \textbf{U}\textbf{D}\textbf{U}^T$ and set $ \textbf{A}=\textbf{U}{\textbf{D}}^{1/2} $ in order to get $ \textbf{Y} = \textbf{U}{\textbf{D}}^{1/2}\tilde{\textbf{Y}}$. This approach may result in spatial discontinuities because the decomposition is non-unique. However, $ \textbf{C} $ can be uniquely decomposed as the product of a positive-diagonal lower triangular matrix by Cholesky decomposition. This is a suitable choice for our purpose: $ \textbf{C} = \textbf{L}\textbf{L}^T$.	

Once we have a continuous decomposition for $\textbf{C}(\textbf{u})$ and the independent variables, the overall process of estimation and simulation becomes straightforward, by working separately on the spatial behavior modeling of each independent variable.

Notice that, once the Cholesky decomposition is applied, the solution can be rotated and still reproduces the correlation: $ \textbf{Y} = \textbf{L}\tilde{\textbf{Y}}'$, with $\tilde{\textbf{Y}}'= \textbf{R}\tilde{\textbf{Y}}$ and $ \textbf{R}$ a rotation matrix. Any decomposition of the form $ \textbf{C} = \textbf{L}\textbf{R}\textbf{R}^T\textbf{L}^T$ is valid. However, this is not a problem since by applying a rotational transformation different than the identity delivers correlated input factors  $ \tilde{Y}'_i(\mathbf{u}) $
and, therefore, the Cholesky decomposition followed by selecting $ \textbf{R}= \textbf{I}_p $ ensures us to obtain cross-semi-variograms with a zero sill.

We focus now on linking the input variables in original units $ {{\textbf{Z}}}= [{Z}_1(\textbf{u}),\dots,$${Z}_p(\textbf{u})]^T $, showing a global non-linear behavior in the attribute space, with the Gaussian vector  $ {{\textbf{Y}}} $.

\subsection{The Nataf Transformation}

We begin by defining a simple but efficient Gaussian transformation for a set of globally non-linear input variables $ {{\textbf{Z}}}= [{Z}_1(\textbf{u}),\dots, {Z}_p(\textbf{u})]^T $ into the Gaussian vector  $ {{\textbf{Y}}} $. Applied locally in a neighborhood of $ \textbf{u}_\alpha $,  this transformation allows us to compute the covariance matrix $ \textbf{C}(\textbf{u}_\alpha) $ locally. After describing the transformation step, we can focus on the spatial modeling of $ {{\textbf{Y}}} $ and how to interpolate the correlated behavior among variables into unsampled locations.

The proposed methodology relies on the assumptions that non-linear multivariate features can be reconstructed by mapping the original $ p $-variate cumulative distribution function (CDF) with a $ p $-variate Gaussian distribution equipped with a proper prior covariance matrix. This procedure is also known as Nataf transformation \citep{nataf1962determination} or NORTA (NORmal To All), and several properties of the transformation have been studied in different contexts, for instance, in \citeauthor{cario1997modeling} (\citeyear{cario1997modeling}); \citeauthor{ayadi2019norta} (\citeyear{ayadi2019norta}); \citeauthor{xie2015quantifying} (\citeyear{xie2015quantifying}); \citeauthor{xiao2014evaluating} (\citeyear{xiao2014evaluating}); \citeauthor{li1975generation} (\citeyear{li1975generation}); and by \citeauthor{bourgault2014revisiting} (\citeyear{bourgault2014revisiting}) in the geostatistical context. We start by a brief motivation and then highlight the relevant theoretical aspects of the transformation.

\begin{figure}[h]
	\begin{center}
		\includegraphics[width=0.48\textwidth]{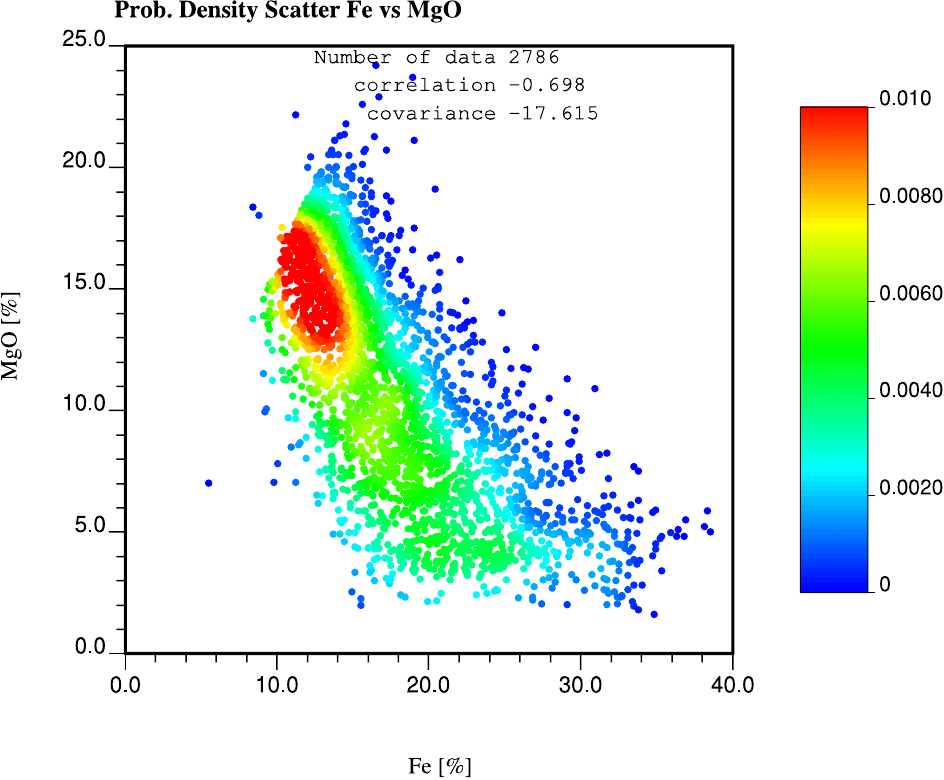}
		\includegraphics[width=0.48\textwidth]{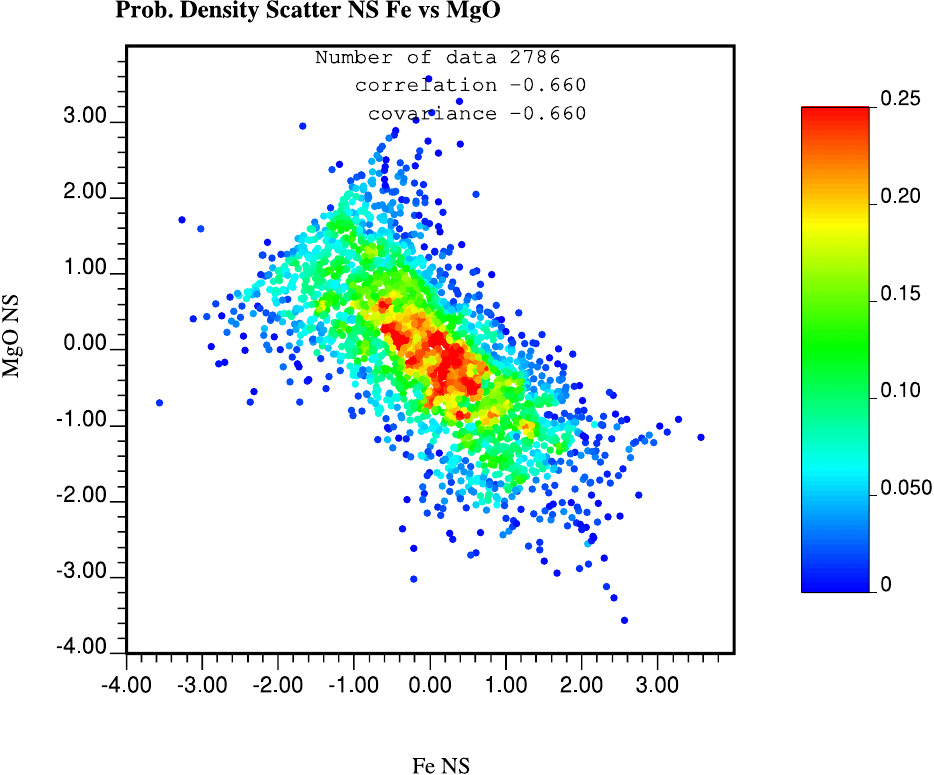}
	\end{center}
	\caption{Correlation of variables before and after applying a normal-score transformation. Variables are still correlated.}
	\label{anamor}
\end{figure}

Let $ {{{\textbf{Z}}}}= [{Z}_1(\textbf{u}),\dots, {Z}_p(\textbf{u})]^T $ be a vector-valued random function (RF) considering $p$ simultaneous RFs  ${Z}_i=\{{Z}_i(\mathbf{u}): \mathbf{u} \in D \subseteq \R^3\}$. The sampling data is given by the multivariate vectors $ {{{\textbf{z}}}}_\alpha = [{z}_1(\textbf{u}_\alpha),\dots,{z}_p(\textbf{u}_\alpha)]^T $, $ \alpha \in \{1,\dots,n\} $. Let $ \phi_i(\cdot) $ be the \textit{anamorphosis} function that transforms the data from Gaussian values to original values, for the \textit{i}-th variable \citep{deutsch1998gslib}, $ \phi_i = F_i^{-1} \circ G $ (with $ F_i $  the CDF of ${Z}_i$ and $ G $ the standard Gaussian CDF). Then, we encounter that the naive procedure of independently transforming the values $ {z}_i $ of the different RV $ {Z}_i $, into univariate Gaussian values $ y_i $, \[ y_i= \phi_i^{-1}({z}_i)=G^{-1}\big(F_i({z}_i)\big) , \]
does not translate into independent Gaussian variables, $ Y_i $. This is depicted in the cross plots of Fig. \ref{anamor}, showing that two originally correlated variables are still correlated after this transformation.

\begin{figure}[h]
	\begin{center}
		\includegraphics[width=0.35\textwidth]{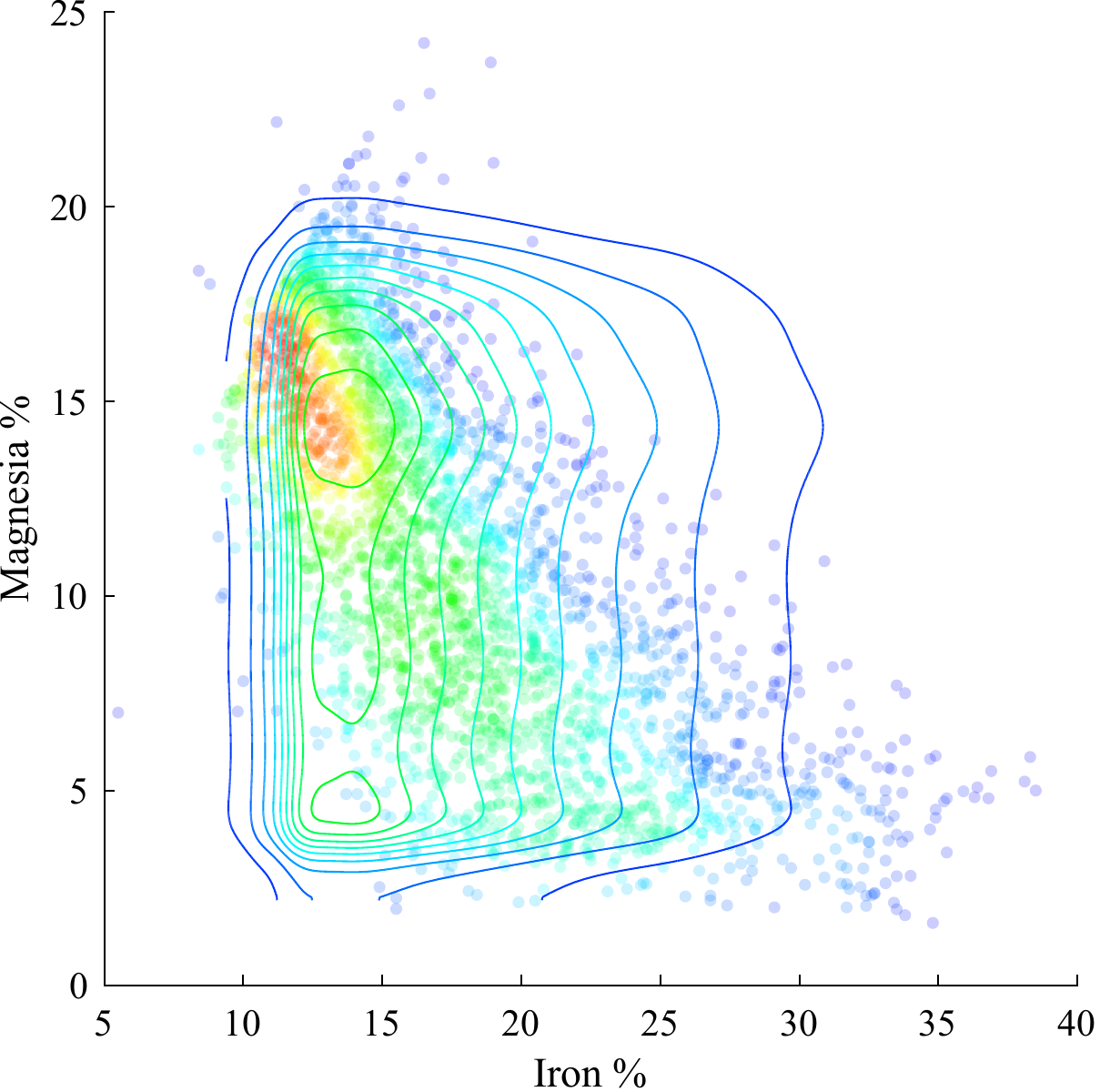}
		\includegraphics[width=0.35\textwidth]{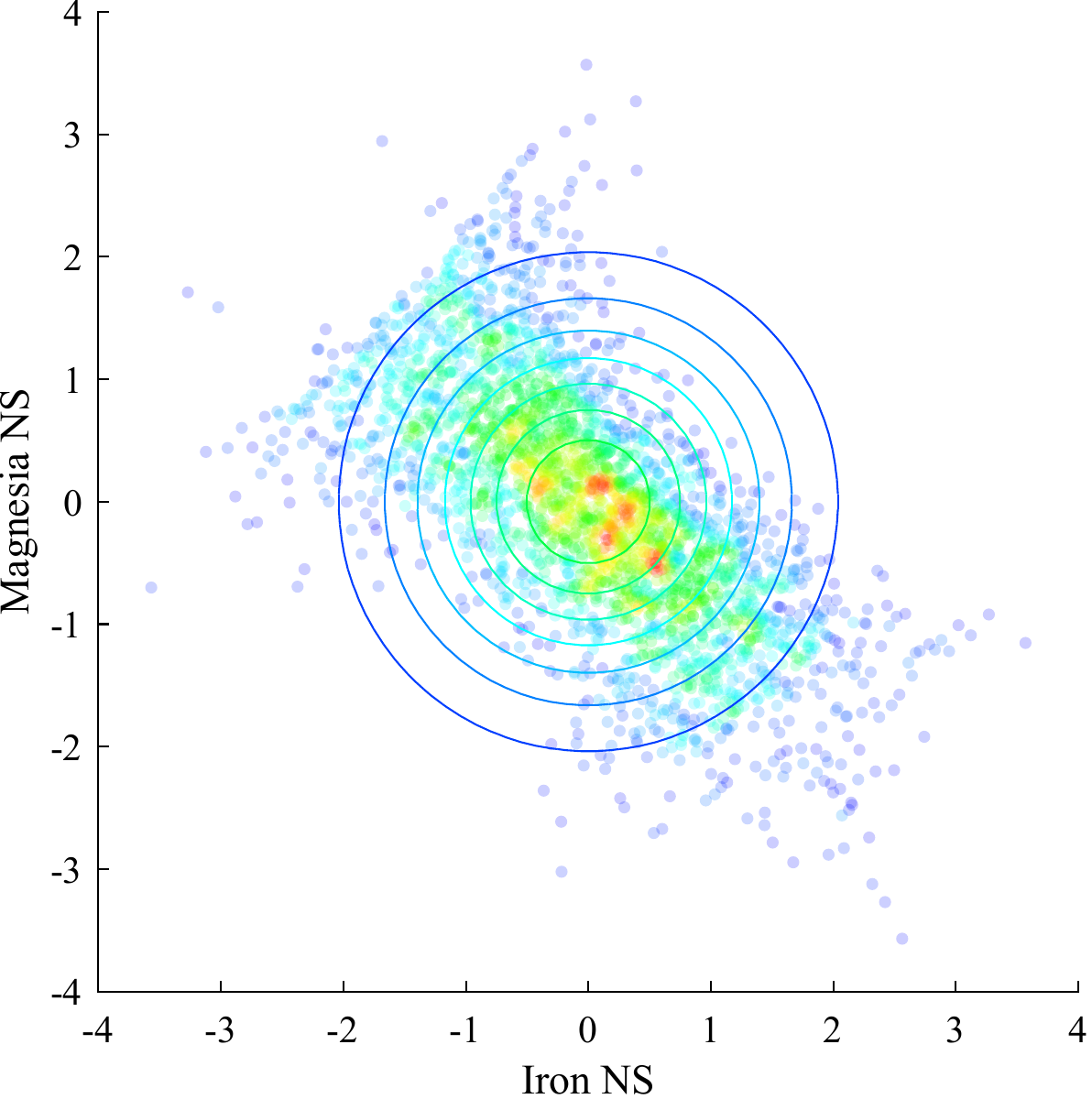}
		\includegraphics[width=0.35\textwidth]{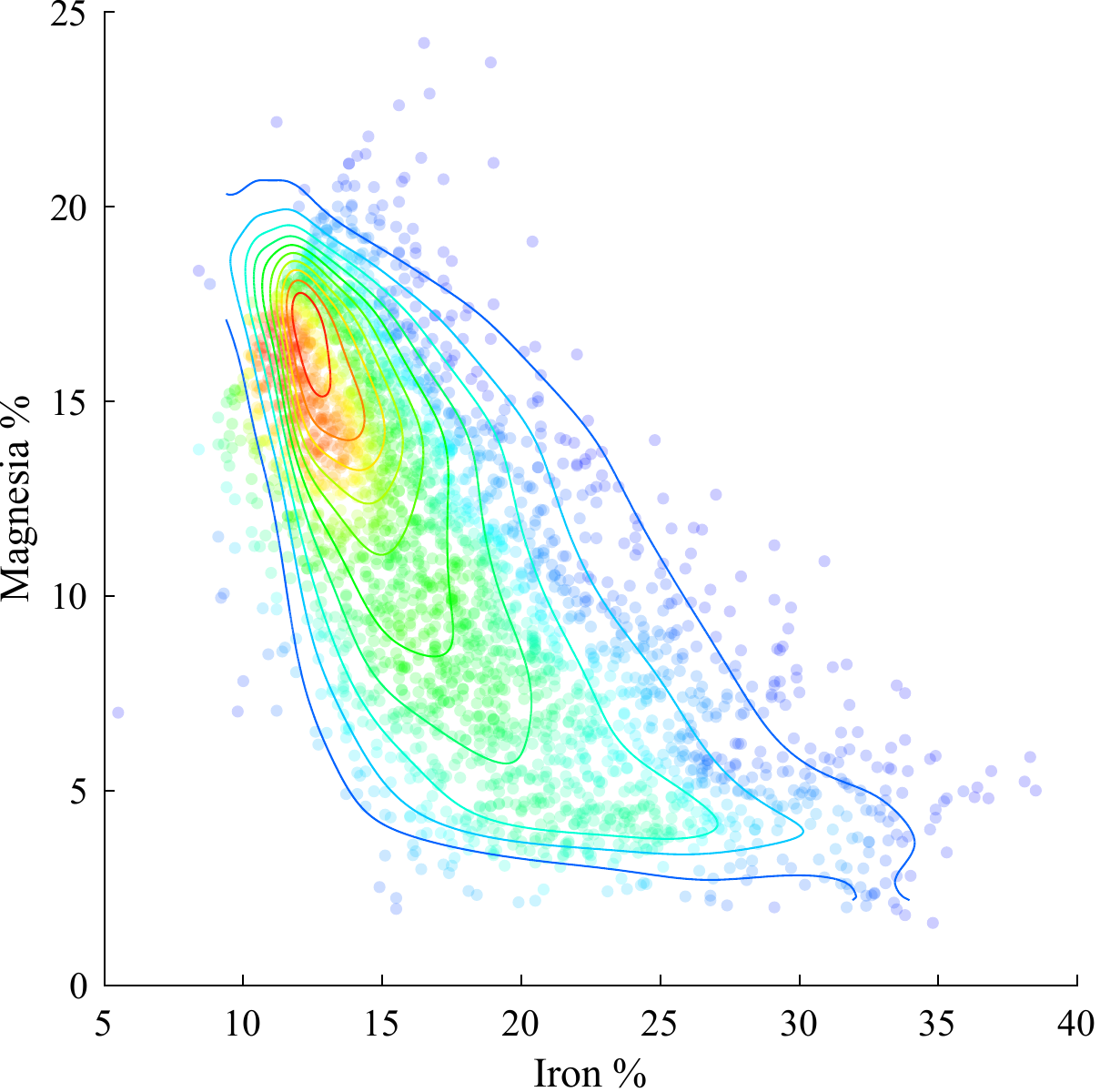}
		\includegraphics[width=0.35\textwidth]{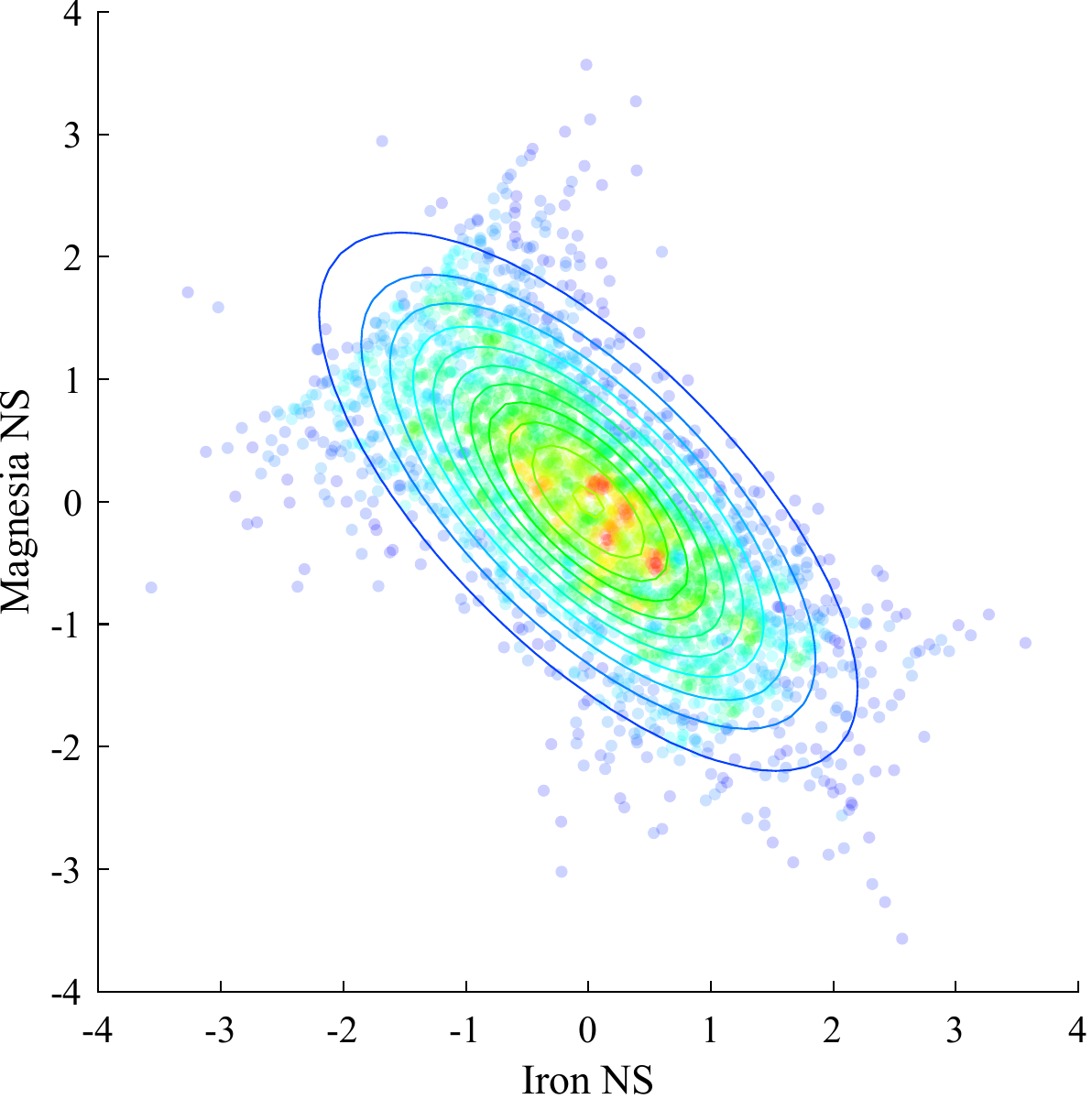}
	\end{center}
	\caption{Probability densities (in contour lines) resulting from the mapping of the raw CDF with a standard Gaussian PDF (top) and the mapping of the raw CDF with a correlated Gaussian PDF (bottom). The contour lines fit better the data in the second case.}
	\label{anamor2}
\end{figure}

Therefore, if two or more correlated raw variables are mapped independently into a non-correlated multi-Gaussian PDF, the procedure entails an incorrect mapping among multivariate probability densities (Fig. \ref{anamor2}, top), as we are not imposing any relation among the independent transformations. In consequence, any estimation done following this path is prone to give bad results.
However, the problem can be fixed if a correlated Gaussian distribution is considered instead for the mapping of the multivariate raw distribution. The method gives a better result for modeling the multivariate PDF  (Fig. \ref{anamor2}, bottom) as now we provide information on how the raw CDFs have to be coupled, by giving the correlation coefficient of the $ Y_j $ variables in the multiGaussian PDF. We formalize this procedure below.

We define the \textit{non-coupled} transformation of the initial multivariate RF ${{{\textbf{Z}}}}$ into a \textit{stationary} $p$-variate Gaussian RF with zero vector mean ${\varvec\mu}= (0,\dots,0)^T=\textbf{0} $ and covariance matrix equal to the identity matrix $\textbf{I}_p$, that is, $\textbf{Y}=[Y_1(\textbf{u}),\dots,Y_p(\textbf{u})]^T \sim \N(\textbf{0},\textbf{I}_p)$, by using the anamorphosis function $\phi_i$ on each of the components of $\textbf{Y}$:
\begin{eqnarray}
	{{{\textbf{Z}}}}=&[{Z}_1(\textbf{u}),\dots,{Z}_p(\textbf{u})]^T&\nonumber\\
	=&(\phi_1[Y(\textbf{u})],\dots,\phi_p[Y(\textbf{u})])^T&={\varvec\upPhi_\textbf{I}}_p(\textbf{Y})
	\text{.}\nonumber
\end{eqnarray}

The \textit{coupled} prior distribution of $\textbf{Y}$, is still a $p$-variate Gaussian distribution ${\N}( \textbf{0},{\textbf{C}})$, with $ \textbf{0} $ mean vector and \textbf{\textit{correlation}} matrix given by
\begin{equation*}
	{\textbf{C}} = \begin{pmatrix}
		1&\quad\rho_{12}&\quad\cdots&\quad\rho_{1p} \\
		\rho_{21}&\quad1&\quad\cdots&\quad\rho_{2p} \\
		\vdots    &\quad\vdots      &\quad\ddots&\quad\vdots \\
		\rho_{p1}&\quad\rho_{p2}&\quad\cdots&\quad1
	\end{pmatrix}.
\end{equation*}

Then, the random variables  $Y_i$ are correlated and their pairwise relationships are quantified by the correlation coefficients $\rho_{ij}$ with $i,j \in I$, which have to be inferred. We do this in the next section. The $p$-variate CDF over the original variables is then retrieved simply as:
\begin{equation}
	\F_{{Z}_1(\textbf{u}),\dots,{Z}_p(\textbf{u})}({z}_1,\dots,{z}_p)=G_\mathbf{0}^{{\textbf{C}}}\big(\phi_1^{-1}({z}_1),\dots,\phi_p^{-1}({z}_p)\big),\label{transnat}
\end{equation}
with $ G_\mathbf{0}^{{\textbf{C}}} $  the $ p $-dimensional Gaussian cumulative distribution of zero vector mean and correlation matrix $ {{\textbf{C}}}$.

\begin{figure}[h]
	\begin{center}
		\includegraphics[width=0.89\textwidth]{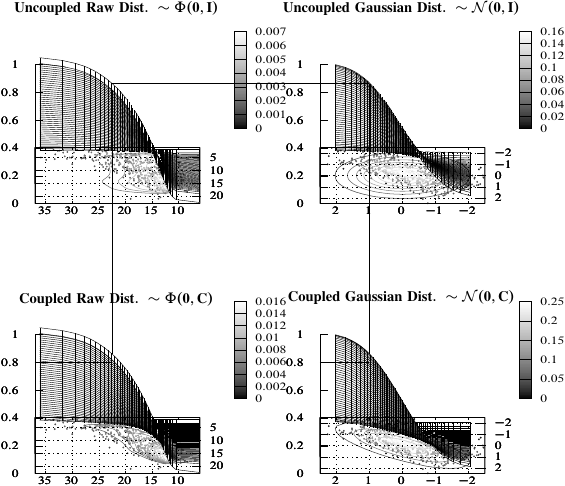}
	\end{center}
	\caption{Conceptual bivariate picture of the adjustment of the correlated behavior of independent raw distributions through the Gaussian coupled anamorphosis. In surface are the corresponding CDFs, and in contour plots the PDFs.}
	\label{GAn3}
\end{figure}

We will say that $\textbf{Z}$ follows a \textit{coupled anamorphosis function}, that is, $\textbf{Z} \sim \varvec\upPhi( {\textbf{0}},{{\textbf{C}}})$. The transformation (or coupling process) is conceptually illustrated, for the bivariate case, in Figure \ref{GAn3}. This transformation is \textit{well-defined} in the sense that the order of variables does not play a role, and a permutation of them just translates into a permutation of the correlation coefficients in $ {{\textbf{C}}} $.

Given the different RVs that describe ore deposits, $ {{{\textbf{Z}}}}= [{Z}_1(\textbf{u}),\dots, {Z}_p(\textbf{u})]^T $, we transform the variables into Gaussian RVs jointly, according to Eq. \ref{transnat}, in order to get the vector ${{{\textbf{Y}}}}=[{Y}_1(\textbf{u}),\dots,{Y}_p(\textbf{u})]$. 

\subsection{Inference of the Correlation Matrix}

Gaussianization of variables may be done in a ``global'' fashion, that is, by running Eq. \ref{transnat} once and using all the data $ \textbf{\textit{z}}_\alpha $, $ \alpha \in \{1,\dots,n\} $.
The previous procedure entails the severe hypothesis that the multivariate behavior of geological attributes can be modeled by assuming Gaussian distribution with a fix correlation matrix on the domain, which coincides with the conventional LMC, assuming a stationary behavior for the correlation among variables, and may be a poor model globally. 

In order to move away from the stationarity path, we take the alternative approach of performing the Gaussianization ``locally'' at a given location \textbf{u}, which means to collect subsets of data in a vicinity to the location under study, $ \mathbbcal{V}(\textbf{u}) $, defined either by fixing a radius of search from the location \textbf{u} or by fixing the cardinality $ |\mathbbcal{V}(\textbf{u})|=l $ of the closest samples to be considered in the vicinity, according to the sampling density and spacing considerations. Unfortunately, when considering scattered data over the geological domain, this is the only parameter available for modification to infer the dependency among variables. This forces us to incorporate a local stationarity assumption (\citeauthor{metheron1971theory}, \citeyear{metheron1971theory}; \citeauthor{wackernagel2013multivariate}, \citeyear{wackernagel2013multivariate}), which  translates in having a random field with both mean vector-value, $ \E\{Z(\textbf{u})\} $, and correlation matrix among attributes, $ \textbf{C}(\textbf{u}) $, smooth functions varying slowly in space with respect to the spatial domain under consideration.

The next step is the inference of a local correlation matrix $\textbf{C}(\textbf{u}_\alpha) $ at the sampling locations, based on the collection $ \{{{\textbf{z}}}_\beta\}_{\beta \in \mathbbcal{V}(\textbf{u}_\alpha)} $, in moving neighborhood fashion.  Gaussianization is then done only on the selected vicinity, and the inference of the correlation matrix is obtained locally. This path is consistent with the traditional methodology for uncertainty modeling, which consists in partitioning the data in stationary domains, and continue the work on each of the domains separately. In our case, there is no need of defining stationary domains as we assume that $\textbf{C}(\textbf{u}) $ is varying smoothly on the domain.

Once the correlation is inferred at a given location $ \textbf{u}_\alpha $, we can gen the vector values of the independent underlying factors $\tilde{{{{\textbf{\textit{y}}}}}}_\alpha = [\tilde{y}_1(\textbf{u}_\alpha),\dots,\tilde{y}_p(\textbf{u}_\alpha)]^T $ first by applying Cholesky decomposition on $\textbf{C}(\textbf{u}_\alpha) = \textbf{L}(\textbf{u}_\alpha)\textbf{L}^T(\textbf{u}_\alpha) $, and then by applying $ \tilde{{{{\textbf{\textit{y}}}}}}_\alpha = \textbf{L}^{-1}(\textbf{u}_\alpha){{{{\textbf{\textit{y}}}}}}_\alpha $. This process is repeated for every sample location $ \alpha \in \{1,\dots,n\} $.

\subsection{Interpolation of Correlation Matrices}

In order to obtain a multivariate simulated value at a given unsampled location \textbf{u}, coherent with the behavior of the correlation shown among the components, it required to work in parallel both on simulating the independent factors $ \tilde{Y}_i $ on the domain and, at the same time, to obtain an estimate of the correlation matrix at \textbf{u}.

Let $ \textrm{Corr}(p) $ be the space of all $ p \times p $ correlation matrices. One could propose, for a set of $ n $ correlation matrices $ \textbf{C}_1,\dots,\textbf{C}_n \in \textrm{Corr}(p)$, the use a weighted mean
\begin{equation*}
	\widehat{\textbf{C}}(\textbf{u})= \sum_{i=1}^{n}\lambda_i\textbf{C}_i.
\end{equation*}
with $ \lambda_i $ a set of weights accounting for geographical information, such as kriging interpolation weights. However, this estimation procedure is not closed, thas it, the procedure entails the problem that the estimate $ \widehat{\textbf{C}} $ may not be a correlation matrix, for instance, if one or more of the used weights are negative.

As $ \widehat{\textbf{C}}(\textbf{u})= \sum_{i=1}^{n}\lambda_i\textbf{C}_i $ is the ``point'' that minimizes the Euclidean distance to the data $ \textbf{C}_1,\dots,\textbf{C}_n$, the previous problem is solved if, instead of using a linear interpolation method (or Euclidean), one changes the metric used to a one that ensures to obtain a correlation matrix. We can represent this metric by $ d_{\textrm{Corr}(p)} $ and define the appropriate estimate, also known as Fr\'echet mean or geometric mean, as the minimization problem
\begin{equation}\widehat{\textbf{C}} = \textmd{arg} \inf_{\substack{\textbf{C}}} \sum_{i=1}^n d^2_{\textrm{Corr}}(\textbf{C}_i,\textbf{C}),\label{int}
\end{equation}
that is, searching for the correlation matrix that minimizes the sum of the square distance $ d_{\textrm{Corr}(p)} $ to the data $\textbf{C}_1,\dots,\textbf{C}_n$. This is discussed in Appendix A.

\subsection{Methodology}
\label{Methodology}

The proposed methodology to extend the LMC, is summarized next, with the first and the last steps being optional and suggested when the data are compositional:

\begin{enumerate}[label=\arabic*)]
	\item (Perform log-ratio transformation on the data if this is compositional.).
	\item At each sample location $ \textbf{u}_\alpha $, find the nearest $ n_\alpha $ samples.
	\item Perform a local Gaussian transformation for each variable $\phi_i^{-1}[ Z_i(\textbf{u}_\alpha)]={Y}_i(\textbf{u}_\alpha)$ using the nearest $  n_\alpha $ samples.
	\item Compute the correlation matrix $ \textbf{C}(\textbf{u}_\alpha) $ of the vector $ {\textbf{Y}}=[{Y}_1(\textbf{u}_\alpha),\dots,{Y}_p(\textbf{u}_\alpha)] $.
	\item Model the variogram of each $ {\tilde{Y}_i} $, $ \forall i \in I $, and simulate.
	\item Apply the Cholesky decomposition of $ \textbf{C}(\textbf{u}_\alpha)= \textbf{L}\textbf{L}^T $ and apply $ \textbf{L}^{-1}{\textbf{Y}}=\tilde{\textbf{Y}} $ to decorrelate the Gaussian variables.
	\item Interpolate $ \textbf{C}(\textbf{u}_\alpha)$ on the domain $ D $ using the weighted Fr\'echet mean and a set of weights $ \lambda_i $, as described in Appendix A. Kriging weights given by the variogram modeling of $ \tilde{\textbf{Y}} $ work appropriately.
	\item At each unsampled location $ \textbf{u} $, take the estimated correlation matrix $ \widehat{\textbf{C}}(\textbf{u}) $, perform Cholesky decomposition, and recover $ {\textbf{Y}}(\textbf{u})=\widehat{\textbf{L}}(\textbf{u})\tilde{\textbf{Y}}(\textbf{u}) $.
	\item At the unsampled location $ \textbf{u} $, find the nearest $  n_\alpha $ samples, perform Gaussian transformation individually for each variable $\widehat{\phi}_i^{-1}[Z_i(\textbf{u})]={Y}_i(\textbf{u})$, and recover the simulated value $ \tilde z_i(\textbf{u})=\widehat{\phi}_i[y_i(\textbf{u})] $.
	\item (Perform log-ratio back-transformation on the data if this is compositional.).
\end{enumerate}

\section{Synthetic Case Study}
\label{Syn Study}

\begin{figure}[h!]
	\raggedright
	\includegraphics[width=0.49\textwidth,trim={0 0cm 0.0cm 0cm},clip]{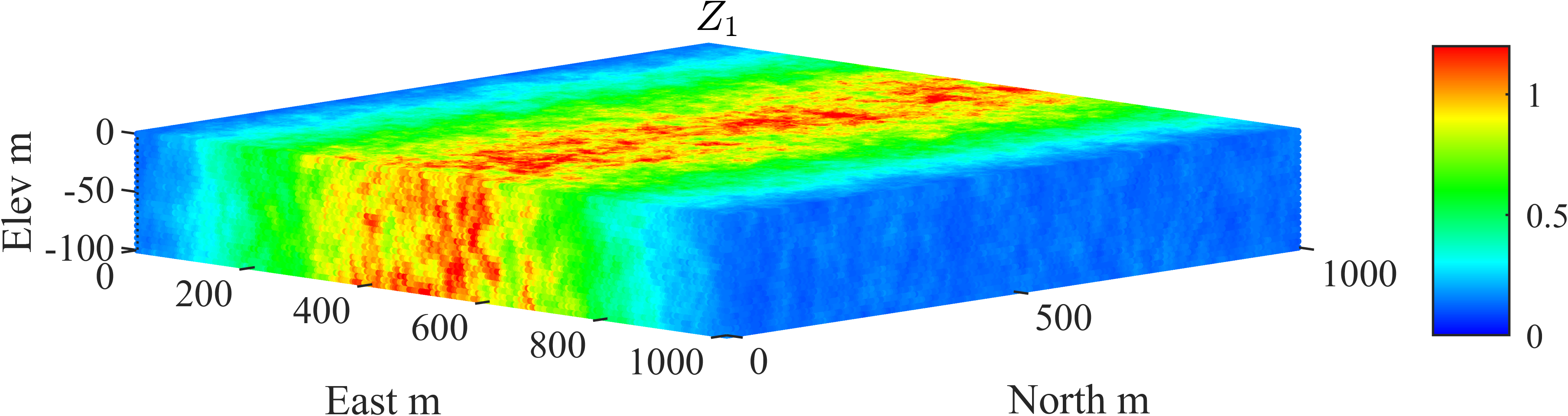}
	\includegraphics[width=0.49\textwidth,trim={0 0cm 0.0cm 0cm},clip]{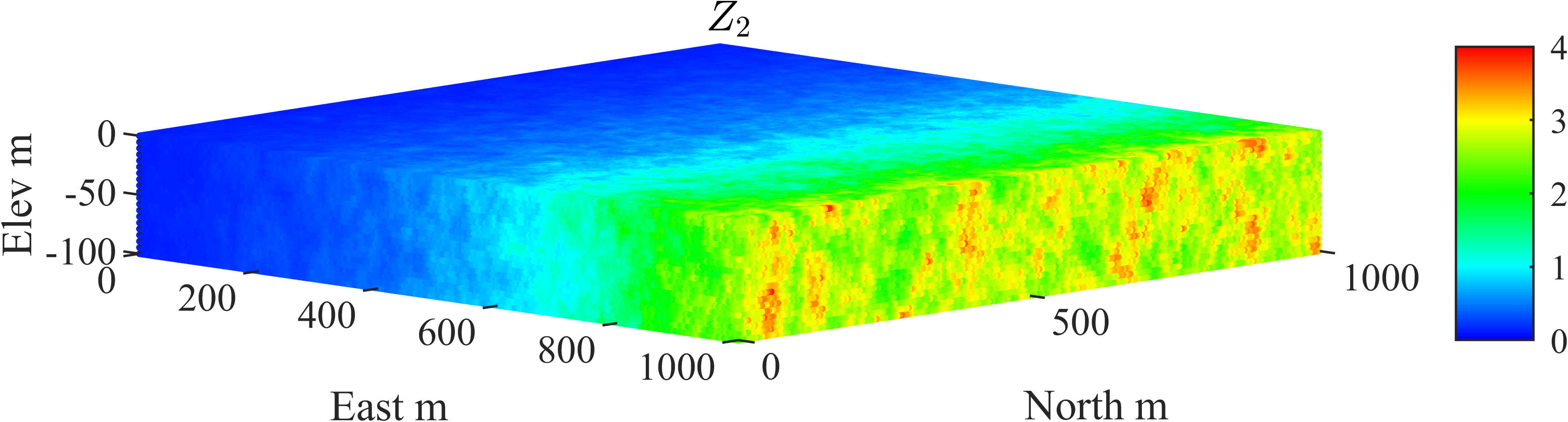}\\
	\includegraphics[width=0.49\textwidth,trim={0 0cm 0.0cm 0cm},clip]{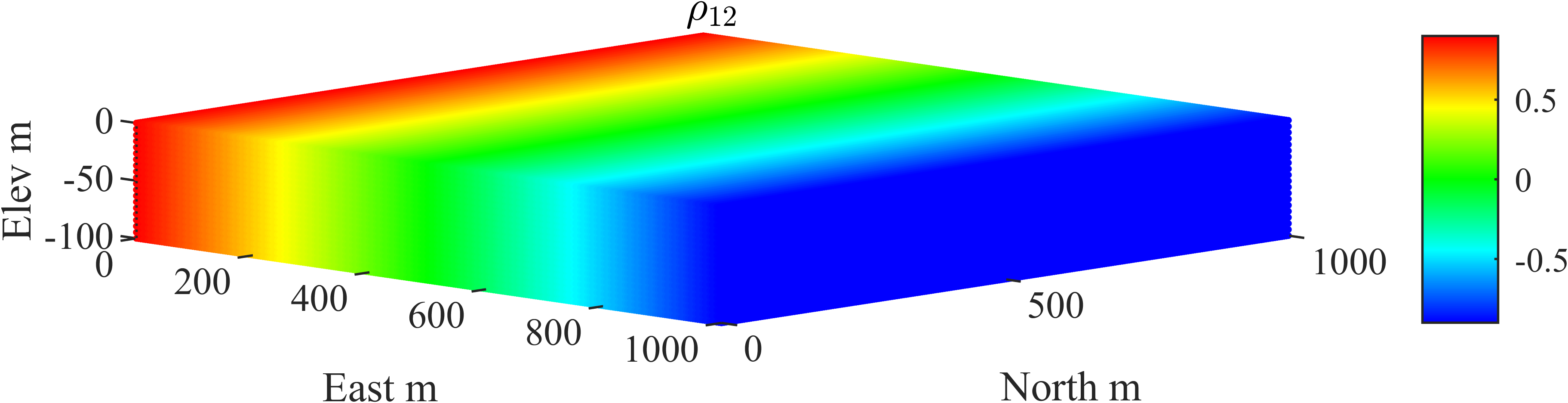}
	\includegraphics[width=0.49\textwidth,trim={0 0cm 0.0cm 0cm},clip]{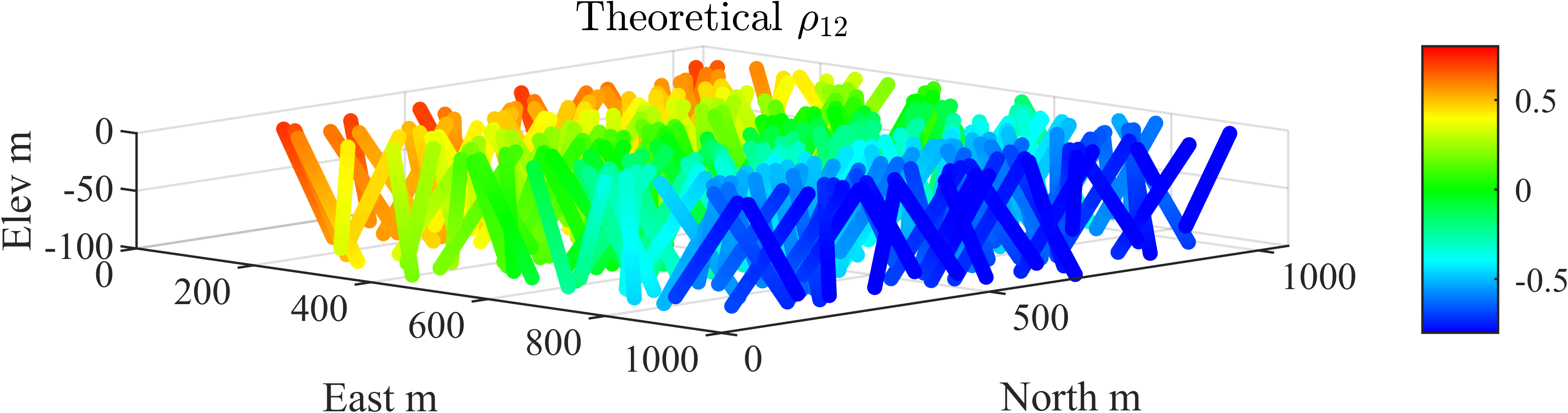}\\
	\centering
	\includegraphics[width=0.45\textwidth,trim={0 0cm 0.0cm 0cm},clip]{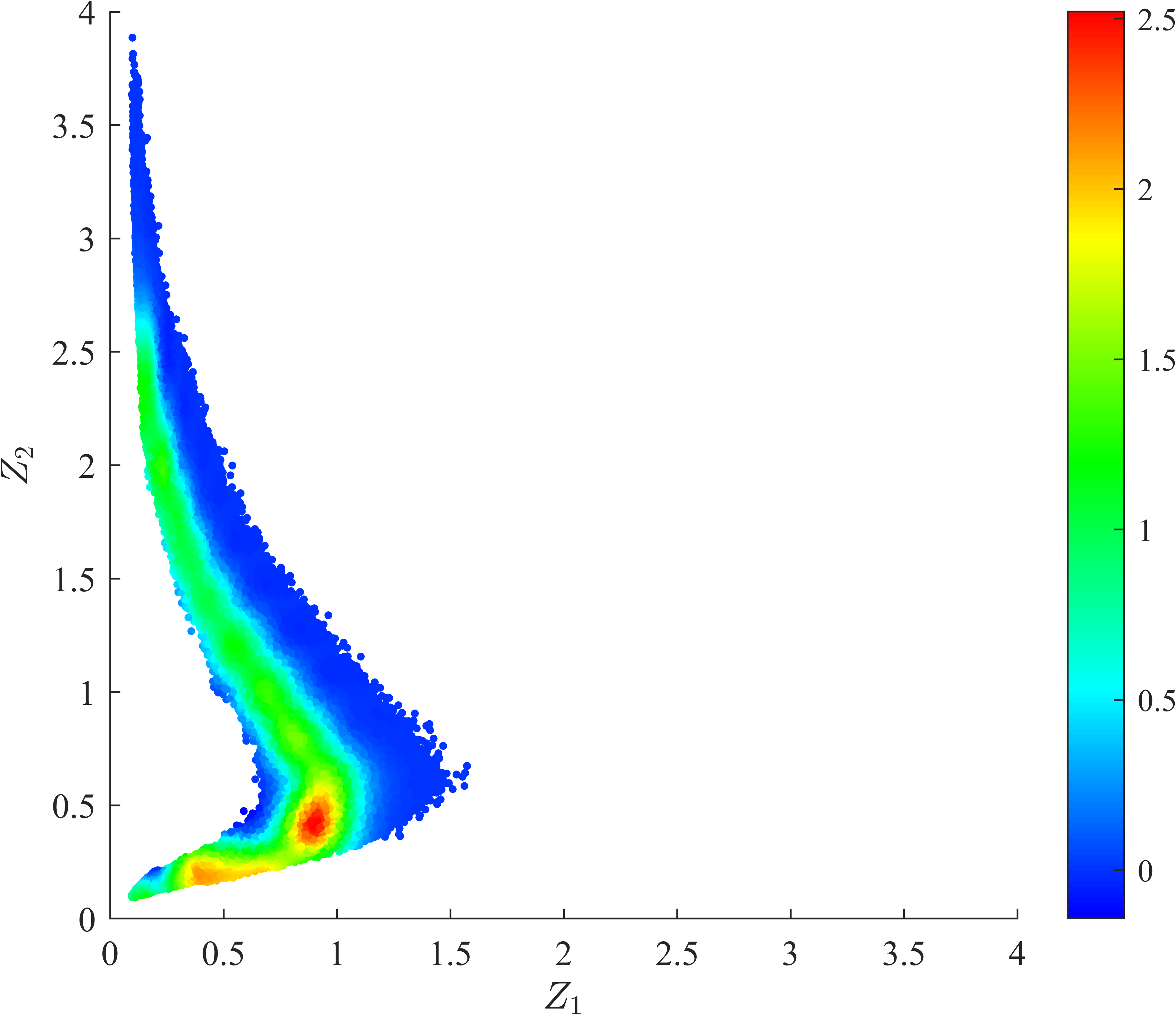}
	\caption{{Set-up of the synthetic experiment for the inference of the correlation at sampled locations: Correlated fields in original values (top),  correlation imposed on the Gaussian values and synthetic drillholes (middle), and non-linearity obtained in the attribute space (bottom).}}
	\label{exp1}
\end{figure}

We build a synthetic case study by starting with an independent vector RF $ \tilde{\textbf{Y}} $ and a correlation field $ {\textbf{C}}(\textbf{u}) $. A global non-linear vector RF $ {\textbf{Z}} $ is then constructed considering the following sequence of transformations
$$ \tilde{\textbf{Y}} \rightarrow \textbf{Y} \rightarrow \textbf{Z},$$
as follows. We take, for simplicity, two correlated Gaussian variables,  $ \tilde{\textbf{Y}}= [\tilde{Y}_1(\textbf{u}),  \tilde{Y}_2(\textbf{u})]^T $, of zero mean and unit variance, that is, $ p=2 $. We simulate one realization of $ \tilde{\textbf{Y}} \sim \N(\textbf{0},\textbf{I}_2) $ with both factors following  an exponential variogram of range 50 m on a $ 1000 $ m $ \times 1000 $ m $\times  100 $ m domain, and generating drillholes that sample that RF with a $ 50 $ m spacing approximately, randomizing the azimuth and dip direction of the synthetic drillholes, resulting in $ 14.646 $ samples. Then applying  $\textbf{Y}(\textbf{u}) = \textbf{L}(\textbf{u})\tilde{\textbf{Y}}(\textbf{u}) \sim \N\big(\textbf{0},\textbf{C}(\textbf{u})\big)$, with $ \textbf{L} $ the lower triangular matrix resulting from Cholesky decomposition, $\textbf{C}=\textbf{L}\textbf{L}^T$. We let $ (\textbf{C})_{12}= \rho_{12}(\textbf{u}) $  varying linearly with east-coordinate from +0.9 to -0.9. A last step is to take the exponential transformation $$ \textbf{Z}(\textbf{u})=\exp\big(\varvec{\mu}(\textbf{u})+\sigma\cdot\textbf{Y}(\textbf{u})\big),$$
with $ \varvec{\mu}(\textbf{u}) $ a mean vector constant with the north-coordinate but varying differently for each factor, and $ \sigma $ a constant value to build our synthetic deposit. The first component of $ \varvec{\mu}(\textbf{u}) $ attains a maximum at the middle of the deposit and the second component grows linearly with the east-coordinate. The RF $ {\textbf{Z}} $ is sampled in a drill-hole campaing fashion, obtaining the sampling data $ {{{\textbf{\textit{z}}}}}_\alpha = [{z}_1(\textbf{u}_\alpha),{z}_2(\textbf{u}_\alpha)]^T $, $ \alpha \in \{1,\dots,N\} $. The situation is depicted in Fig. \ref{exp1}, showing the non-linearity obtained in the attribute space.

We proceed to infer the original independent values of $ \tilde{\textbf{Y}} $ at sampled locations, compare with the initial values and assess if spatial behavior is preserved. A neighborhood is set by iteration and having under consideration, first, to obtain independent factors and, second, not to exceed the search beyond the range of the variogram of these factors. This considerations, together with allowing enough data to apply normal-score transformation locally, are also taken later in the real case study for defining the vicinity $ \mathbbcal{V} $. We search for the closes 300 samples at each location.

\begin{figure}[h!]
	\raggedright
	\includegraphics[width=0.49\textwidth,trim={0 0cm 0.0cm 0cm},clip]{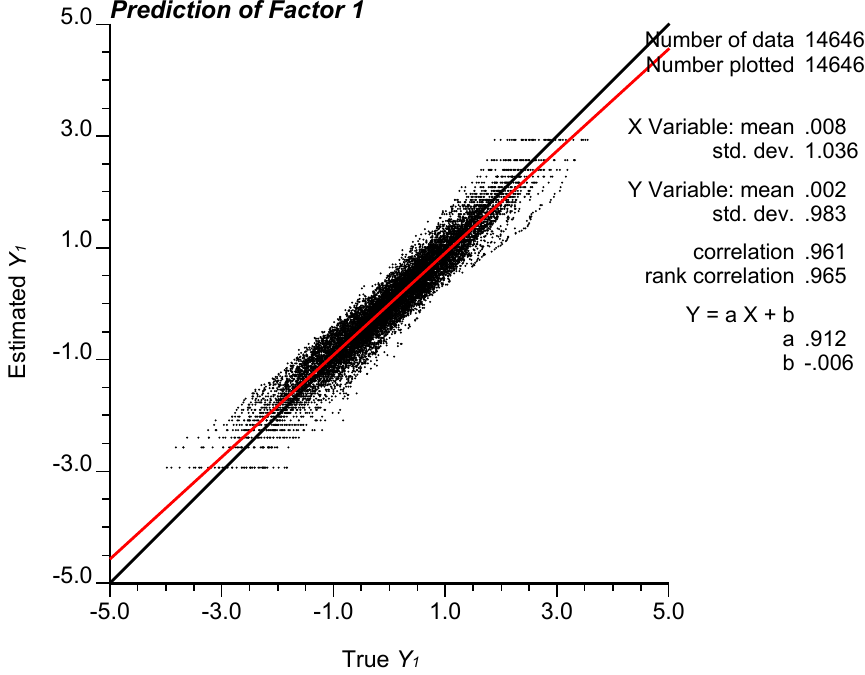}
	\includegraphics[width=0.49\textwidth,trim={0 0cm 0.0cm 0cm},clip]{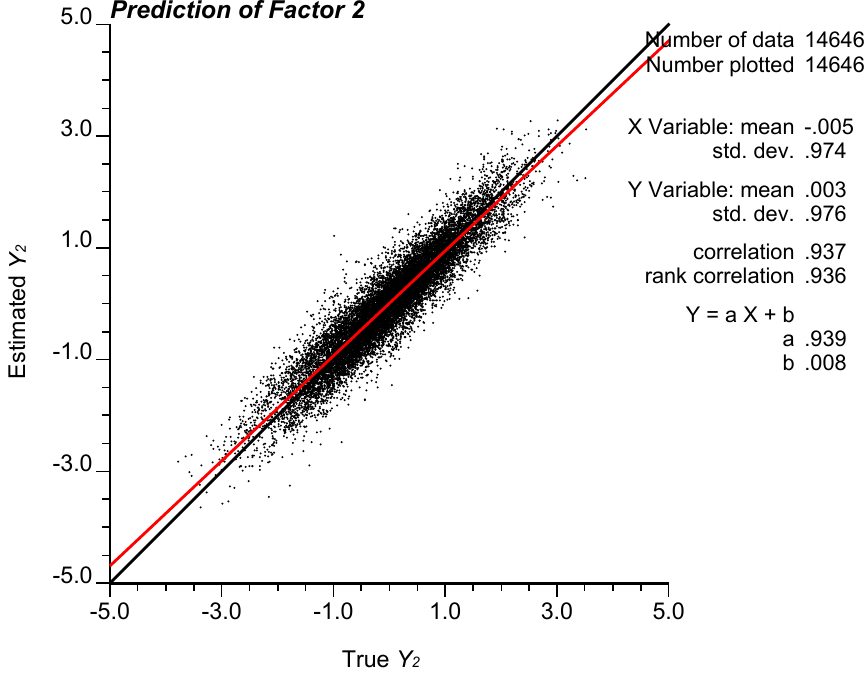}\\
	\includegraphics[width=0.49\textwidth,trim={0 0cm 0.0cm 0cm},clip]{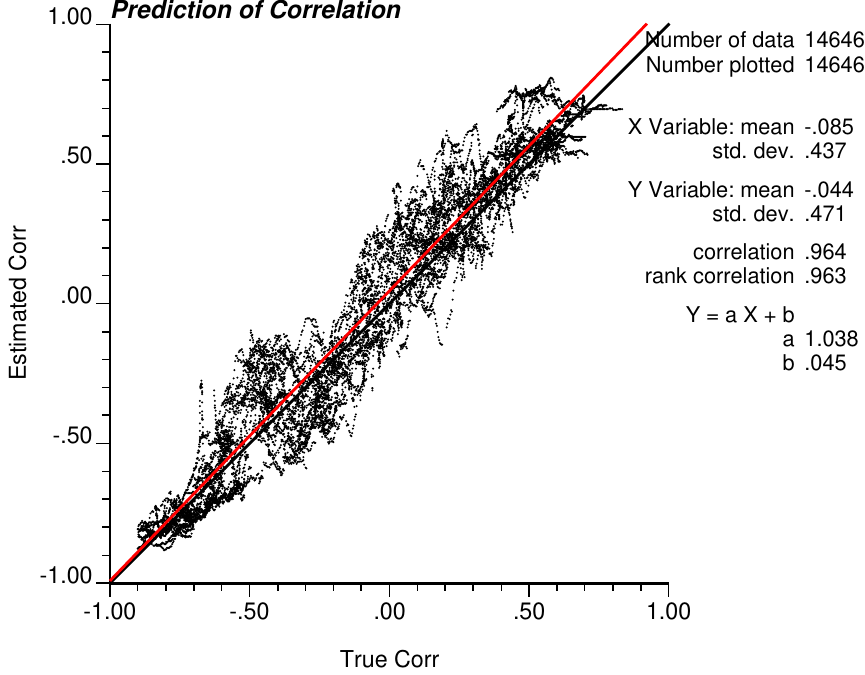}
	\includegraphics[width=0.49\textwidth,trim={0 0cm 0.0cm 0cm},clip]{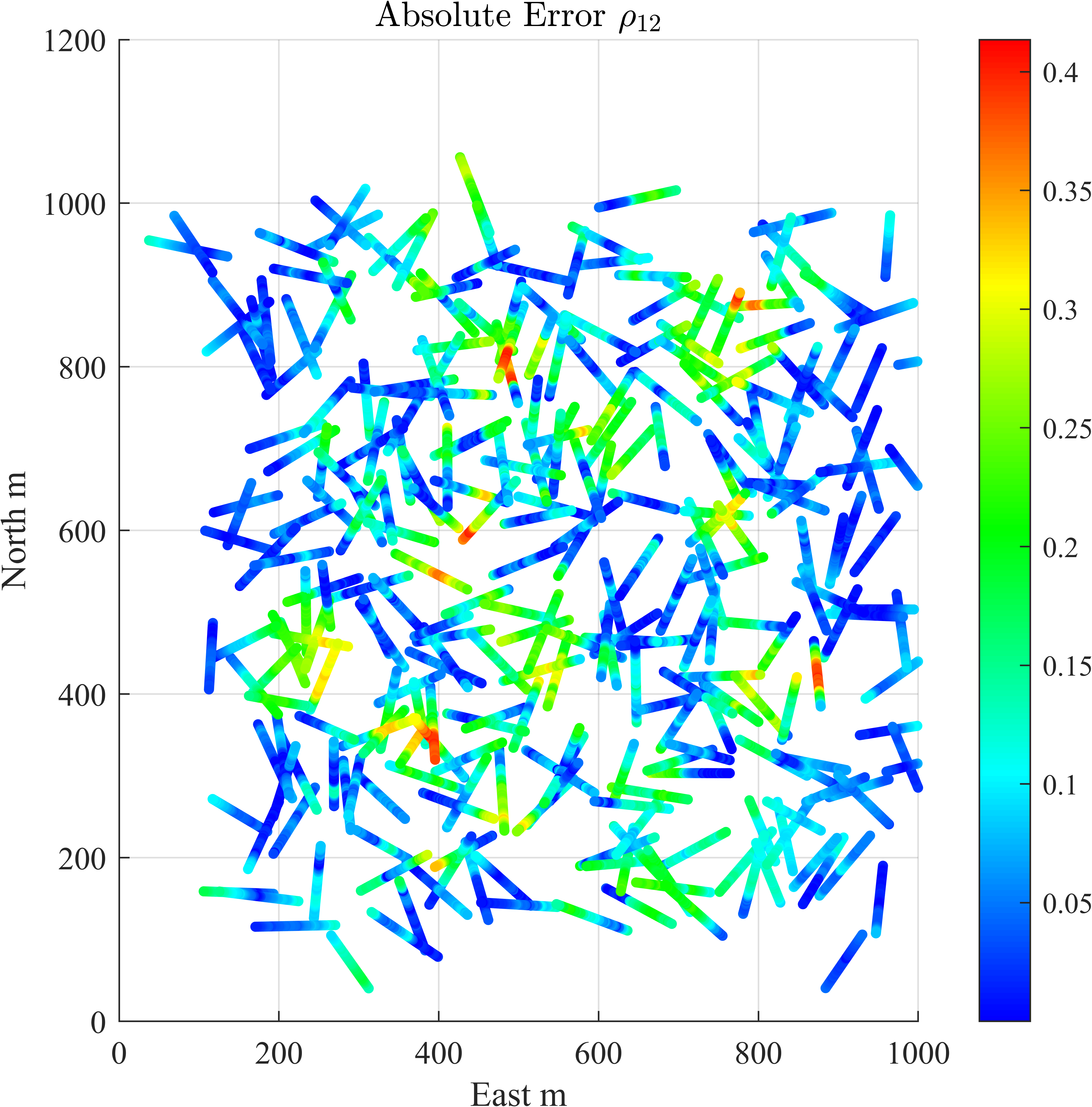}\\
	\caption{{Results of the synthetic experiment: prediction of independent factors versus true values (top), and inferred correlation versus the imposed at sampled location (bottom left). Absolute errors at sampled locations are shown in the location map (bottom right).}}
	\label{exp3}
\end{figure}

Results (Fig. \ref{exp3}) show, in scatter-plots, the comparison of the inferred values versus the original values, for the independent factors, with high accuracy. Factor 1 scatter shows artifacts due to the nature of normal-score transformation \citep{deutsch1998gslib} and the fixed neighborhood of samples selected (the same quantiles of the standard normal CDF are considered at each location). Factor 2 do not show the same artifact due to the nature of Cholesky decompositon for correlation matrices and the matrix $ \textbf{L} $ and $ \textbf{L}^{-1} $  (the upper-left corner of $ \textbf{L} $ is 1, which may not be the case for the rest of the entries of $ \textbf{L} $). A location-map showing the absolute error of the inferred correlation at sample location is shown, indicating that the error is homegenous on the domain and apparently not related with the mean of $ \textbf{Z} $ nor $ \varvec{\mu}(\textbf{u}) $. Differences between the imposed correlation and the inferred one are not critical for invalidating the proposed methodology, as shown on the scatter-plot.

\begin{figure}[h!]
	\centering
	\includegraphics[width=0.89\textwidth,trim={0 0cm 0.0cm 0cm},clip]{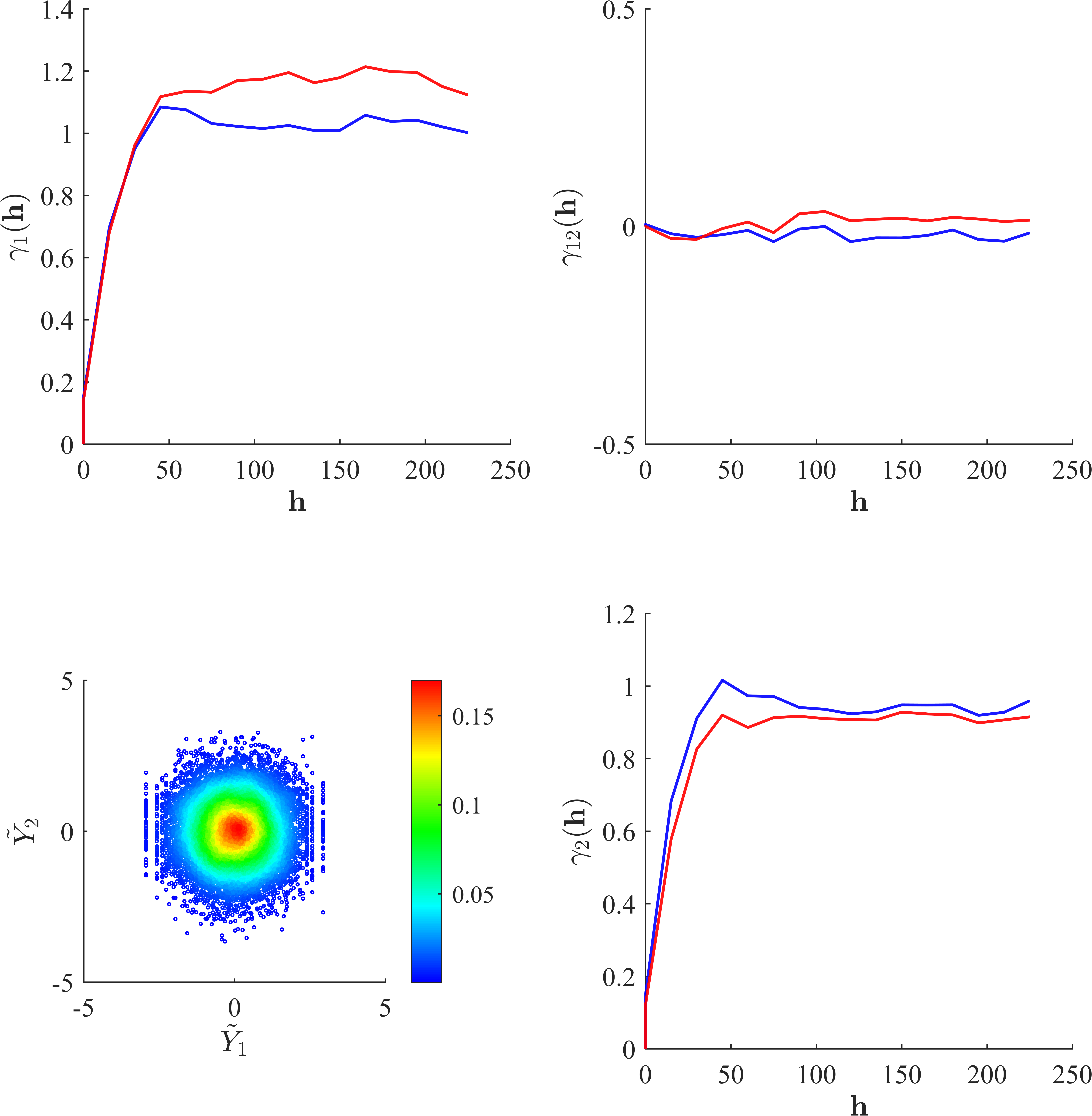}
	\caption{{Results of the synthetic experiment: comparison between realization direct- and cross- variograms (red) and the retrieved from inference of independent components (blue), at sampled locations. Scatter between the inferred factors shows independent Gaussian behavior after local transformation.}}
	\label{exp4}
\end{figure}

Finally, a comparison among the variogram of original values and the inferred  independent factors are shown on Fig. \ref{exp4}, demonstrating that the proposed method is able to retrieve the underlying variogram structures and  correlations among variables.

\section{Real Case Study}
\label{Case Study}

\subsection{The Data}

In order to demonstrate the application of the methodology, a data set from a blast hole campaign of a Nickel-Laterite deposit is considered and six cross-correlated variables  isotopically assayed at each sample point \citep{wackernagel2013multivariate}: Fe, Ni, MgO, SiO$_2$, Al$_2$O$_3$, and Cr. The case study includes 9990 samples available with a very dense sampling pattern. 

\begin{figure}[h!]
	\centering
	\includegraphics[width=0.45\textwidth,trim={0 0cm 0.0cm 0cm},clip]{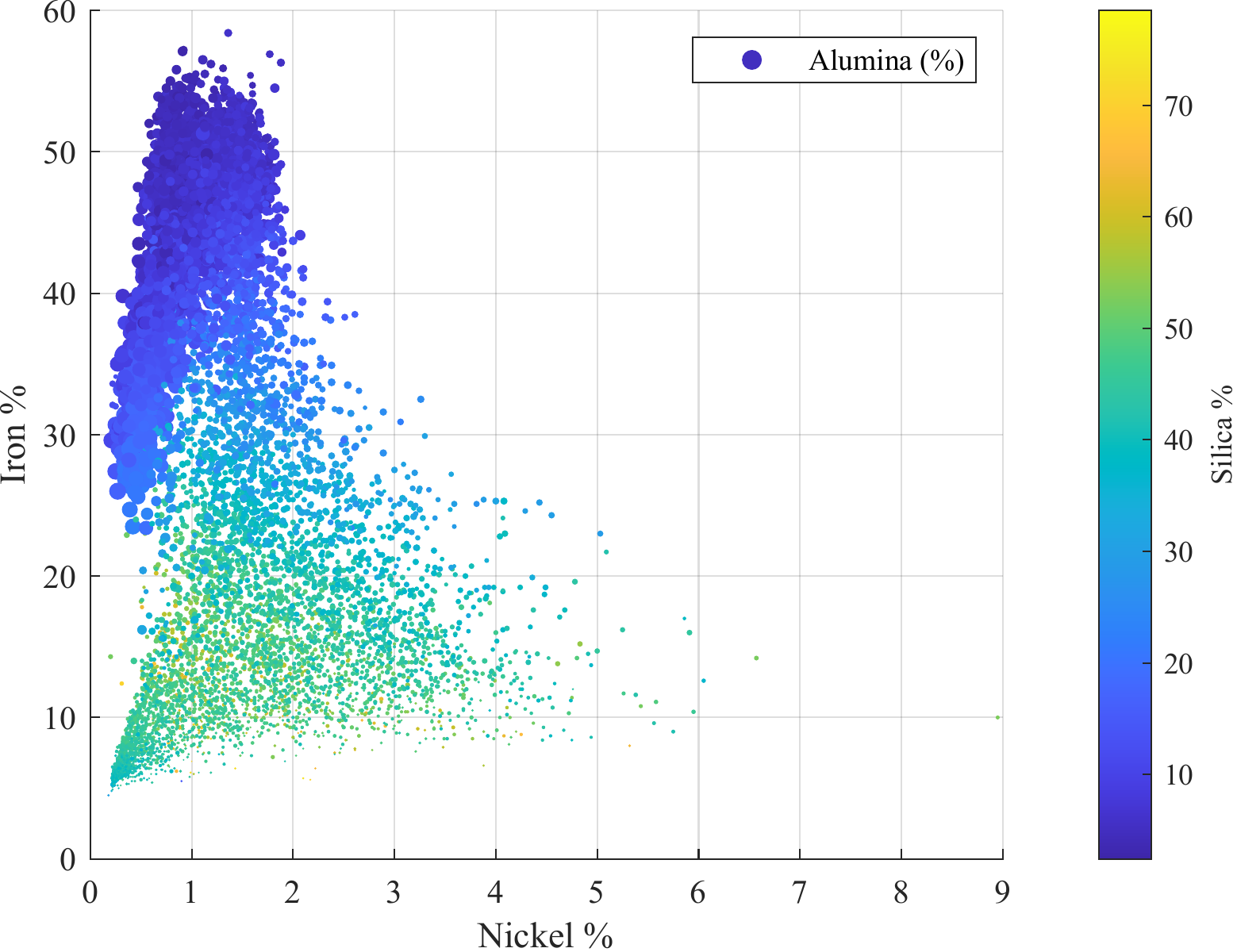}\,
	\includegraphics[width=0.45\textwidth,trim={0 0cm 0.0cm 0cm},clip]{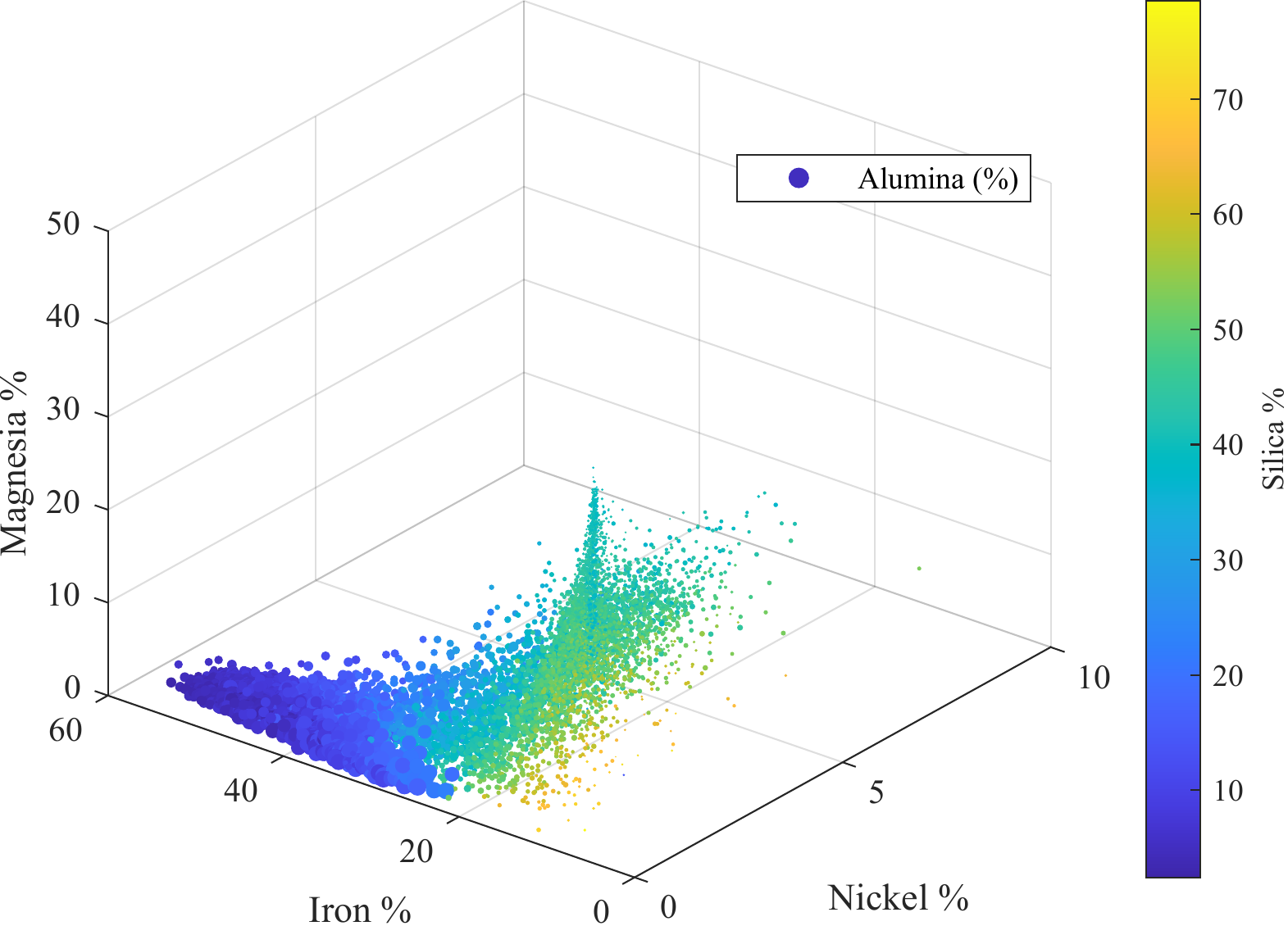}
	\includegraphics[width=0.45\textwidth,trim={0 0cm 0.0cm 0cm},clip]{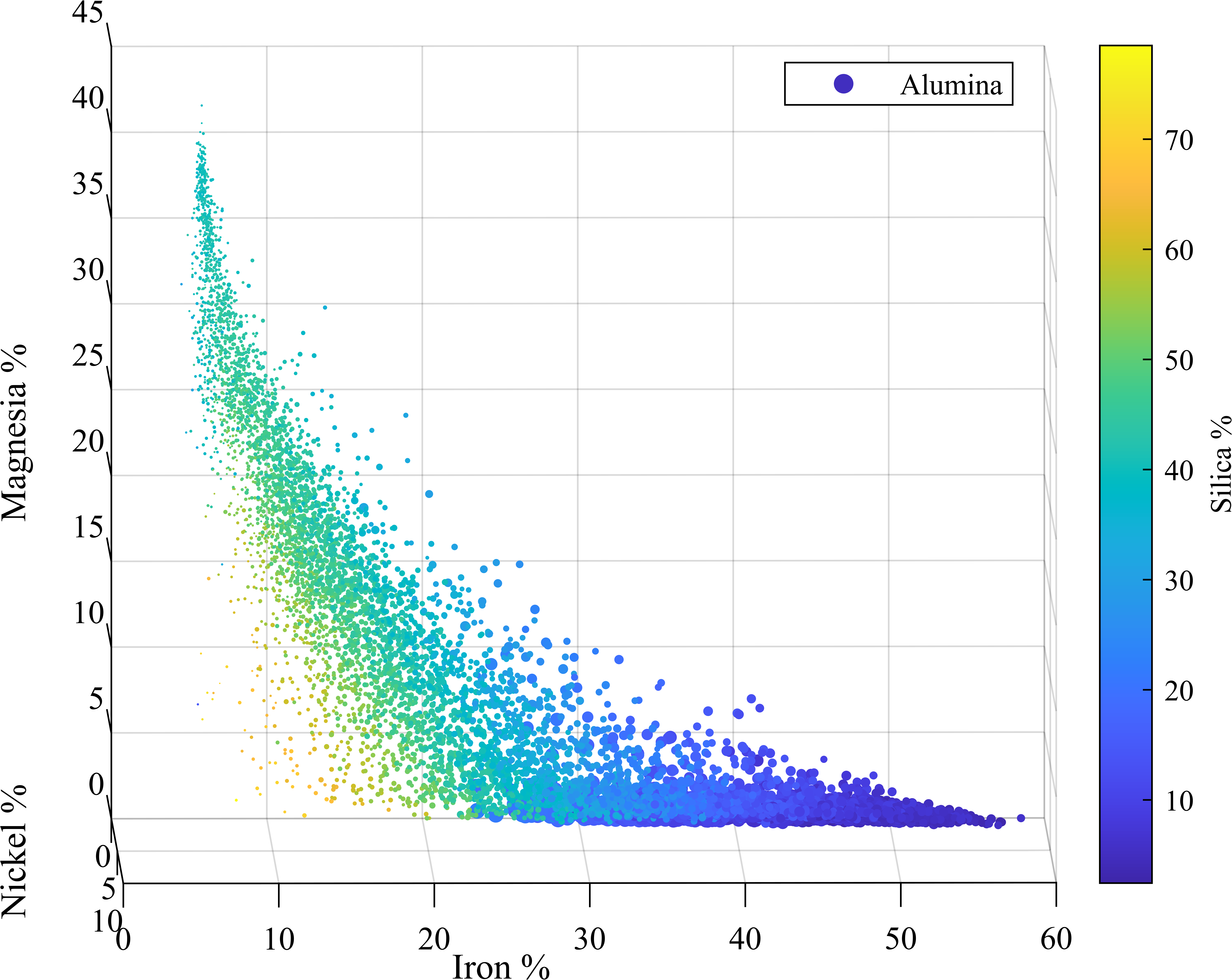}
	\caption{Display of multivariate features on sampling data of Nickel-Laterite Deposit. 5 out of 6 variables can be seen on the scatter plots, by adding color and a variable diameter to the bullets, proportional to the amount of alumina.}
	\label{loc2}
\end{figure}

\begin{figure}[h!]
	\centering
	\includegraphics[width=0.49\textwidth,trim={0 0cm 0.0cm 0cm},clip]{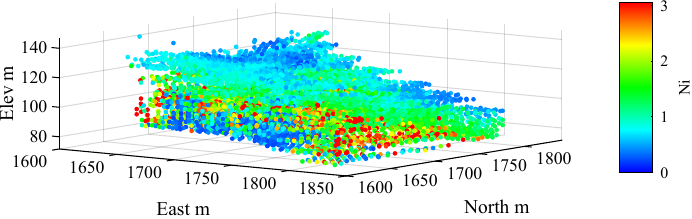}
	\includegraphics[width=0.49\textwidth,trim={0 0cm 0.0cm 0cm},clip]{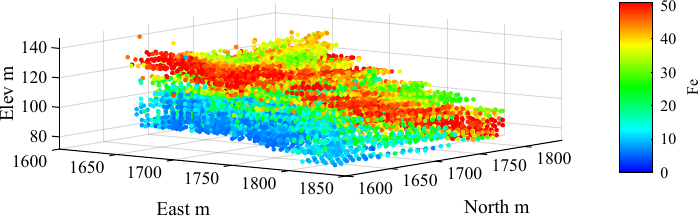}
	\includegraphics[width=0.49\textwidth,trim={0 0cm 0.0cm 0cm},clip]{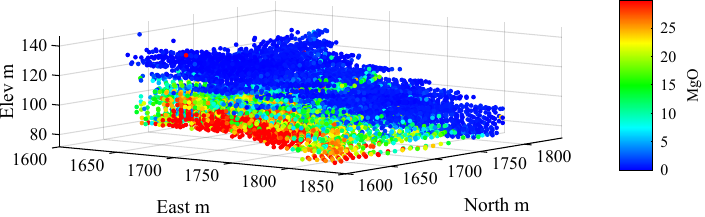}
	\includegraphics[width=0.49\textwidth,trim={0 0cm 0.0cm 0cm},clip]{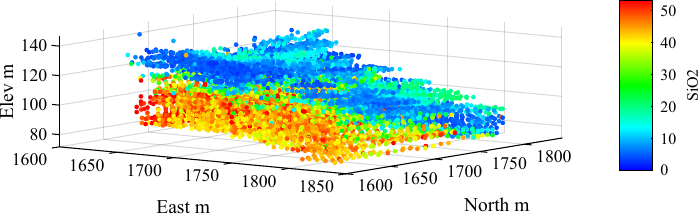}
	\includegraphics[width=0.49\textwidth,trim={0 0cm 0.0cm 0cm},clip]{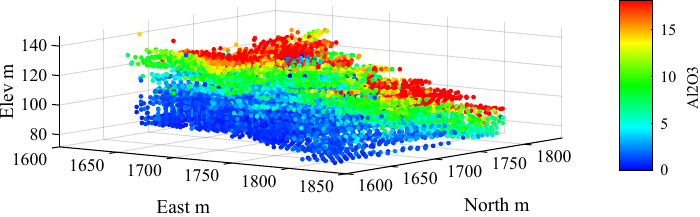}
	\includegraphics[width=0.49\textwidth,trim={0 0cm 0.0cm 0cm},clip]{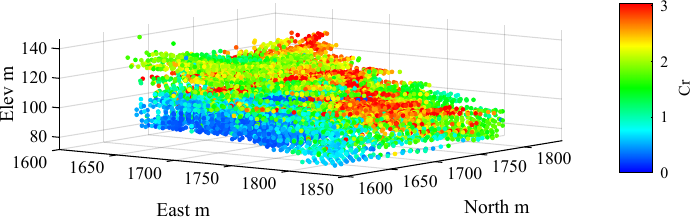}
	\caption{Isometric view showing the sampling grade information.}
	\label{loc14}
\end{figure}

A primary inspection of multivariate relations (scatter-plots shown in Fig. \ref{loc2}) exposes many aspects of complexity such as non-linearity and heteroscedasticity. A map of the samples for each variable is presented in Fig. \ref{loc14}. In order to show the predictability of the proposed methodology, 30\% of samples (2997) are randomly selected and removed for testing purposes, leaving 6993 samples for analysis.

\subsection{Variography}

\begin{figure}[h!]
	\centering
	\includegraphics[width=0.65\textwidth,trim={0 0cm 0.0cm 0cm},clip]{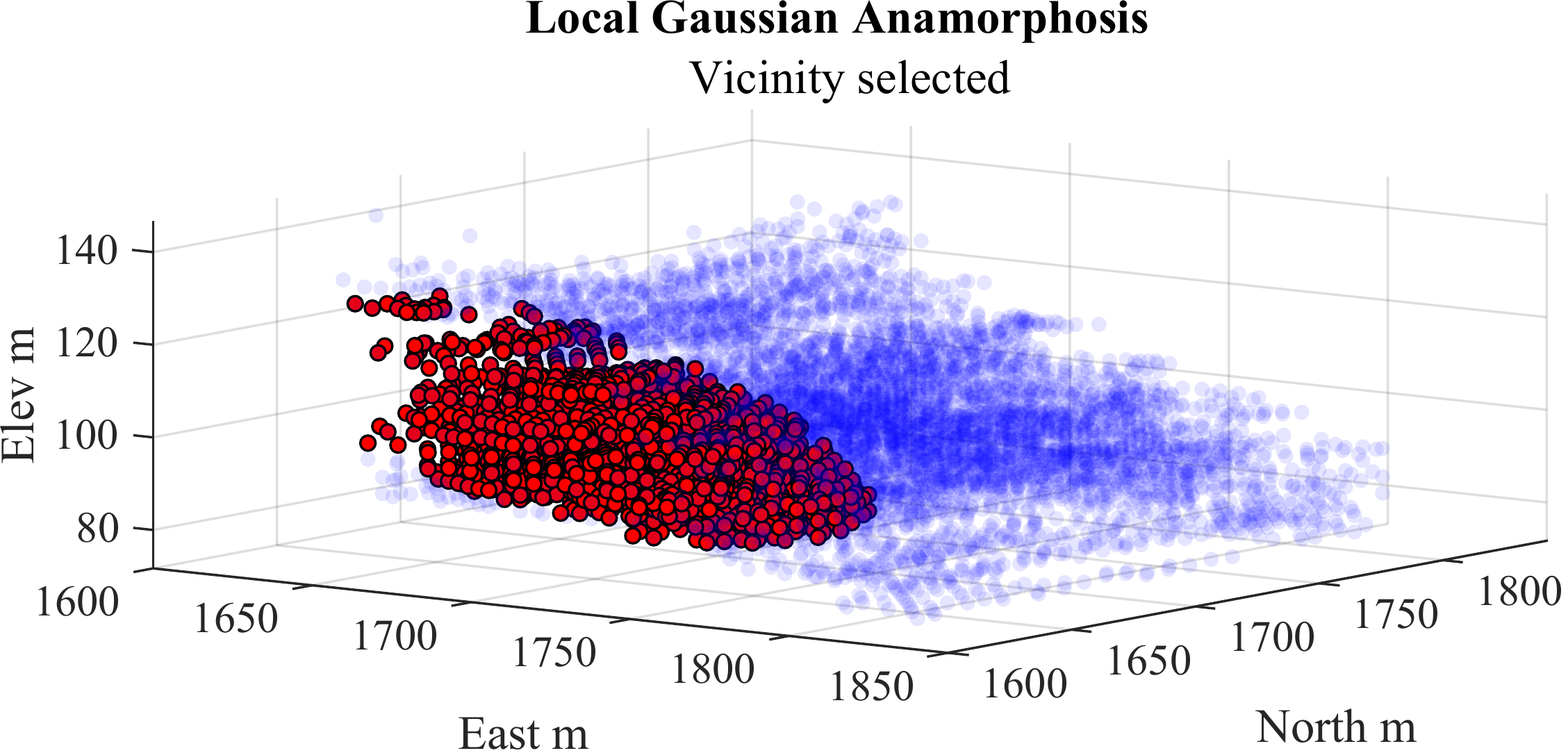}\\
	\includegraphics[width=0.45\textwidth,trim={0 0cm 0.0cm 0cm},clip]{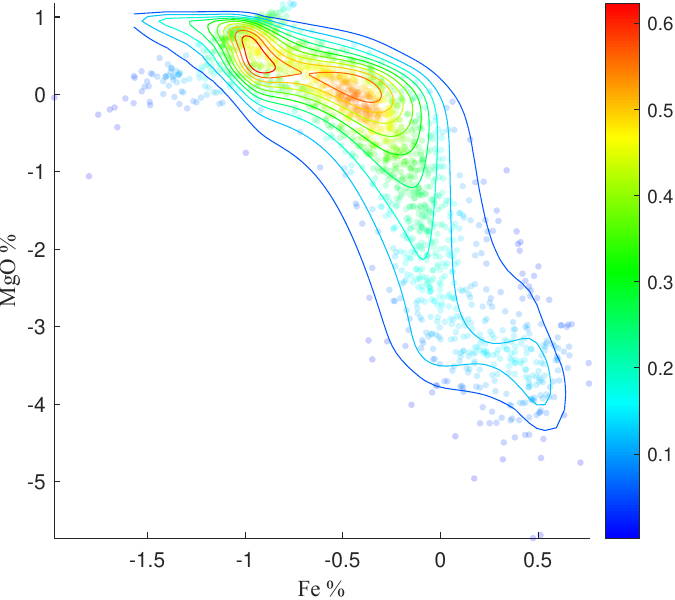}
	\includegraphics[width=0.45\textwidth,trim={0 0cm 0.0cm 0cm},clip]{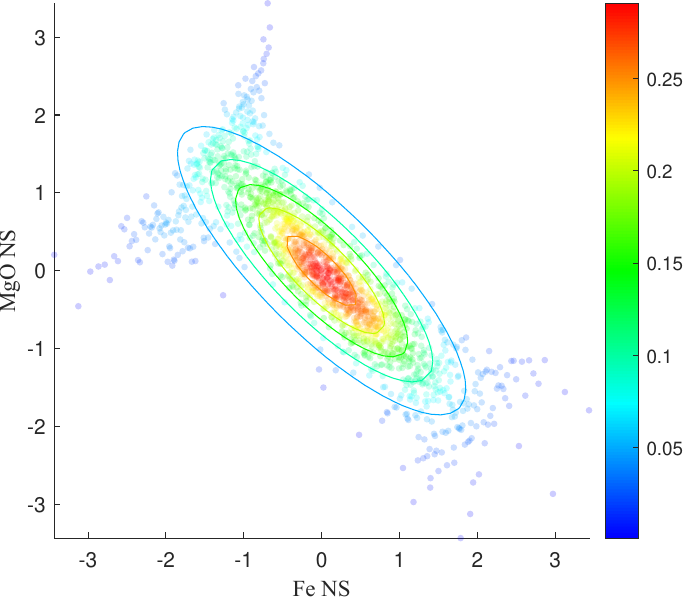}
	\caption{{Vicinity selected for the inference of correlation (800 samples) and the PDF adjustment for original log-ratio values based on the inference of correlation of Gaussian values.}}
	\label{nataf_local}
\end{figure}

We apply the additive log-ratio transformation to the data, taken with respect to the Rest variable (Rest$ = 100$ \% $ - $ Ni\% $-\dots-$ Cr\%). Gaussian transformation is applied at each sample location, by selecting a neighborhood of the closest $  n_\alpha=800 $ samples with isotropic search. This parameter was calibrated, showing that working with less data reduces the capabilities for  reproduction of the multivariate behavior drastically, as the correlation matrix gets distorted with a lower amount of data. In Fig. \ref{nataf_local}, one of the neighborhoods used for applying the local Gaussianization is displayed, indicating that local linearity assumption is a good approximation for retrieving the non-linear behavior in original values, as shown by the local scatter-plot. We note that tail values are still being captured through the univariate anamorphosis $ \phi_i $, but showing a low multivariate fitting.

Once the data is gaussianized and decorrelated after obtaining the correlation matrix, the experimental direct and cross omni-directional variograms are calculated. Variogram analysis in different directions was not considered as the amount of data in the vertical direction is less than horizontally. This aspect is accounted for later when defining the search radii for estimation.

\begin{figure}[h!]
	\centering
	\includegraphics[width=0.95\textwidth,trim={0 0cm 0.0cm 0cm},clip]{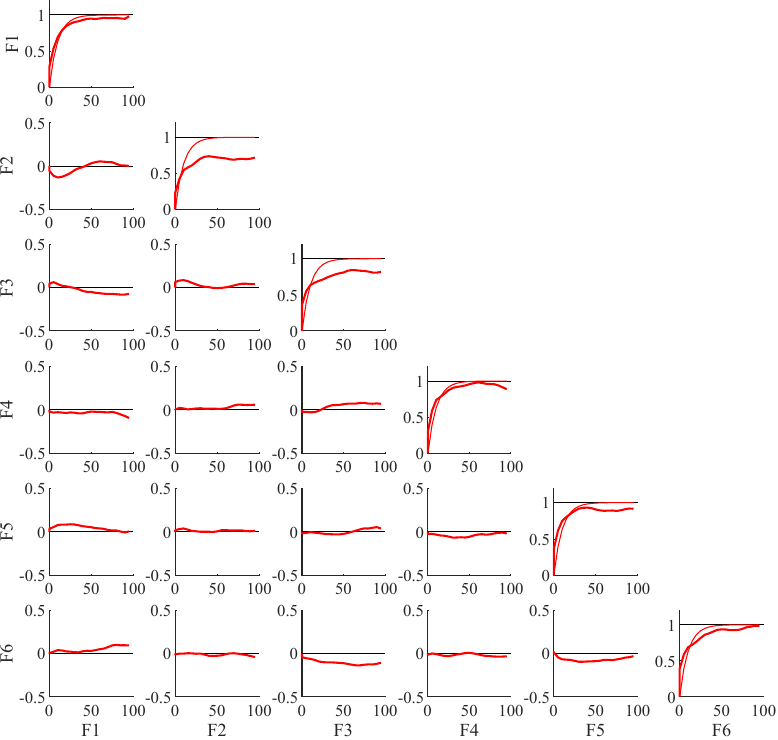}
	\caption{Experimental direct and cross variograms of the gaussian factors, and the final model used.}
	\label{variogram}
\end{figure}

Variogram analysis and calibration is the weakest point of the methodology. A first complication is that decorrelation breaks the marginal gaussianity on the factors $ \tilde{\textbf{Y}}=\textbf{L}^{-1}\textbf{Y} $, suggesting that the assumption of multi-gaussianity on $ \textbf{Y} $ is not a perfect model at every location (as shown in  Fig. \ref{nataf_local}). As a consequence, the experimental variances of the factors $ \textbf{Y} $ do not reach the value of 1. This fact can be seen on the sill of the experimental  variograms in Fig. \ref{variogram}. However, a single exponential variogram with 10 m of range is fit as the final model, matching relatively well most of the direct variograms. Cross variograms show low correlation among variables, as expected.

\subsection{Results}

\begin{figure}[h!]
	\centering
	\includegraphics[width=0.95\textwidth,trim={0 0cm 0.0cm 0cm},clip]{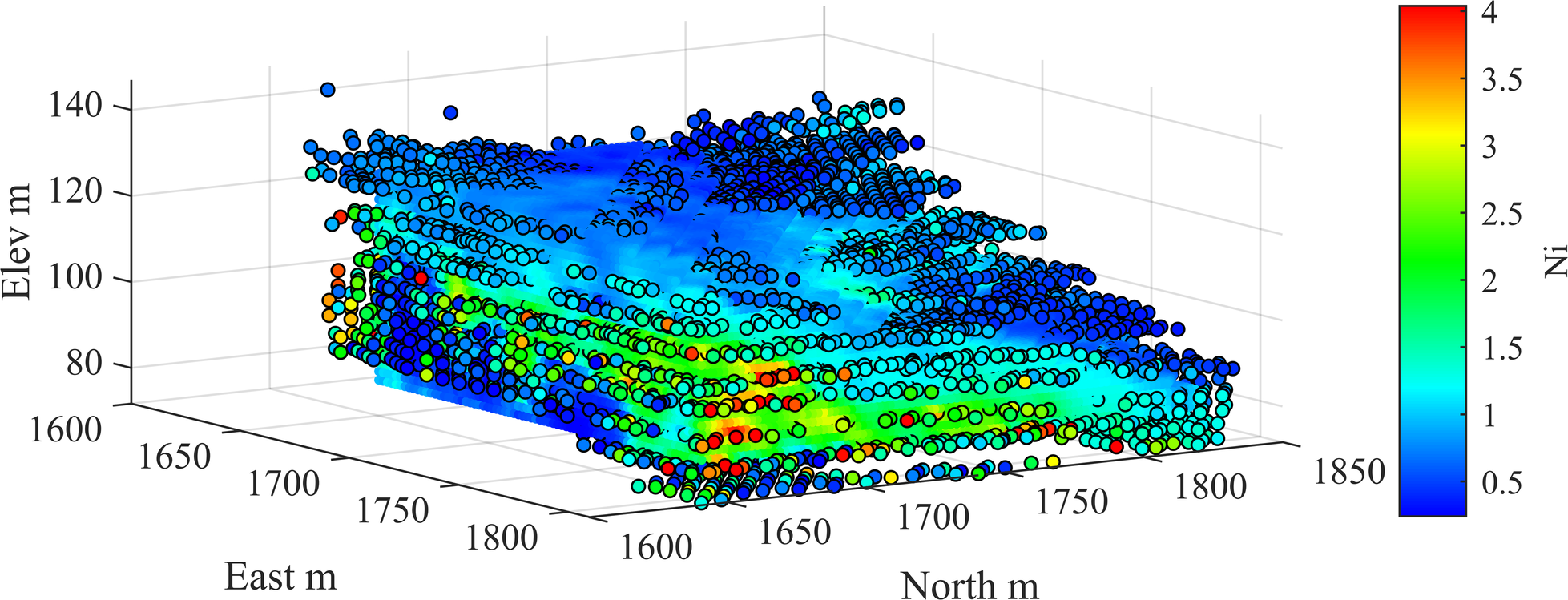}
	\includegraphics[width=0.95\textwidth,trim={0 0cm 0.0cm 0cm},clip]{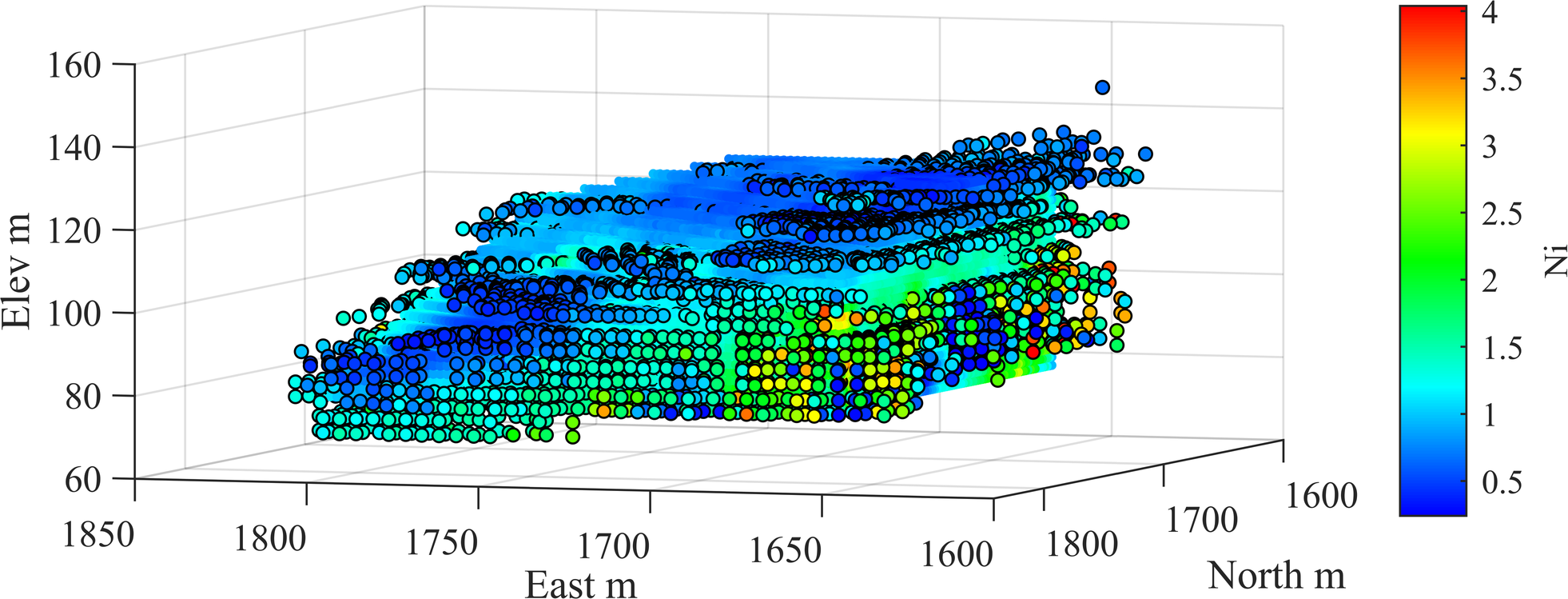}
	\caption{Two viewpoints of the grid used in the case study, showing the estimated mean and sampling data, for the Nickel case.}
	\label{grid}
\end{figure}

Once the single variogram model is derived, each factor can be treated separately. An initial grid with node spacing of 2×2×2 (in meters) and 75, 90 and 25 nodes along East, North, and elevation coordinates, respectively, is considered. We generate 1000 simulations by using turning bands (\citealt{emery2006tbsim}; \citealt{Chiles};  \citealt{marcotte2016spatial}) (1200 directions used). A moving neighborhood is used with a search neighborhood of 100 m with up
to 25 samples and without considering a requirement for samples per octant. The simulated factors are correlated according to the estimated correlation, which is interpolated by ordinary kriging at each location of the grid, and using the same variogram model used for the factors. Results are then back-transformed from gaussian values and from log-ratios into the raw distribution. The mean of the simulations in the Nickel case, excluding the nodes far from sample data, is shown in Fig. \ref{grid}.

\begin{figure}[h!]
	\centering
	\includegraphics[width=0.9\textwidth,trim={0 0cm 0.0cm 0cm},clip]{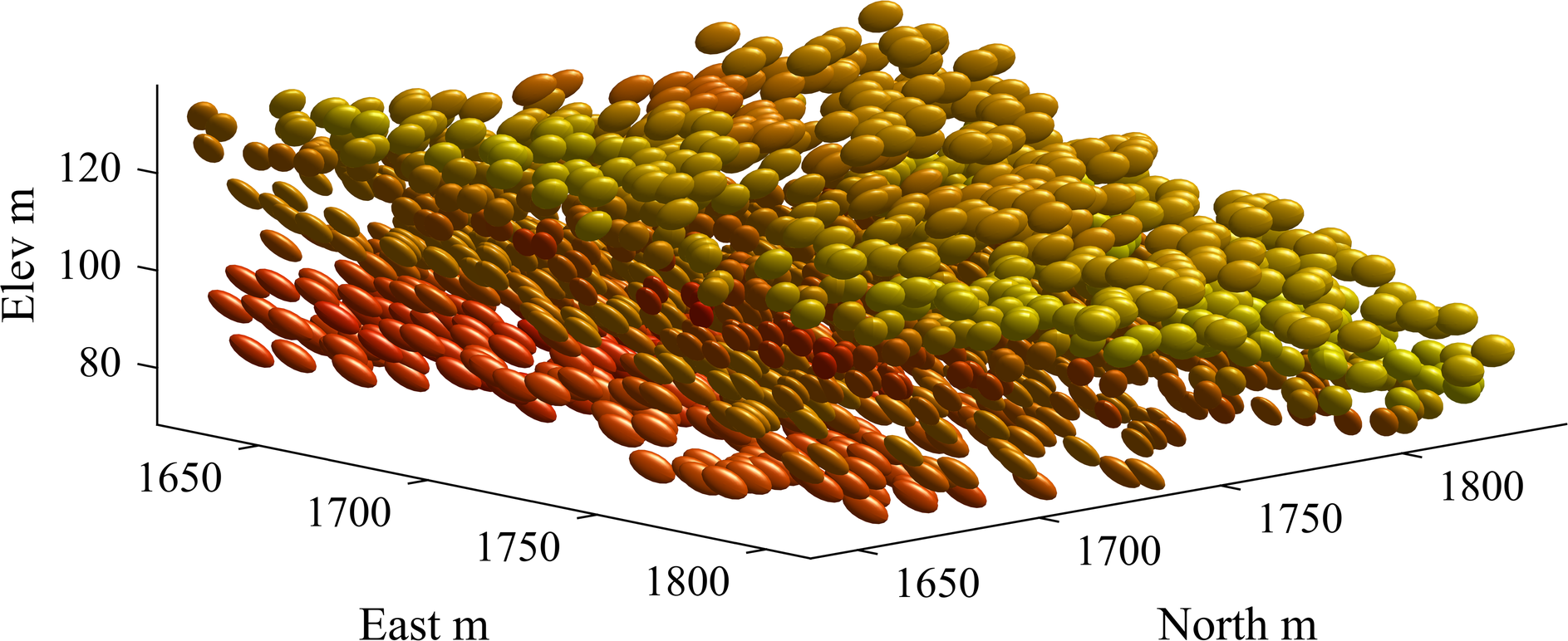}
	\includegraphics[width=0.9\textwidth,trim={0 0cm 0.0cm 0cm},clip]{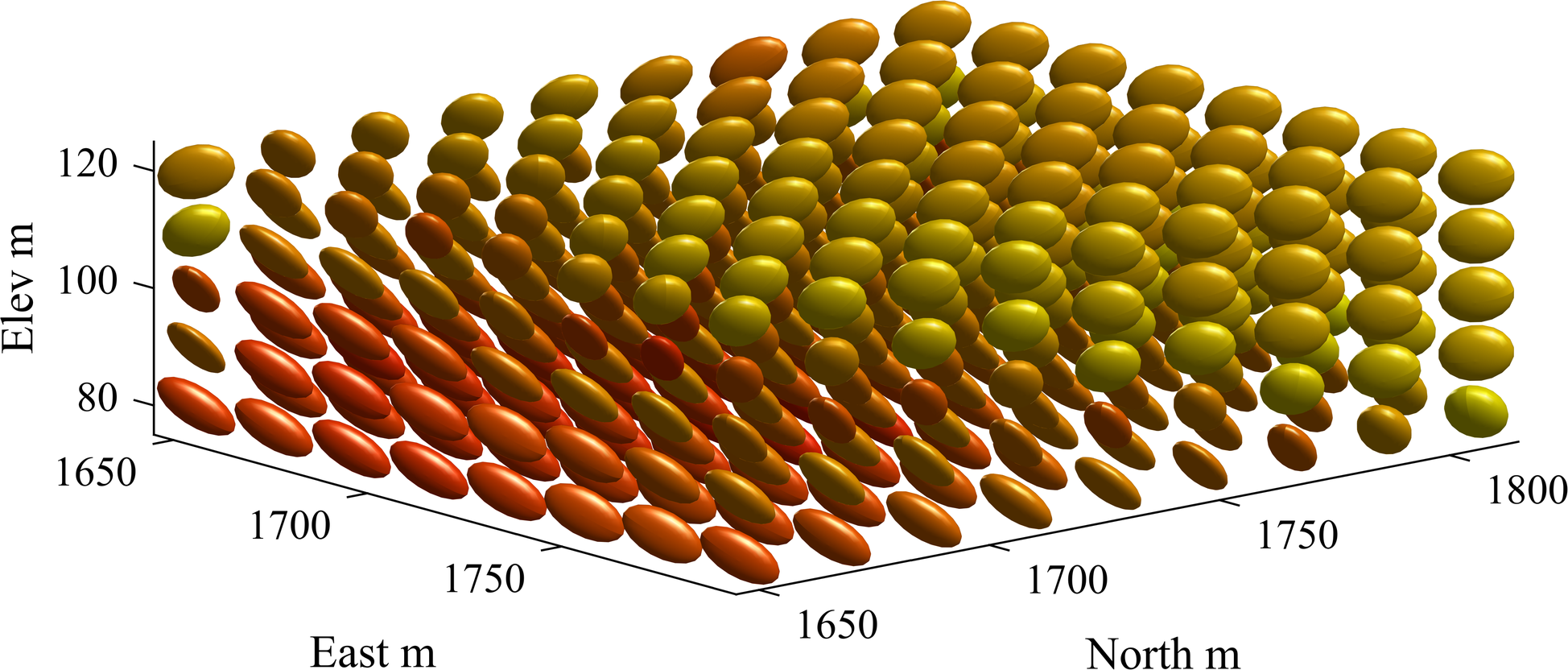}\\
	\includegraphics[width=0.15\textwidth,trim={0 0cm 0.0cm 0cm},clip]{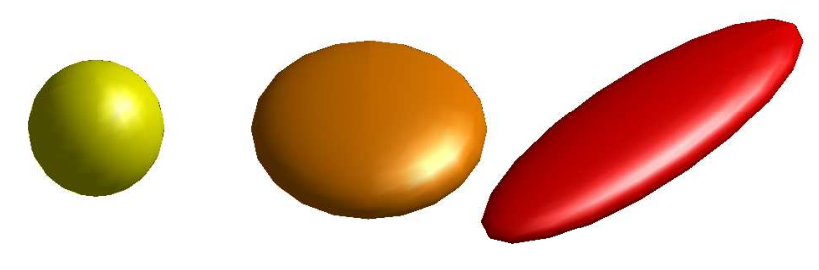}
	\caption{Estimated correlation matrices among independent factors 1, 2 and 3 at sample positions, represented by ellipses (left). Interpolation of correlation matrices on a regular grid (right).The color of ellipsoids is related to their anisotropy (bottom). From left to right: isotropic tensor, planar tensor (flat ellipsoid) ($ \lambda_1 \simeq \lambda_2 > \lambda_3 $), elongated ellipsoid ($ \lambda_1 \gg \lambda_2 \geq \lambda_3 $)}
	\label{gridelli}
\end{figure}

\begin{figure}[h!]
	\raggedright
	\includegraphics[width=0.32\textwidth,trim={0 0cm 0.0cm 0cm},clip]{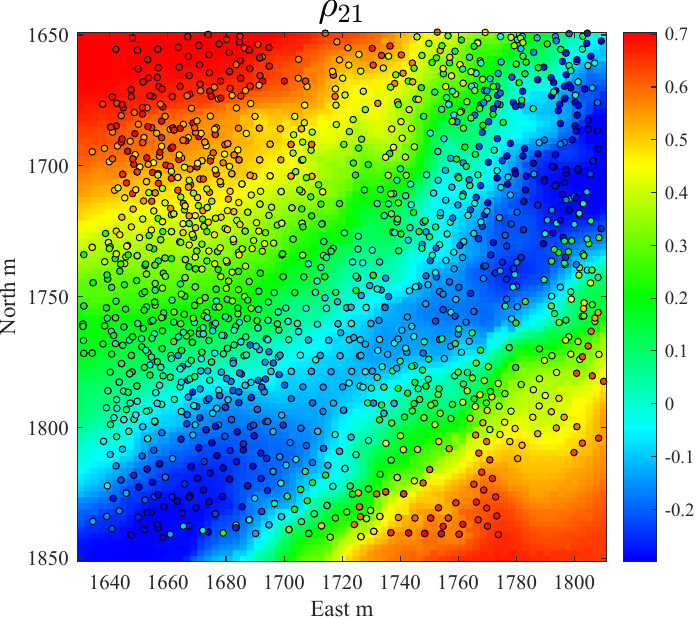}\\
	\includegraphics[width=0.32\textwidth,trim={0 0cm 0.0cm 0cm},clip]{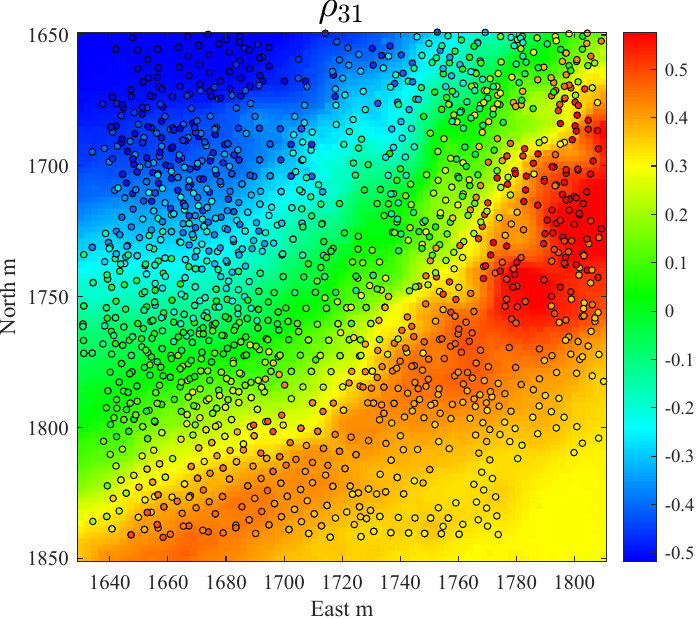}
	\includegraphics[width=0.32\textwidth,trim={0 0cm 0.0cm 0cm},clip]{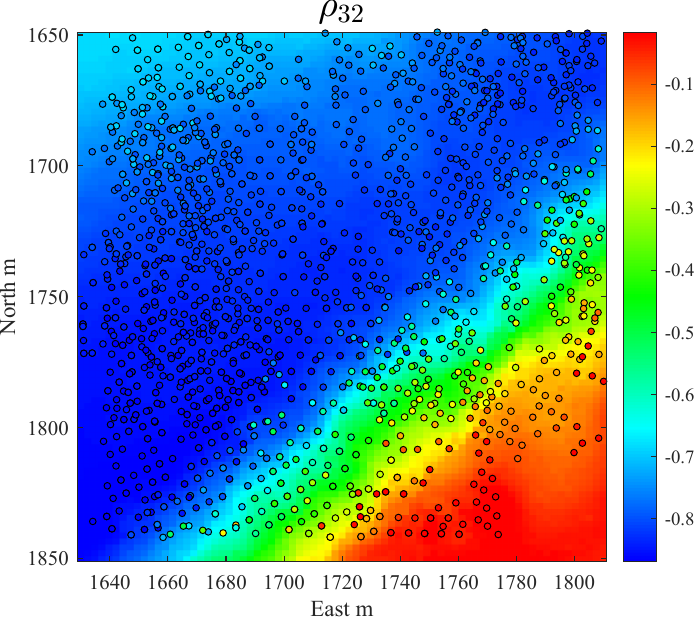}\\
	\includegraphics[width=0.32\textwidth,trim={0 0cm 0.0cm 0cm},clip]{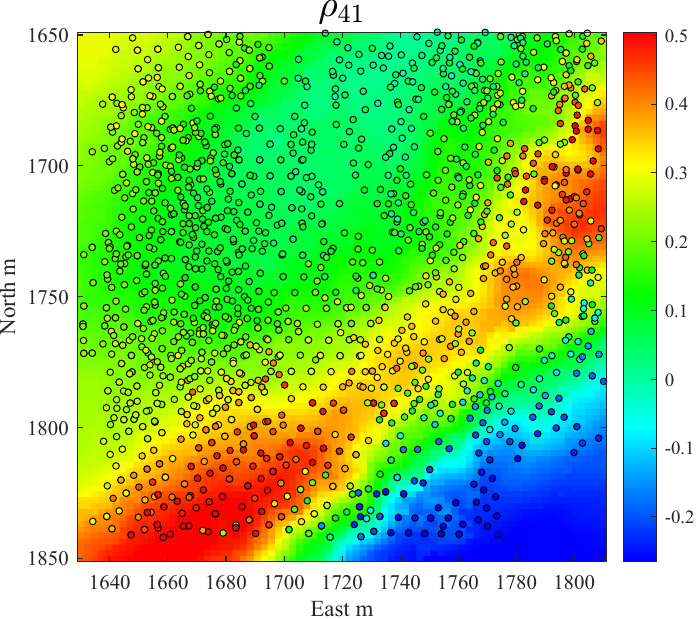}
	\includegraphics[width=0.32\textwidth,trim={0 0cm 0.0cm 0cm},clip]{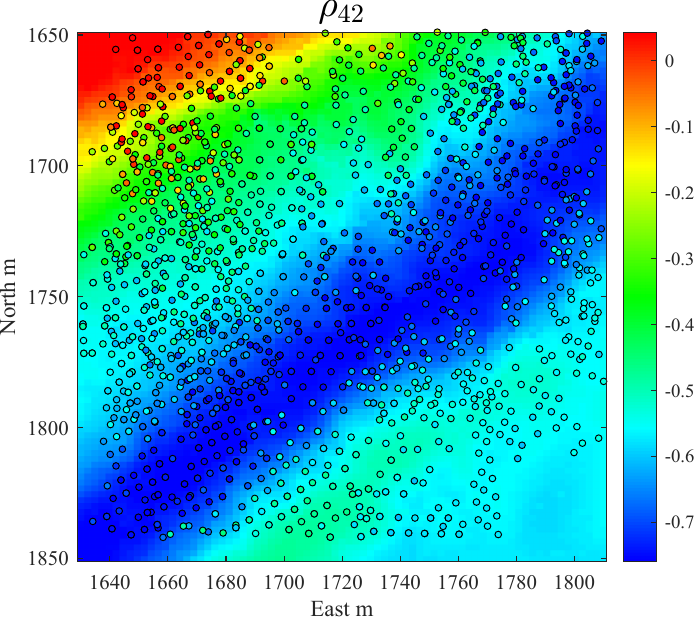}
	\includegraphics[width=0.32\textwidth,trim={0 0cm 0.0cm 0cm},clip]{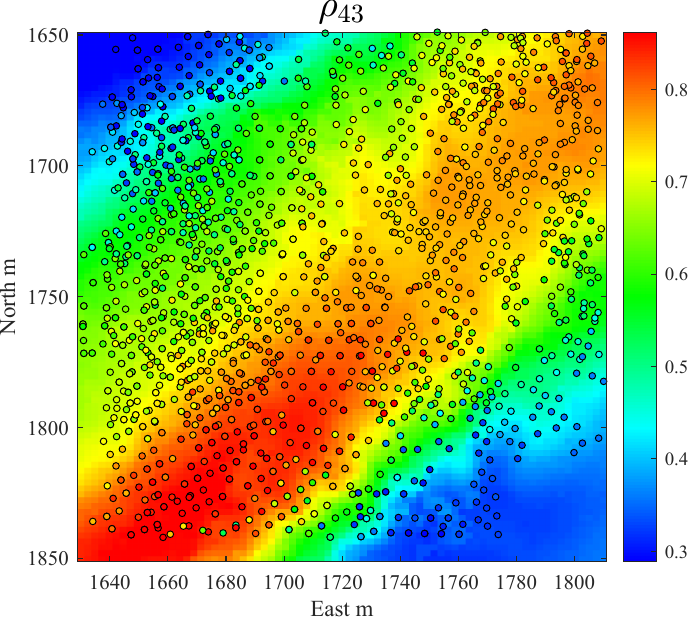}
	\caption{Inferred correlation at sampled locations and grid interpolation around level 95 m, using the weighted Fr\'echet mean implementation for correlation matrices. 6 out of 15 entries are shown.}
	\label{corr}
\end{figure}

{The inferred correlation at sample locations and the interpolated correlations on the regular grid are represented by ellipsoids in Fig. \ref{gridelli}. Part of the individual components of the matrices, inferred at sample locations and then interpolated, are displayed in Fig. \ref{corr}, showing that the interpolation is robust enough to address the problem of discontinuities in the inferred values of the correlation matrix, as data locations are
included and excluded from the neighborhood when the target moves. As a result, the map of individual components is smooth.}

\begin{figure}[h!]
	\centering
	\includegraphics[width=0.48\textwidth,trim={0 0cm 0.0cm 0cm},clip]{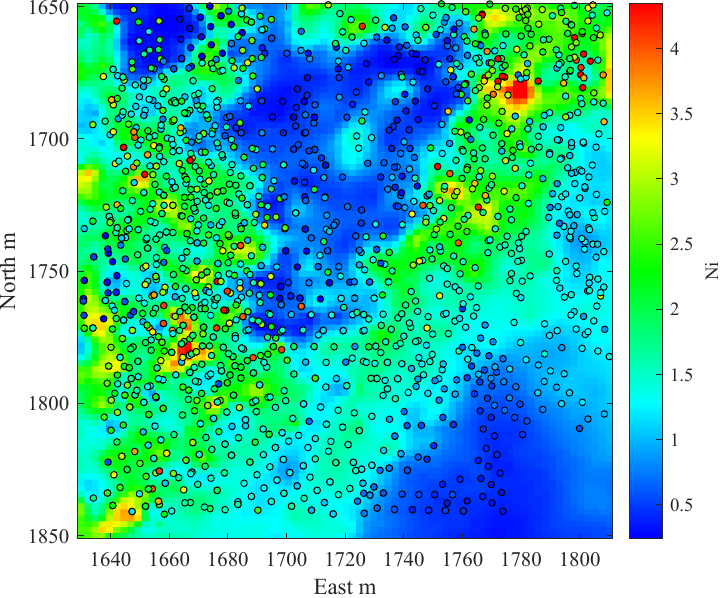}
	\includegraphics[width=0.48\textwidth,trim={0 0cm 0.0cm 0cm},clip]{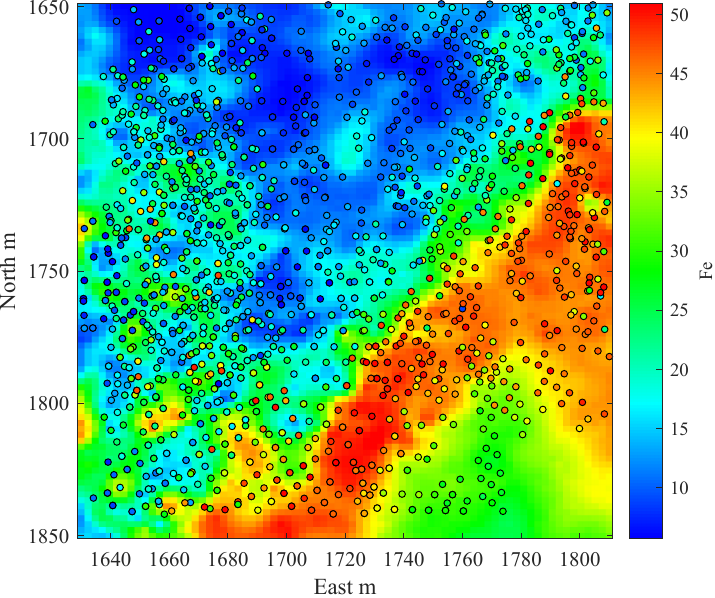}
	\includegraphics[width=0.48\textwidth,trim={0 0cm 0.0cm 0cm},clip]{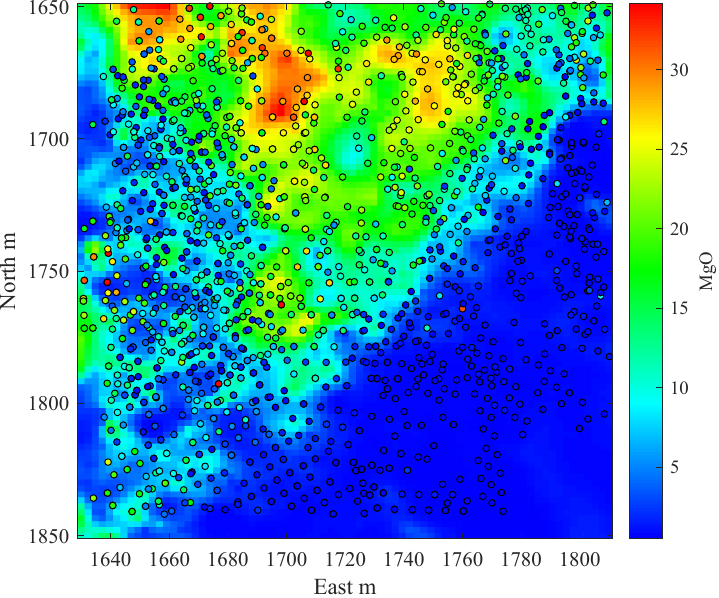}
	\includegraphics[width=0.48\textwidth,trim={0 0cm 0.0cm 0cm},clip]{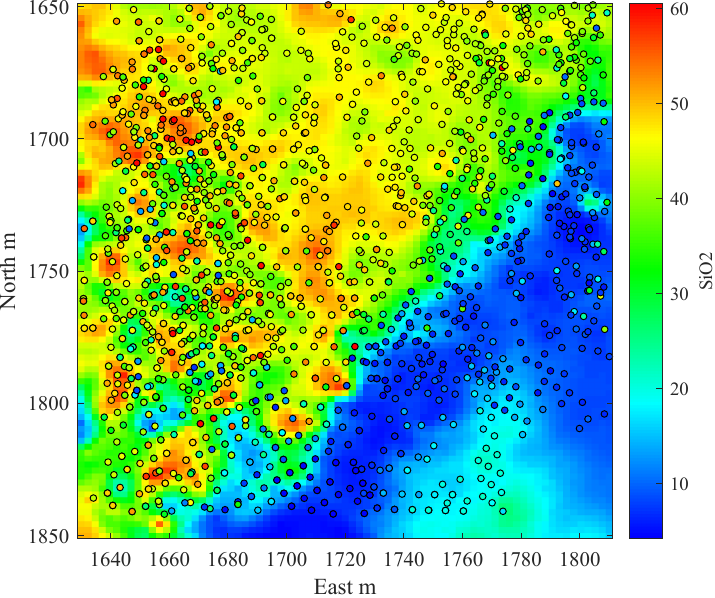}
	\includegraphics[width=0.48\textwidth,trim={0 0cm 0.0cm 0cm},clip]{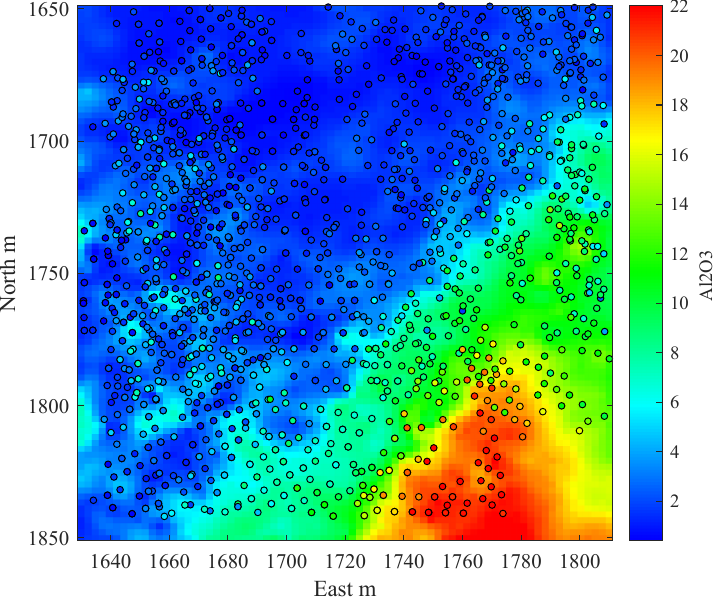}
	\includegraphics[width=0.48\textwidth,trim={0 0cm 0.0cm 0cm},clip]{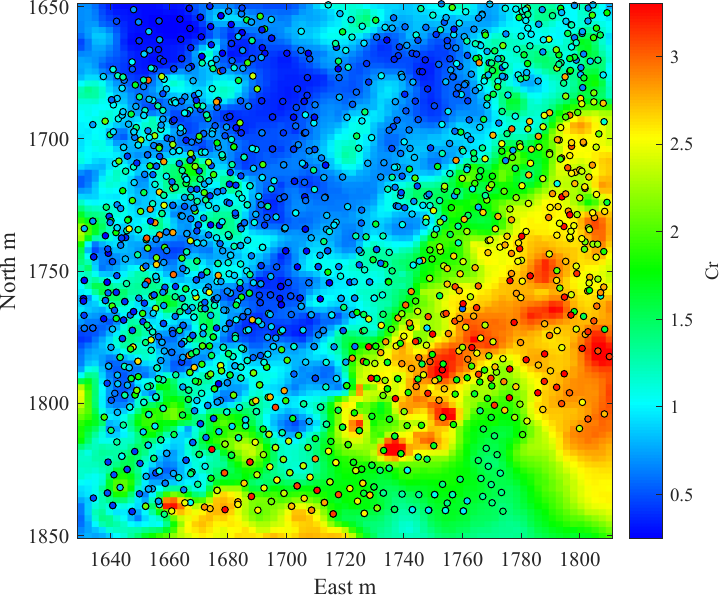}
	\caption{Plan view showing the estimated mean of the simulations at level 95 m and sampling data.}
	\label{mean}
\end{figure}

The produced maps showing the mean of the simulations, at level 95 m, are given in Fig. \ref{mean}, for the six back-transformed cross-correlated variables. The results reproduce cross-correlation trends in the maps. For instance, there is a strong negative correlation between Fe and MgO, which can be corroborated from visual inspection.

\begin{figure}[h!]
	\centering
	\includegraphics[width=0.45\textwidth,trim={0 0cm 0.0cm 0cm},clip]{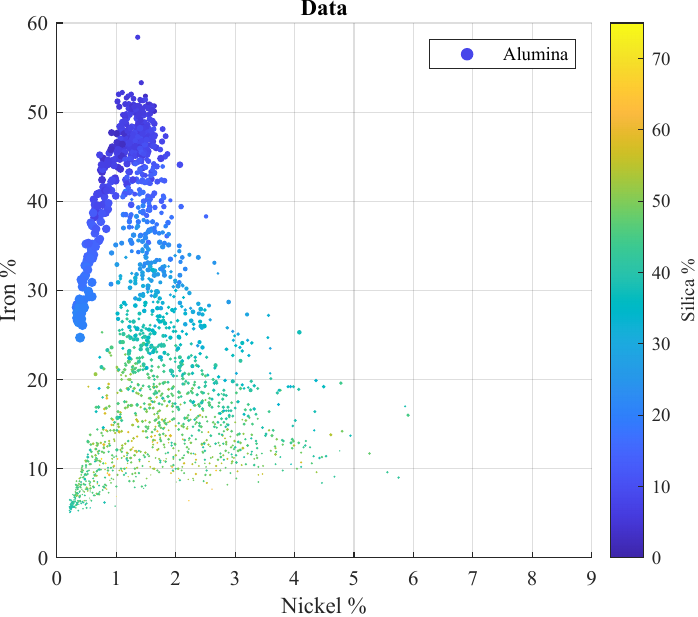}\\
	\includegraphics[width=0.45\textwidth,trim={0 0cm 0.0cm 0cm},clip]{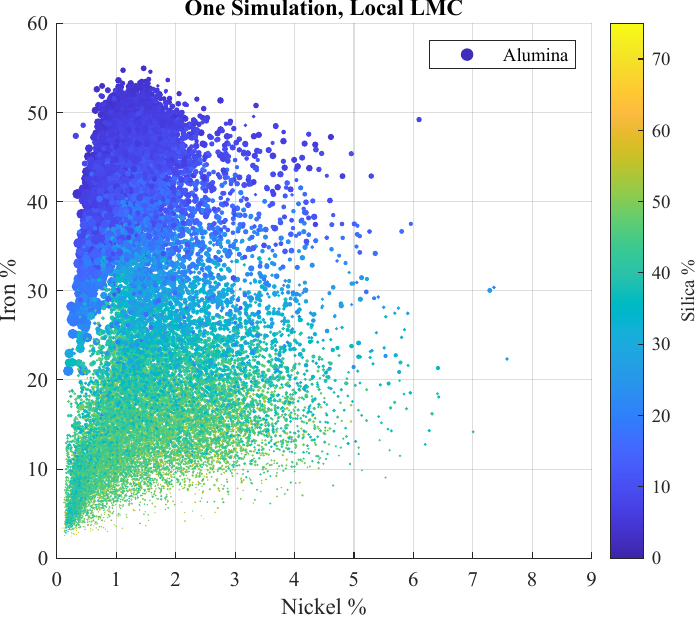}
	\includegraphics[width=0.45\textwidth,trim={0 0cm 0.0cm 0cm},clip]{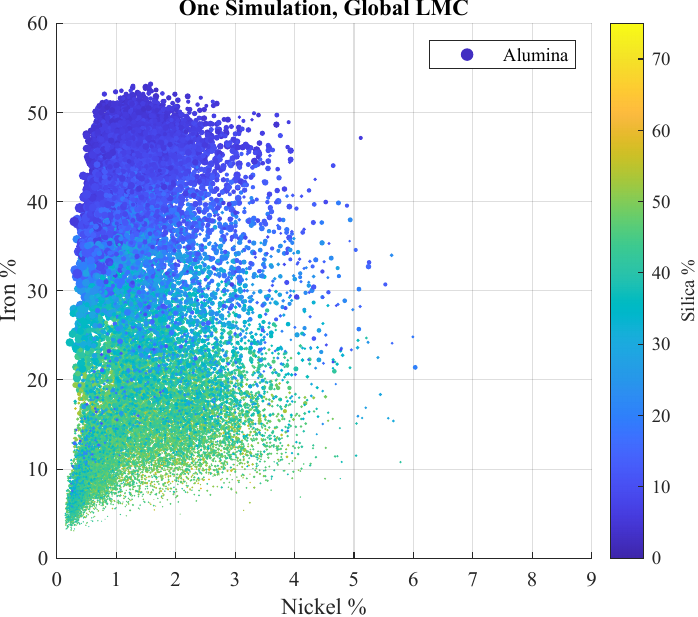}
	\includegraphics[width=0.45\textwidth,trim={0 0cm 0.0cm 0cm},clip]{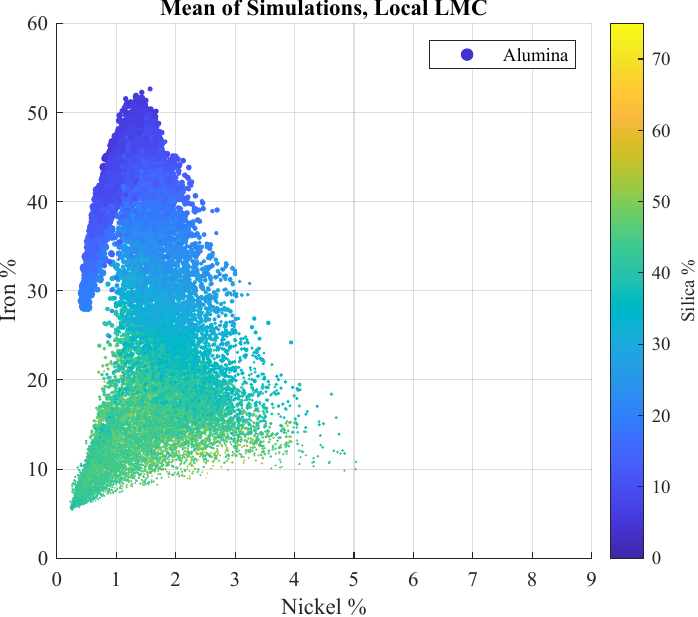}
	\includegraphics[width=0.45\textwidth,trim={0 0cm 0.0cm 0cm},clip]{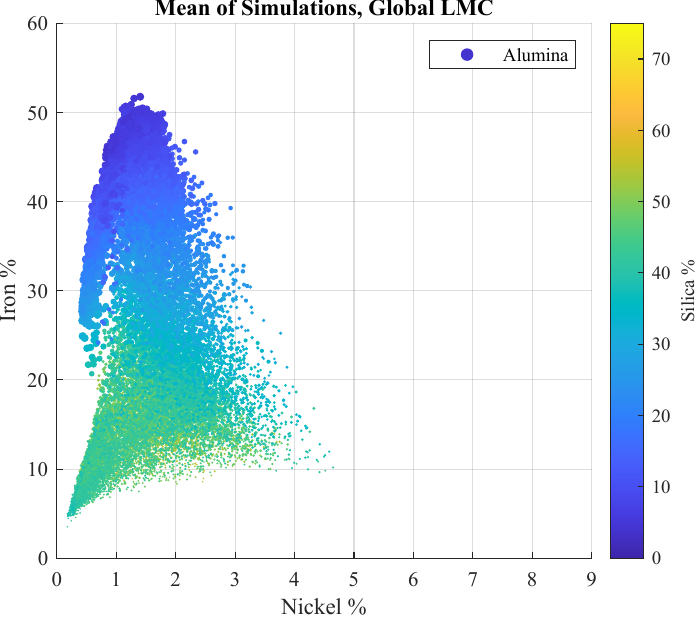}
	\caption{Scatter plot for sampling data (top left), the mean values on the grid (top right) and one simulation (bottom), around level 95 m.}
	\label{scatt2}
\end{figure}

The non-linear behavior among variables is well reproduced. This is shown on Fig. \ref{scatt2} in the case of the mean of the simulations and for one particular realization, around level 95 m. We include the same results given by running the case study under a classical LMC approach (one global neighborhood for correlation adjustment), as a manner  of comparison, showing some artifacts in the non-linear behavior for the mean of the simulations, manifested by short circuits presented in the attribute space.

\begin{figure}[h!]
	\centering
	\includegraphics[width=0.95\textwidth,trim={0 0cm 0.0cm 0cm},clip]{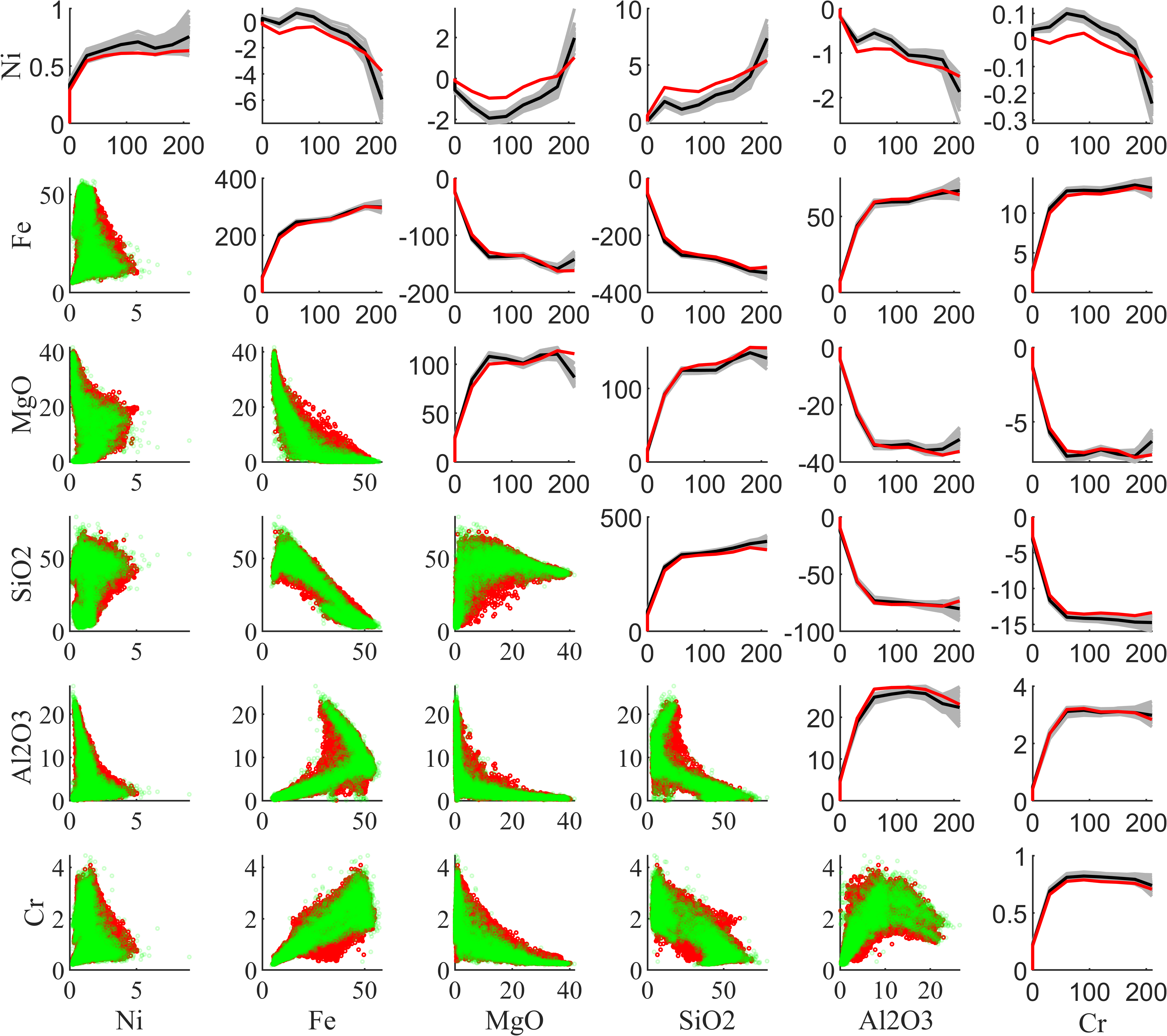}
	\caption{Scatter plots and variograms comparing input data values and mean of simulations. Variograms from simulations are shown in light gray lines, the mean of the variogram in black, and the variogram of the input data in red. The scatter plot showing bivariate relations of the original samples is represented by green dots, superposed to the mean of simulations as red dots.}
	\label{bivar}
\end{figure}

Scatter plots showing all bivariate relations for the mean of the simulations are shown in Fig. \ref{bivar}, together with the results of the variogiaphy. Variograms are well reproduced, besides the issues commented previously. It is quite impressive how well-fitted are most of the direct and cross variograms, given the fact that only one variogram model was considered for the purpose of the presented methodology.

\subsection{Testing}

In order to test the predictability and the uncertainty assessment capabilities of the methodology, we bring back the testing data left out from the first part of the case study. Each testing data was migrated to the closest node on the grid, retaining only the data within less than 2.5 m to the corresponding node. This resulted in 2313 samples to be considered from the initial 2997 in an uncertainty analysis.

\begin{figure}[h!]
	\centering
	\includegraphics[width=0.95\textwidth,trim={0 0cm 0.0cm 0cm},clip]{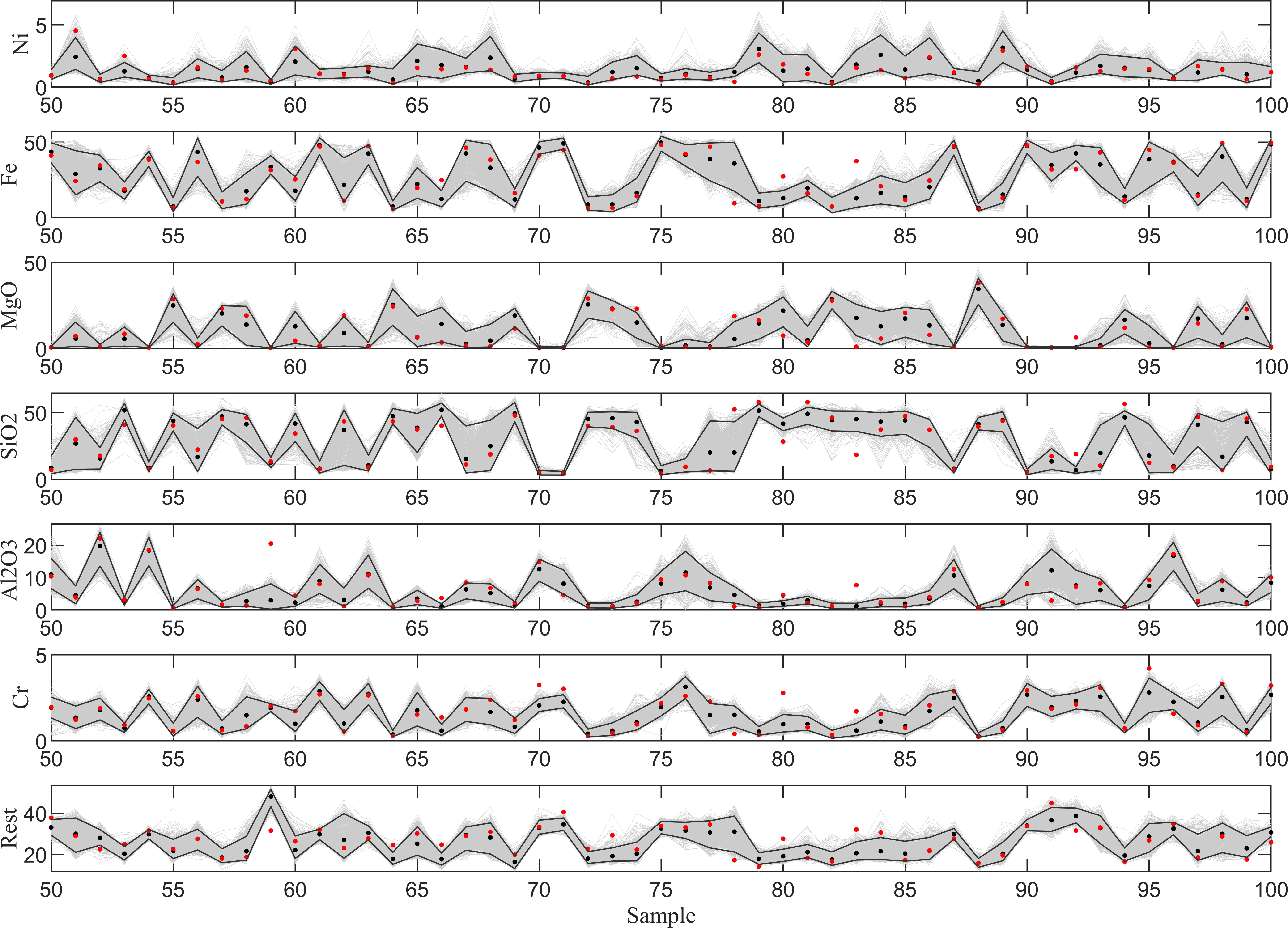}
	\caption{Uncertainty assessment for 50 samples taken from testing data, showing 1000 simulations in gray lines, the lower 5\% and the upper 95\% percentile as confidence boundary in  black lines, the estimated mean in black dots and, in red dots, the ground truth.}
	\label{samples2}
\end{figure}

The resulting PDFs from the simulations are shown in Fig. \ref{samples2} for 50 samples. We inspect results in detail for a small range of samples. The realizations are displayed in light gray lines, and the mean estimation of the simulation (in black dots) is shown for the seven variables. Red dots represent the true grade of the samples. A 5\% and 95\% percentile lines are displayed in black to give a 90\% confidence interval.

\begin{figure}[h!]
	\centering
	\includegraphics[width=0.45\textwidth,trim={0 0cm 0.0cm 0cm},clip]{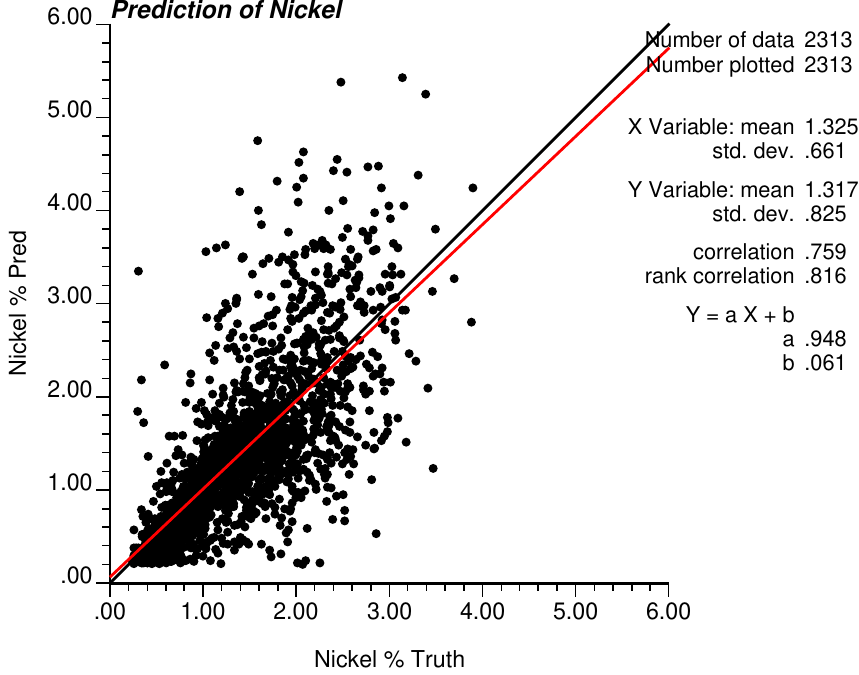}
	\includegraphics[width=0.45\textwidth,trim={0 0cm 0.0cm 0cm},clip]{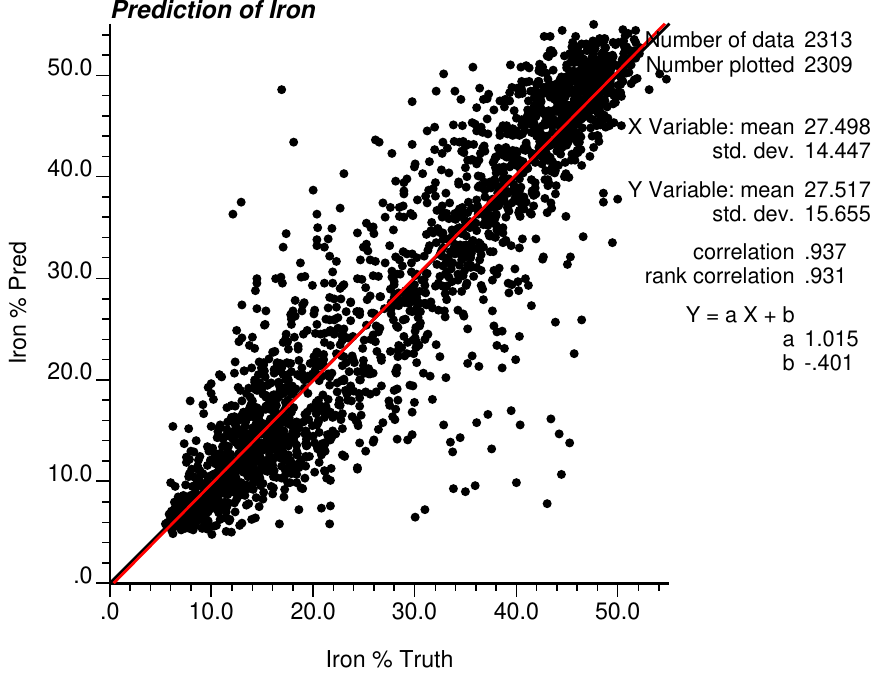}
	\includegraphics[width=0.45\textwidth,trim={0 0cm 0.0cm 0cm},clip]{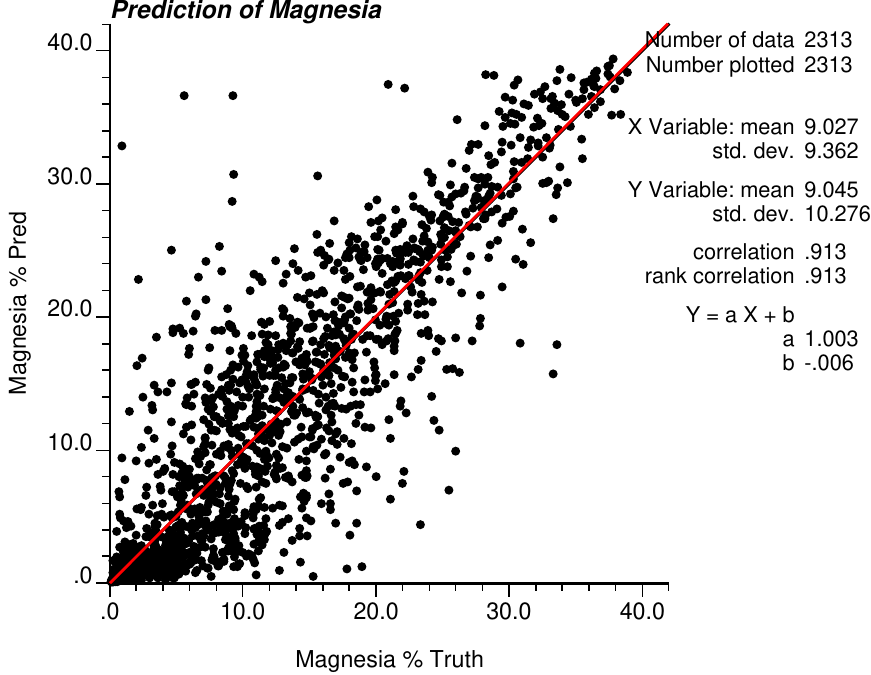}
	\includegraphics[width=0.45\textwidth,trim={0 0cm 0.0cm 0cm},clip]{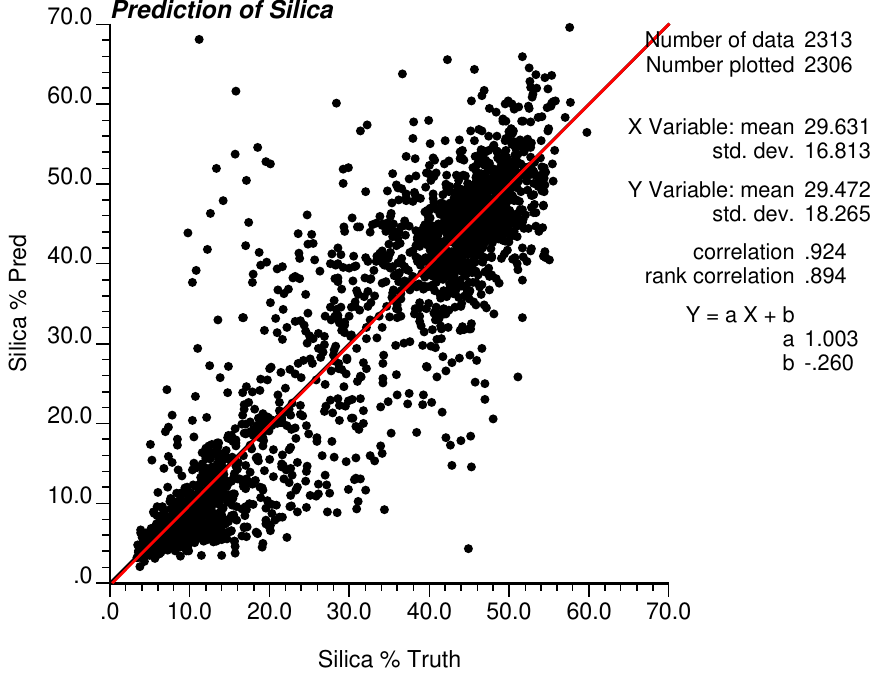}
	\includegraphics[width=0.45\textwidth,trim={0 0cm 0.0cm 0cm},clip]{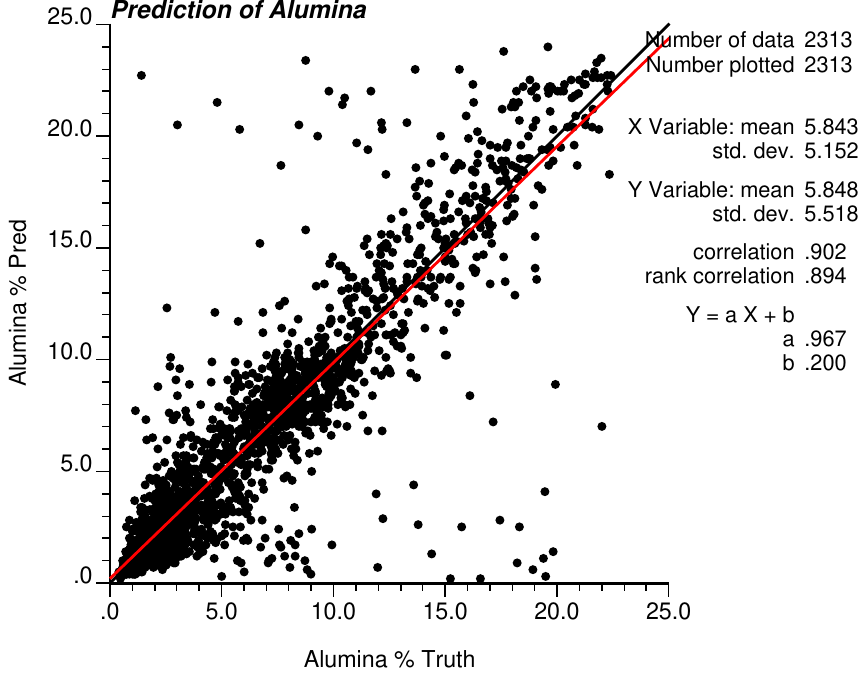}
	\includegraphics[width=0.45\textwidth,trim={0 0cm 0.0cm 0cm},clip]{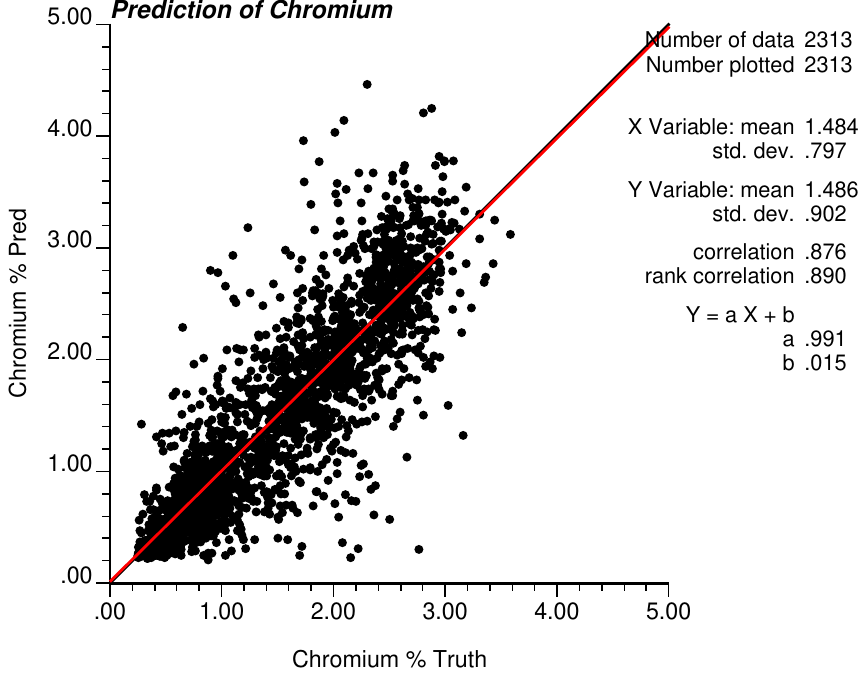}
	\caption{Scatter plots comparing the estimated mean of the simulations on locations close to testing data with the ground truth.}
	\label{scatt1}
\end{figure}

Figure \ref{scatt1} shows the scatter plots comparing the estimated mean of nodes versus their ground truth value. Low bias on the prediction and high correlation values are obtained, varying from a lowest value of 0.75 in the case of Nickel to 0.93 in the case of Iron.

\begin{figure}[h!]
	\centering
	\includegraphics[width=0.49\textwidth,trim={0 0cm 0.0cm 0cm},clip]{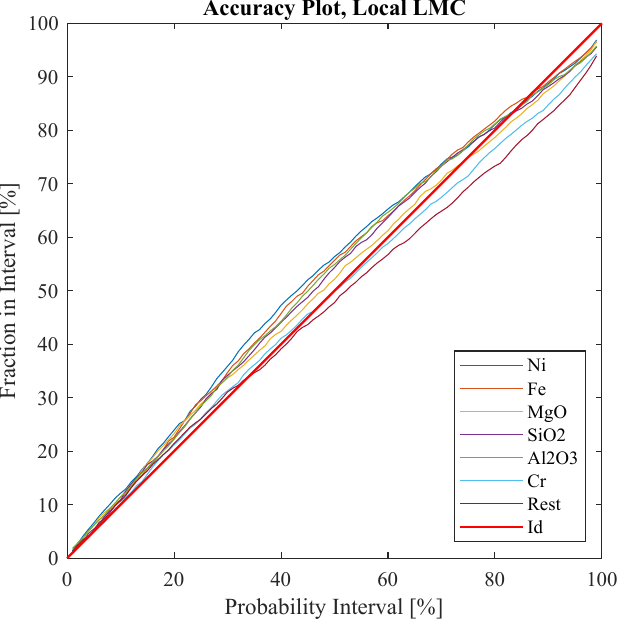}
	\includegraphics[width=0.49\textwidth,trim={0 0cm 0.0cm 0cm},clip]{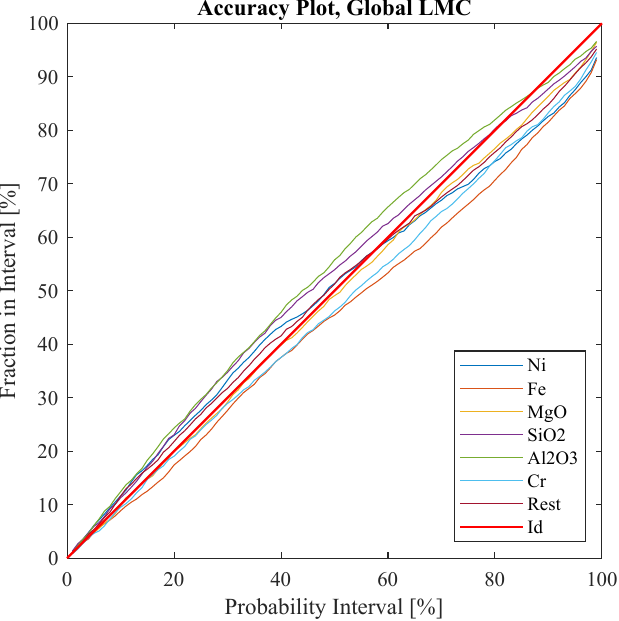}
	\caption{Accuracy plot, which calculates the proportion of locations where the true value falls within symmetric $ p $-probability interval.}
	\label{acc}
\end{figure}

The main advantage of simulating is that we can validate if the decision made on previous steps was correct. The validation is completed with the generation of an accuracy plot to check that the uncertainty given by the PDFs effectively represents the experimental frequencies on the ground truth of testing data (Fig. \ref{acc}).

Finally, a comparison table (Table \ref{tab:becn}) evaluating the predictive performances of the locally varying methodology relatively to the classical LMC is carried out using discrepancy measures calculated on the testing set: mean error (MAE), mean absolute error (MAE), and root mean square error (RMSE). Results show consistent less error for the proposed locally varying methodology in most of the variables evaluated. We note that better performance of the proposed method is observed for some variables (such as Nickel and Chromium) than others (such as Iron and Silica) due to a good spatial continuity of the first ones and a bimodal complexity in the distribution of the last ones, which induces further biases due to regression toward the mean (see Fig. \ref{scatt1}).

\renewcommand{\arraystretch}{1.5}
\begin{table}[h!]
	\centering
	\caption{Comparison in the predictability performance of LVLMC against LMC.}
	\label{tab:becn}
	\resizebox{1\textwidth}{!}{\begin{tabular}{l l l l l l l}
			\hline
			& \multicolumn{2}{c}{ME}                                         & \multicolumn{2}{c}{MAE}                                      & \multicolumn{2}{c}{RMSE}                                      \\   
			Variable                      & LVLMC $ \quad $ $ \quad $                         & LMC $ \quad $ $ \quad $                          & LVLMC$ \quad $$ \quad $                          & LMC$ \quad $$ \quad $                          & LVLMC$ \quad $$ \quad $                          & LMC $ \quad $$ \quad $                          \\ \hline
			\multicolumn{1}{l}{Nickel}   & \textbf{0.008} & -0.013 & \textbf{0.341} & 0.344 &  \textbf{0.538} & 0.547 \\
			\multicolumn{1}{l}{Iron}     & \textbf{-0.018} & -0.133 & \textbf{3.751} & 3.794 &  \textbf{5.475} & 5.533 \\
			\multicolumn{1}{l}{Magnesia} & \textbf{-0.017} & 0.084 & \textbf{2.609} & 2.702 &  \textbf{4.182} & 4.297 \\
			\multicolumn{1}{l}{Silica}   & \textbf{0.158}  & 0.359  & \textbf{4.589} & 4.883 &  \textbf{7.001} & 7.269 \\
			\multicolumn{1}{l}{Alumina}  & \textbf{-0.005} & -0.127 & \textbf{1.292} & 1.363 &  2.383 & \textbf{2.322} \\
			\multicolumn{1}{l}{Chromium} & \textbf{-0.002} & -0.016 & \textbf{0.300} & 0.302 & 0.434 & \textbf{0.430} \\
			\multicolumn{1}{l}{Rest}     & \textbf{-0.123} & -0.152 & \textbf{2.506} & 2.546 &  \textbf{3.744} & 3.752 \\   \hline
	\end{tabular}}
\end{table}
\renewcommand{\arraystretch}{1}
\section{Conclusions}
\label{Conclusions}

We have shown how multivariate data can be understood and modeled as a RF with varying step-zero correlation, with the purpose of reproducing global complexities shown in the attribute space. At every sample location, the dependency among variables is inferred from computing the sample correlation matrix on the neighborhood of each sample. The collection of these matrices is mapped into a correlation manifold with suitable distance properties. In this space, we can proceed with the weighted interpolation of the different known correlation matrices, where the spatial information is carried in the weights of the interpolation. The inference of the local correlation allows us to decorrelate the observed Gaussianized variables and proceed with their spatial modeling, which can be done independently. After the interpolation step of correlation matrices and variables simulation, the model is carried forward, coupling the independent simulated variables locally using the inferred correlation matrix. The procedure demonstrates that the purpose of reproducing the non-linear multivariate features of data is achieved.

The methodology step requiring the inference of the local correlation matrix at sample locations is done by selecting a neighborhood with the closest samples, either by fixing a radius or a number of samples, according to the sampling density and availability at the location under study. The neighborhood will allow to perform the transformation of data into Gaussian values. This is an important aspect of the methodology as a correct neighborhood affects other aspects of modeling: (\textit{i}) a proper standardization of data and obtaining unit-sill variograms and zero-sill cross-variograms after decoupling of factors; (\textit{ii}) a proper inference of linear correlation parameters; and (\textit{iii}) Gaussianity in bi-variate distributions among pairs of variables locally in space. Once the neighborhood is selected and the inference of correlation is done, it is possible to model the spatial behavior of independent factors under a single variogram, as a simplifying step.
After simulating individual factors, it is possible to perform the linear combination of these to get correlated Gaussian variables, followed by applying back-transformation to get values in their original units.

Among the limitations, we can mention that the proposed methodology only works when enough isotopic data are available  to estimate the correlations locally. As with other methodologies that try to handle non-stationarity, when limited data is available, it is better to simplify the problem and assume stationarity on the data, as calibration of hyperparameters, such as the correlation matrices at different locations, may become difficult. The variography is theoretically challenging to handle and interpret under the assumption of different underlying structures. Working with different models of spatial continuity for the different structures would add a ``rotation'' of the structures as a free parameter and this would be a valid model that also fits the spatial correlation among variables. This is why we chose to work with a single variogram model for all the factors. A third issue is that the definition of stationary geological domains beforehand may replace the presented methodology. If the multivariate behavior changes ``continuously'', the proposed methodology may be a promising approach for handling non-stationary.

\section{Acknowledgments}

The authors acknowledge the funding provided by the Natural Sciences and Engineering Research Council of Canada (NSERC), funding reference number RGPIN-2017-04200 and RGPAS-2017-507956, and by the International Association for Mathematical Geosciences (IAMG) student grant, funding reference number MG-2020-14. The authors are grateful to three anonymous reviewers for their valuable comments on an earlier version of this paper.

\section{Conflict of Interest}

The authors declare having no conflict of interest that could influence the work reported in this paper.

\bibliographystyle{spbasic} 
\bibliography{LVCM}

\section*{Appendix A: Interpolation of Correlation Matrices}
\label{appA}

Our algorithm for interpolation of correlation matrices relies on basic concepts from Riemannian geometry. For this purpose, we present a brief summary of the theory of Riemannian manifolds. We refer the reader to \citeauthor{do2016differential} (\citeyear{do2016differential}, \citeyear{carmo1992riemannian}) and \citeauthor{lee2018introduction} (\citeyear{lee2018introduction}) for more details. Then we introduce  some useful notation and provide a brief review of the Riemannian geometry of Symmetric Positive Definite (SPD) manifolds, from where the  Correlation manifold is derived by projection.

\subsection{Geometric Background of SPD and Correlation Matrices}
\label{Background}

As a part of our methodology, we are interested in interpolating the known correlation matrices over the domain. The space $ \textrm{Corr}(p) $ of all $ p \times p $ correlation matrices, $ \textbf{C} $, satisfies both properties of having a diagonal of ones and being a symmetric positive definite matrix, that is, satisfying the property: $ \textbf{v}^T\textbf{C}\textbf{v} > 0 \textrm{ for all nonzero } \textbf{v} \in \R^p $. This space is not a vector space since, when multiplied by a negative scalar, a correlation matrix $ \textbf{C} $ is no longer in $ \textrm{Corr}(p) $. Therefore, the use of linear interpolation (or an Euclidean metric) is no longer suitable in this geometry.

\subsubsection{Review of Riemannian Manifolds}
A {differentiable manifold} $ M $ of dimension $ p $ generalizes the notion of a two dimensional surface to a dimension $ p $. Formally, it is a topological space  that is locally similar to an Euclidean space, with every point on the manifold having a neighborhood for which there exists a homeomorphism (a continuous bijection whose inverse is also continuous) mapping the neighborhood to $ \R^{p} $. Differentiable manifolds allows us to define derivatives of curves lying on the manifold. The {tangent space} $ T_\textbf{x}M $ at $ \textbf{x} $ is the vector space that contains the derivatives at a point $ \textbf{x} $ to all $ 1 $-D curves on $ M $ passing through $ \textbf{x} $. A {Riemannian metric} on a manifold $ M $ is an inner product $ \langle \cdot,\cdot \rangle^{}_{\!\textbf{x}} $ on the tangent space $ T_\textbf{x}M $ at $ \textbf{x} $, which varies smoothly from point to point. The length of a tangent vector $ \textbf{v} \in T_\textbf{x}M $, induced by the norm, is denoted by $ \norm{\textbf{v}}^2_\textbf{x} = \langle \textbf{v},\textbf{v} \rangle^{}_{\!\textbf{x}} $. The minimum length curve connecting two points $ \textbf{x}_i $ and $ \textbf{x}_j $ on the manifold is called {geodesic curve} $ \gamma $, and the length of this curve gives us the {Riemannian distance}, $ d(\textbf{x}_i, \textbf{x}_j) $.

\begin{figure}[h!]
	\begin{center}
		\includegraphics[width=0.5\textwidth]{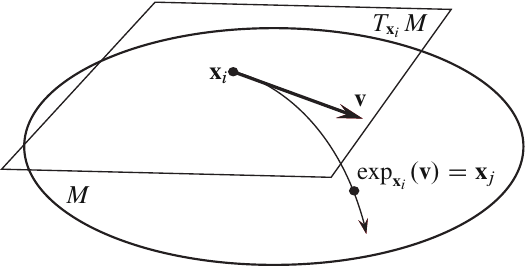}
	\end{center}
	\caption{A two-dimensional manifold $ M $, the tangent space at a point $ {\textbf{x}_i} \in  M$, $T_{\textbf{x}_i}M$, and the exponential map, $ \textmd{exp}_{\textbf{x}_i}(\cdot) $.}
	\label{tange}
\end{figure}

Given a tangent vector $ \textbf{v} \in T_\textbf{x}M $, there exists a
unique geodesic $ \gamma_\textbf{v}(t) $ starting at $ \textbf{x} $ with initial velocity $ \textbf{v} $, noticing that this speed remains constant and equal to $ \norm{v}^2_\textbf{x} $. The {exponential map}, $ \textmd{exp}_\textbf{x}: T_\textbf{x}M  \rightarrow M$ maps a tangent vector $ \textbf{v} $ to the point on the manifold reached at time $ 1 $ by the geodesic $ \gamma_\textbf{v}(t) $. Generally, the exponential map is only one-to-one in a neighborhood of $ \textbf{x} $. The inverse mapping of $ \textmd{exp}_\textbf{x} $ known as the logarithm map $ \textmd{log}_\textbf{x}: M  \rightarrow T_\textbf{x}M$ is, therefore, uniquely defined only around a small neighborhood of the point. For any two points $ \textbf{x}_i $ and $ \textbf{x}_j $ on the manifold $ M $, the tangent vector to the geodesic curve from $ \textbf{x}_i $ to $ \textbf{x}_j $ is defined as $ \textbf{v} = \textmd{log}_{\textbf{x}_i}(\textbf{x}_j) $, and the exponential map takes $ \textbf{v} $ to the point $ \textbf{x}_j  = \textmd{exp}_{\textbf{x}_i}\big(\textmd{log}_{\textbf{x}_i}(\textbf{x}_j)\big)$. In addition, $ \gamma_\textbf{v}(0) = \textbf{x}_i $
and $ \gamma_\textbf{v}(1) = \textbf{x}_j $. The Riemannian distance between $ \textbf{x}_i $ and $ \textbf{x}_j $ is defined as $ d(\textbf{x}_i, \textbf{x}_j)=\norm{\textmd{log}_{\textbf{x}_i}(\textbf{x}_j)}_{\textbf{x}_i} $. Figure \ref{tange} shows an example of a two-dimensional manifold, and illustrates the notion of tangent space and exponential map. 

Given the data $\textbf{x}_1 ,\dots,\textbf{x}_n \in M$, the {{geometric}} or {{Fr\'echet mean}}$ \hat{\mathbf{x}} $ is defined as a minimizer of the sum of squared distances:
\begin{equation}
	\hat{\mathbf{x}} = \textmd{arg}\inf_{\substack{\bar{\mathbf{x}} \in \mathbbcal{M}}} \sum_{i=1}^k d^2(\bar{\mathbf{x}},\textbf{x}_i),\nonumber
	\label{freme}
\end{equation} 
We consider the use of Fr\'echet mean later when computing the mean of correlation matrices.

\subsubsection{Notation}
Let $ \varvec{\mathcal{M}}(p) $ denote the set of $ p \times p $ real-valued matrices. For any $ \textbf{P} \in \varvec{\mathcal{M}}(p) $, let $ \textbf{P}^T$ denote the transpose of $ \textbf{P}$, and let $ \textbf{I}_p \in\varvec{\mathcal{M}}(p)$ denote the  $ p \times p $ identity matrix. The following sets of matrices will be of interest:
\begin{flushleft}
	\begin{tabular}{  p{0.15\textwidth} p{0.85\textwidth} } 
		GL($ p $)& the {{general linear group}}, set of $ p \times p $ invertible matrices \\ 
		& $ \textrm{GL}(p) =\{ \textbf{A} \in \varvec{\mathcal{M}}(p) : \textrm{det}(\textbf{A})\ne 0\} $. \\
		Sym($ p $) & the set of $ p \times p $  invertible symmetric matrices\\
		& $ \textrm{Sym}(p) =\{ \textbf{A} \in \textrm{GL}(p) : \textbf{A}=\textbf{A}^T\} $. \\  
		$ \textrm{Sym}^+(p)$ & the set of $ p \times p $ symmetric  positive definite (SPD) matrices  \\
		& $ \textrm{Sym}^+(p) =\{ \textbf{A} \in \textrm{Sym}(p) : \textbf{v}^T\textbf{A}\textbf{v} > 0 \textrm{ for all } \textbf{v} \in \R^p\} $. \\  
		O($ p $) & the set of $ p \times p $   orthonormal matrices\\ 
		&$ \textrm{O}(p) =\{ \textbf{A} \in \textrm{GL}(p) : \textbf{A}^T\textbf{A}=\textbf{AA}^T=\textbf{I}_p\} $. \\
		SO($ p $) & the set of $ p \times p $   rotation matrices\\ 
		&$ \textrm{SO}(p) =\{ \textbf{A} \in \textrm{O}(p) : \textrm{det}(\textbf{A})=1\} $. \\   
		Diag($ p $) & the set of $ p \times p $  diagonal matrices with real entries\\ 
		&$ \textrm{Diag}(p) =\{ \textbf{A} \in \varvec{\mathcal{M}}(p) : \textbf{A}_{ij}=0 \textmd{ for } i \ne j\} $. \\
		$ \textrm{Diag}^+(p) $ & the set of $ p \times p $  diagonal matrices with positive entries\\ 
		&$ \textrm{Diag}^+(p) =\{ \textbf{A} \in \textrm{Diag}(p) : 	\textbf{A}_{ii}> 0 \textmd{ for all } i= 1,\dots,p\} $. \\ 
	\end{tabular}
\end{flushleft}

In the following sections, $ \textbf{W}, \textbf{V} $ and $ \textbf{P} $ will represent positive definite matrices, while $ \textbf{X} $ will represent a real symmetric matrix.

\subsubsection{The Riemannian Manifold of SPD Matrices}
Let $ \textrm{Sym}^+(p) $
denote the set of symmetric, positive definite matrices of size $ p \times p $, that is the set of all symmetric $ p \times p $ matrices $ \textbf{W} $ such that the quadratic form
$ \textbf{v}^T\textbf{W}\textbf{v} > 0 \textrm{ for all } \textbf{v} \in \R^p $. The set $ \textrm{Sym}^+(p) $ is not a vector space but forms a cone-shaped space \citep{hiriart2012fresh}.

One can consider several non-Euclidean metrics when working within $ \textrm{Sym}^+(p) $ (\citeauthor{dryden2009non} \citeyear{dryden2009non}). We focus on the classical {affine-invariant Riemannian metric} (AIRM), which has been a thoroughly studied geometric structure on $ \textrm{Sym}^+(p) $ (\citeauthor{moakher2005differential} \citeyear{moakher2005differential}; \citeauthor{pennec2006riemannian} \citeyear{pennec2006riemannian}), assigning as inner product, for any two tangent vectors $ \textbf{X}_1 $ and $ \textbf{X}_2 $ at a point $ \textbf{P} \in \textrm{Sym}^+(p) $, the value
$$ \langle \textbf{X}_1,\textbf{X}_2 \rangle^{}_{\!\textbf{P}} =\textmd{tr}(\textbf{X}_1\textbf{P}^{-1} \textbf{X}_2\textbf{P}^{-1}). $$
Note that the tangent space $ T_{\textbf{P}}\textrm{Sym}^+(p) $ is the space $ \textrm{Sym}(p) $ of symmetric matrices.

Starting from this definition of metric, it is possible to derive the operations that connect the manifold $ \textrm{Sym}^+(p) $ with its tangent space by analytical formulae (\citeauthor{lang2012fundamentals} \citeyear{lang2012fundamentals}). Given a tangent vector $ \textbf{X} \in T_\textbf{P}\textrm{Sym}^+(p) $ at a point $ \textbf{P} \in \textrm{Sym}^+(p) $, the Riemannian exponential map $\textmd{exp}_{\textbf{P}}: T_{\textbf{P}}\textrm{Sym}^+(p) \rightarrow \textrm{Sym}^+(p)$ is given by 
\begin{equation}
	\textbf{V}=\textmd{exp}_{\textbf{P}}(\textbf{X})=\textbf{P}^{1/2}\textmd{Exp}(\textbf{P}^{-1/2}\textbf{X}\textbf{P}^{-1/2})\textbf{P}^{1/2}\text{.}	\label{exp}
\end{equation}
where $ \text{Exp}(\cdot) $ denotes the exponential of a matrix $$ \textmd{Exp}(\textbf{A})=\sum_{k=0}^{\infty}\frac{1}{k!}\textbf{A}^k.$$
Given two positive definite matrices $\textbf{P}, \textbf{V} \in\textrm{Sym}^+(p)$, the Riemannian logarithmic map $\textmd{log}_{\textbf{P}}: \textrm{Sym}^+(p) \rightarrow T_{\textbf{P}}\textrm{Sym}^+(p)$, of $ \textbf{V}$  in relation to $ \textbf{P}$ is given by
\begin{equation}
	\textbf{X}=\textmd{log}_{\textbf{P}}(\textbf{V})=\textbf{P}^{1/2}\textmd{Log}(\textbf{P}^{-1/2}\textbf{V}\textbf{P}^{-1/2})\textbf{P}^{1/2}\text{,}\label{logmap}
\end{equation}
with $ \textmd{Log}(\textbf{A})=\textbf{B} $ any $ p \times p $ matrix $ \textbf{B} $ such that $ \textmd{Exp}(\textbf{B})=\textbf{A} $ (the matrix logarithm).

The geodesic distance now represents the length of the shortest curve connecting two matrices and is defined over the manifold structure of $ \textrm{Sym}^+(p) $. It is given by an analytical expression that uses the logarithmic map. For two SPD matrices $ \textbf{V} $ and $\textbf{W} $, it can be computed as:
\begin{eqnarray}
	\label{dist}
	d_{\textrm{Sym}^+}^2(\textbf{V}, \textbf{W})&=& 
	\langle \textmd{log}_{\textbf{V}}(\textbf{W}),\textmd{log}_{\textbf{V}}(\textbf{W}) \rangle^{}_{\!\textbf{V}}\nonumber\\
	&=&\textmd{tr}\big(\textmd{Log}^2(\textbf{V}^{-1/2}\textbf{W}\textbf{V}^{-1/2})\big)\nonumber\\
	&=&\norm{\textmd{Log}(\textbf{V}^{-1}\textbf{W})}^2_F,
\end{eqnarray}
with $ \norm{\cdot}_F$ denoting the Frobeniuous matrix norm
$$ \norm{\textbf{A}}_F = \sqrt{\text{tr}(\textbf{A}^{T}\textbf{A})}. $$

Finally, the geodesic curve passing through \textbf{V} in the direction of $ \textbf{X}$ is uniquely given by
\begin{equation}
	\gamma_\textbf{V}(t;\textbf{X}) =  {\textbf{V}^{1/2}}\textmd{Exp}( {\textbf{V}^{-1/2}}\textbf{X} {\textbf{V}^{-1/2}}t) {\textbf{V}^{1/2}}\text{.}	\label{geod}
\end{equation}

SPD matrices considering non-Euclidean metrics have been successfully employed in medical imaging and machine learning applications (\citeauthor{pennec2006riemannian} \citeyear{pennec2006riemannian}; \citeauthor{moakher2011riemannian} \citeyear{moakher2011riemannian};  \citeauthor{goh2008clustering} \citeyear{goh2008clustering}; \citeauthor{jayasumana2015kernel} \citeyear{jayasumana2015kernel}), as computational methods simply relying on the Euclidean distances between SPD matrices are generally sub optimal and show low performance \citep{4479482}.

\subsubsection{The Riemannian Manifold of Correlation Matrices}
Now we turn our focus into the Riemannian structure of correlation matrices. First we gain some intuition about the relation among SPD matrices and correlation matrices. We then derive explicitly the correlation manifold from the SPD manifold, to define distances among correlation matrices.

\subsubsubsection{Visualizing $ \textrm{Corr}(2) $ and $ \textrm{Corr}(3) $}
The affine-invariant structure for $ \textrm{Sym}^+(p) $ is intrinsically linked with $ \textrm{Corr}(p) $ and imposes symmetry on its structure as a quotient manifold, which in this context can be understood as a generalization of the notion of projection, applied in the context of manifolds.

Let us begin by visualizing $ \textrm{Corr}(2) $ as a subset of $ \textrm{Sym}^+(2) $:
\begin{equation*}		
	\textrm{Corr}(2):=\Biggl\{
	\begin{pmatrix}
		1 &\quad x\\
		x &\quad 1\\
	\end{pmatrix} \quad : \quad x \in (-1,1)\Biggr\}\text{.}
\end{equation*}
We can see that this space is a manifold of dimension 1 parameterized by the map $ \varphi : (-1, 1) \rightarrow \textrm{Corr}(2) $ given by
\begin{equation*}		
	\varphi(x) = 
	\begin{pmatrix}
		1 &\quad x\\
		x &\quad 1\\
	\end{pmatrix}\text{.}
\end{equation*}
This is a smooth map into the symmetric matrices \big(containing $ \textrm{Corr}(2) $\big) whose inverse is simply given by projection onto one of the off-diagonal entries. We can visualize any $ \textbf{C} \in \textrm{Corr}(2) $ in the $ x$-$y$ plane by associating the ellipsoid parameterized by the equation $ \textbf{v}^T\textbf{C}^{-1}\textbf{v} =1$, with $ \textbf{v}=[x \, y]^T $.

\begin{figure}[h!]
	\centering
	\includegraphics[width=0.65\textwidth,trim={0 0cm 0.0cm 0cm},clip]{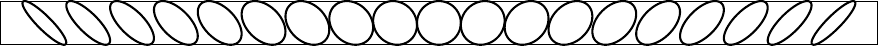}
	\caption{The manifold $\textrm{Corr}(2) $.}
	\label{corrman}
\end{figure}

\begin{figure}[h!]
	\centering
	\includegraphics[width=0.65\textwidth,trim={0 0cm 0.0cm 0cm},clip]{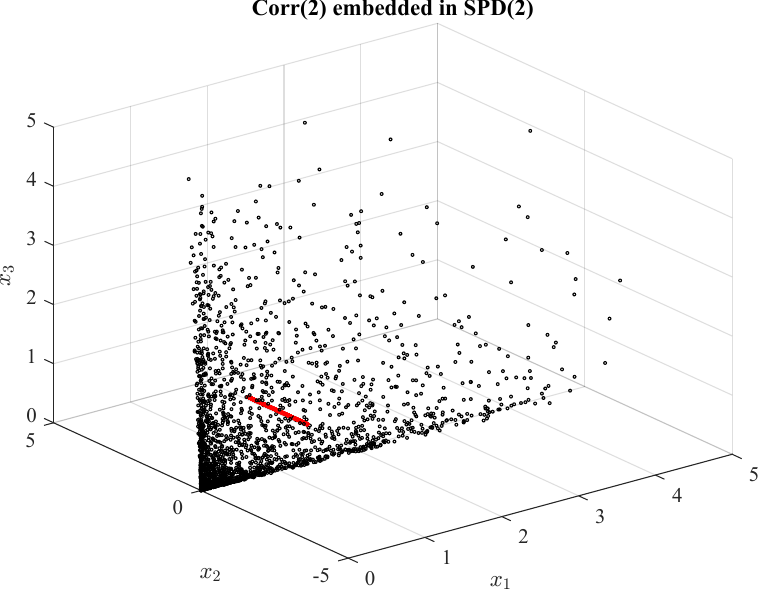}
	\caption{The $ \textrm{Sym}^+(2) $ cone, characterized as the set $ \{(x_1, x_2, x_3) \in \R^3 : x_1 > 0, x_2 > 0, x_1x_2 - x_3^2 > 0\} $, and the embedded submanifold $\textrm{Corr}(2) $ characterized as the set $ \{(1, x_2, 1) \in \R^3 : x_2 \in (-1,1)\} $. Points in $ \textrm{Sym}^+(2) $ (black) are sampled independently of those in $\textrm{Corr}(2) $	(red).}
	\label{corrman23}
\end{figure}

Because of the global parametrization $ \varphi : (-1, 1) \rightarrow \textrm{Corr}(2) $, we can visualize the manifold $\textrm{Corr}(2) $ as the interval $ (-1, 1) $, attaching to each point in the interval the
corresponding ellipsoid to the positive-definite form associated to the matrix. We see this in Fig. \ref{corrman}. Another visualization we will consider is to see the correlation matrices embedded inside the symmetric positive-definite matrices (Fig. \ref{corrman23}).

\begin{figure}
	\centering
	\includegraphics[width=0.55\textwidth,trim={0 0cm 0.0cm 0cm},clip]{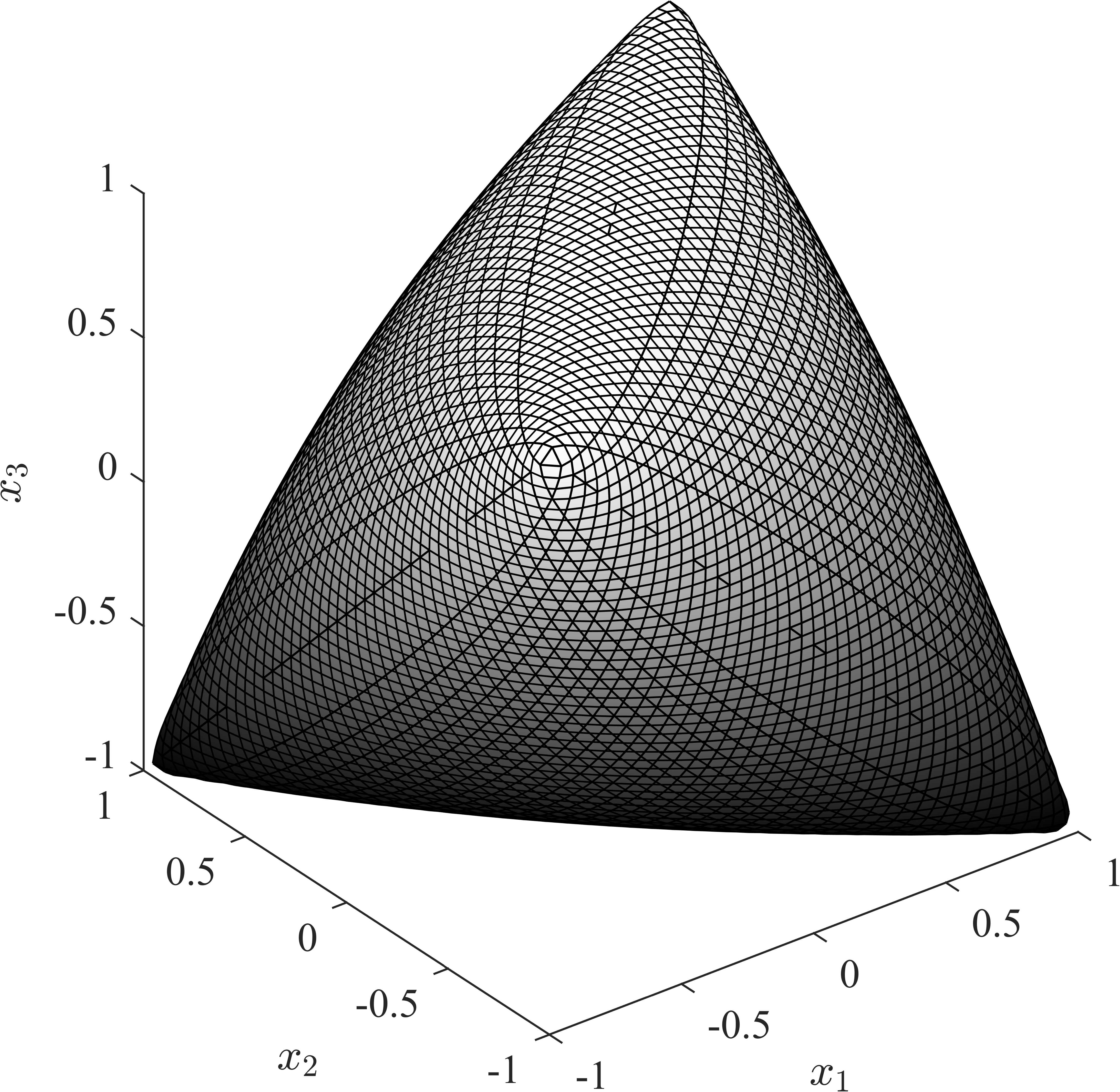}
	\caption{The boundary of the elliptope.}
	\label{elip}
\end{figure}

In the case of correlation matrices of dimension $ 3 $, the shape formed by the set is named the 3-dimensional elliptope \citep{Janas}, which can be represented by the following linear matrix inequality

\begin{equation*}		
	\textrm{Corr}(3):=\Biggl\{
	\begin{pmatrix}
		1 &\quad x_1&\quad x_2\\
		x_1 &\quad 1&\quad x_3\\
		x_2 &\quad x_3&\quad 1\\
	\end{pmatrix} \quad : x_1, x_2, x_3 \in (-1,1), \textmd{ and } \det \begin{bmatrix} 1 &\quad x_1 &\quad x_2\\x_1 &\quad 1 &\quad x_3\\ x_2 &\quad x_3 &\quad 1\end{bmatrix} = 
\end{equation*}
\[ 1 + 2 x_1 x_2 x_3 - x_1^2 - x_2^2 - x_3^2 > 0\Biggr\}\text{.} \]
The boundary of the elliptope (Fig. \ref{elip}) is the cubic surface defined by
\begin{equation*}		
	1 + 2 x_1 x_2 x_3 - x_1^2 - x_2^2 - x_3^2 = 0.
\end{equation*}

\subsubsubsection{Quotient Geometry}

\begin{figure}[h!]
	\centering
	\includegraphics[width=0.63\textwidth,trim={0 0cm 0.0cm 0cm},clip]{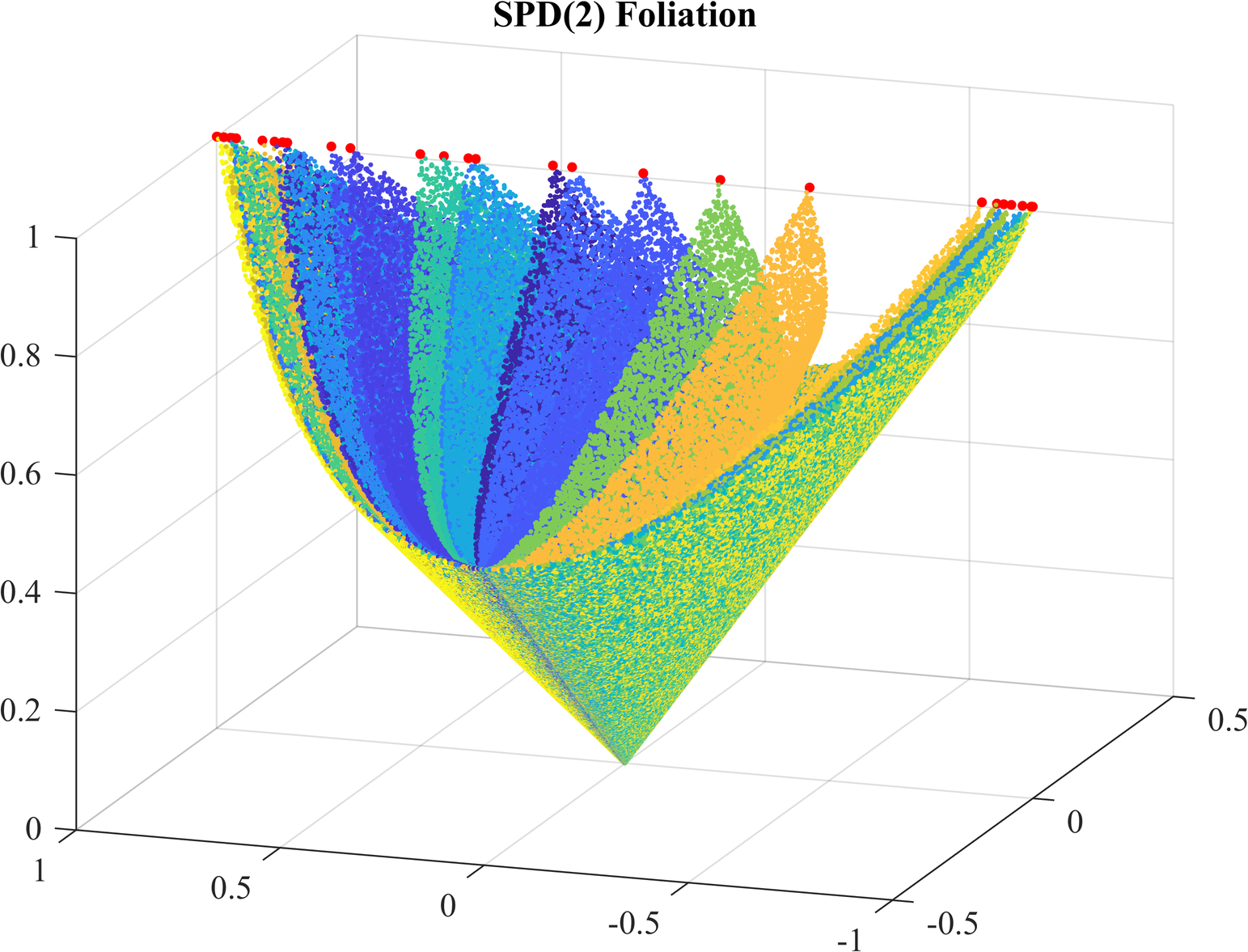}
	\caption{A foliation of the cone $\textrm{Sym}^+(2) $. Each leaf is an embedded two-dimensional submanifolds, obtained by translating a given correlation matrix $ \textbf{C} $ (red dot) by the action $ \textbf{DCD} $.}
	\label{fol}
\end{figure}

Consider an element ${\varvec{\upSigma}} \in \textrm{Sym}^+(p) $. The orbit of ${\varvec{\upSigma}}$, that is, the set of images of ${\varvec{\upSigma}}$ when considering the action of a group diagonal matrices with positive entries $ \textmd{Diag}^{+}(p) $ on it, $\textmd{Diag}^{+}(p) \times \textrm{Sym}^+(p) \rightarrow \textrm{Sym}^+(p)$, given by $ (\textbf{D},{\varvec{\upSigma}}) \mapsto \textbf{D}{\varvec{\upSigma}}\textbf{D} $:
\[ \textmd{Diag}^{+}(p)\, \varvec\cdot\, \varvec{\upSigma} = \big\{\textbf{D}\,\varvec\cdot\, \varvec{\upSigma} : \textbf{D} \in  \textmd{Diag}^{+}(p)\big\}\text{,} \qquad \textbf{D}\,\varvec\cdot\, \varvec{\upSigma}:=  \textbf{D}{\varvec{\upSigma}}\textbf{D}\text{,}\]
is a smooth manifold of dimension equal to $ \textrm{dim Diag}^+(p) = p $. This can be seen explicitly in the case of taking an element $ \textbf{C} \in \textrm{Corr}(2) \subset \textrm{Sym}^+(2) $, and sampling the orbit space by applying $ \textbf{DCD} $, where $ D \in \textrm{Diag}^+(p) $ is generated randomly, resulting in smooth surfaces (Fig. \ref{fol}).

Subsequently the quotient manifold $\textrm{Sym}^+(p)/\textmd{Diag}^{+}(p) $, that is, the manifold resulting from taking $\textrm{Sym}^+(p)$ but considering the elements spanned by the action of $\textmd{Diag}^{+}(p) $ within $\textrm{Sym}^+(p)$ as a same element, in an equivalence relation fashion, results in a smooth manifold. Intuitively, this correspond to a ``retraction'' along the leaves (or just a curved projection) to the one dimensional line $\textrm{Corr}(2)$, for the case of $ \textrm{Sym}^+(2) $. (The notion of quotient manifold is similar to the one used when considering an equivalence relation $ \{(x,y,z) : z \in (-\infty,\infty)\} $ inside $ \R^3 $ in order to obtain $ \R^2 $.) The resulting dimension of $ \textrm{Corr}(p) $ is  $ \textrm{dim Corr}(p) =\textrm{dim Sym}^+(p)-\textrm{dim Diag}^+(p)$  \citep{david2019riemannian}.

One can take, as representative of the equivalence relation, a correlation matrix to generate $ \textrm{Corr}(p) $. Explicitly, the representative that we take on $\textrm{Sym}^+(p)/\textmd{Diag}^{+}(p) $ corresponds to the element given by the projection\[ \pi : \textrm{Sym}^+(p) \rightarrow \textrm{Corr}(p) \qquad  \varvec{\upSigma}\mapsto(\textbf{D}_{\varvec{\upSigma}},\varvec{\upSigma}) = \textbf{C}_{\varvec{\upSigma}}\text{,}\] where $ \textbf{D}_{\varvec{\upSigma}} = (\textbf{I}_p \circ \varvec{\upSigma})^{-1/2} $ and $ \circ $ is the element-wise product $(\textbf{A}\circ\textbf{B})_{ij}=(\textbf{A})_{ij}(\textbf{B})_{ij} $. Since more than one element can be projected into the same correlation matrix, we call to the leave $\pi^{-1}(\textbf{C}_{\varvec{\upSigma}}) $ projected into the correlation matrix $ \textbf{C}_{\varvec{\upSigma}} $, the fiber of $\textbf{C}_{\varvec{\upSigma}} $:
\[ \pi^{-1}(\textbf{C}_{\varvec{\upSigma}}) = \{\varvec{\upSigma} \in \textrm{Sym}^+(p) : (\textbf{D}_{\varvec{\upSigma}},\varvec{\upSigma}) = \textbf{C}_{\varvec{\upSigma}}\}\text{.} \]

\subsubsubsection{Accounting for a Distance in $\textrm{Corr}(p)$}

While the quotient manifold structure of $\textrm{Corr}(p)$ is meaningful in itself, this fact alone does not yield results for computing distances on $\textrm{Corr}(p)$ through closed form expressions. Thus, one must rely only on the Riemannian structure that $\textrm{Corr}(p)$ inherits from $ \textrm{Sym}^+(p) $ in order to obtain an algorithm that computes distances through an optimization procedure.

In order to come up with such an algorithm, a theorem  proved by \cite{huckemann2010intrinsic} is used, showing that the geodesic connecting two points in the quotient can be expressed as the geodesic in the ambient manifold from the starting point to an
optimal representative of the end point, lying on the inverse image (the fiber) over the desired endpoint. We state this theorem as in \citet{david2019riemannian}:

\begin{theorem}
	(Huckemann 2010). Let $ M $ be a Riemannian manifold with an isometric
	action of a Lie group $ G $. Then a geodesic $ \gamma $ in the quotient $ M/G $ with end points $ a, b \in M/G $ can be obtained from the projection of a geodesic $ \tilde\gamma $ on M (that is, $ \gamma = \pi \circ \tilde\gamma$) such that
	\begin{itemize}
		\item $\tilde\gamma $ has end points $ p$, $ q $ with $ \pi(p) = a $, $ \pi(q) = b $, and
		\item $ q $ is the solution to the problem \[ \text{min}\, d_M(p,c) \quad\text{such that } c\in  \pi^{-1}(b)\text{.}\]
	\end{itemize}
	This last point can be rephrased for fixed $ c \in  \pi^{-1}(b) $ as \[ \text{min}\, d_M(p,g \varvec\cdot c) \quad\text{such that } g\in  G\text{.}\] 
\end{theorem}

Let us consider two points $\textbf{C}_1,\textbf{C}_2 \in \text{Corr}(n). $ Then, by adapting equation \ref{geod} to the current scenario, the geodesic and corresponding distance in $ \textrm{Sym}^+(p) $ connecting these two points are given by:

\begin{eqnarray*}
	\gamma_{\textrm{Sym}^+}(t) = \textbf{C}_1^{1/2}\textmd{Exp}\big(\textmd{Log}(\textbf{C}_1^{-1/2}{\textbf{C}}_2\textbf{C}_1^{-1/2})t\big)\textbf{C}_1^{1/2}\text{,} \\
	d^2_{\textrm{Sym}^+}=\big|\big|\textmd{Log}(\textbf{C}_1^{-1/2}{\textbf{C}}_2\textbf{C}_1^{-1/2})\big|\big|^2 \text{.}
\end{eqnarray*} 

In order to adapt this Riemannian structure to $\textrm{Corr}(p)$ we need to find the optimal
representative of $ \textbf{C}_2 $ with respect to the starting point $ \textbf{C}_1 $. This is done by finding the unique element $ \widetilde{\textbf{C}}_2 $ in the fiber $ \pi^{-1}(\textbf{C}_2) $ which minimizes the $ {\textrm{Sym}^+}$-distance between $ \textbf{C}_1 $
and $ \widetilde{\textbf{C}}_2 $. This can be written as

$$ d^2_{\textmd{Corr}}(\textbf{C}_1,\textbf{C}_2)= \inf_{\substack{\textbf{D} \in \textmd{Diag}^{+}(p)}}d^2_{\textrm{Sym}^+}(\textbf{C}_1,\textbf{DC}_2\textbf{D})$$
Using this equation we then aim to solve the following minimization problem:
\begin{equation}
	\text{minimize}\,d^2_{\textrm{Sym}^+}(\textbf{C}_1,\textbf{DC}_2\textbf{D})\quad\text{subject to}\quad\textbf{D}\in\textmd{Diag}^{+}(p)\text{.}	\label{mincorr}
\end{equation}
Assuming $ \textbf{D}^*$  is a sufficient solution to the above problem, we define as $ \widetilde{\textbf{C}}_2 $ this element in the fiber $ \pi^{-1}(\textbf{C}_2) $ which minimizes the $ {\textrm{Sym}^+}$-distance between $ \textbf{C}_1 $ and $\widetilde{\textbf{C}}_2$
$${\widetilde{\textbf{C}}}_2 = \textbf{D}^*\textbf{C}_2\textbf{D}^*\text{.}$$
The corresponding geodesic can be taken as the projection of the $ {\textrm{Sym}^+}$-geodesic connecting $ \textbf{C}_1 $ and $ \widetilde{\textbf{C}}_2$  
$$ \gamma_{\textmd{Corr}}(t) = \pi \Big(\textbf{C}_1^{1/2}\textmd{Exp}\big(t\textmd{Log}(\textbf{C}_1^{-1/2}\widetilde{\textbf{C}}_2\textbf{C}_1^{-1/2})\big) \Big)\text{.}$$

\subsection{Interpolation of the Correlation Matrices}
\label{interp}

Now we analyze the interpolation of different known correlations matrices both in the $\textrm{Corr}(p) $ and then in the geo-spatial setting.

We present a fixed point and a gradient descent algorithm which seek to minimize the mean-squared distances of $ \textrm{Sym}^+(p) $ and $ \textrm{Corr}(p) $-valued observations, respectively, with respect to the affine-invariant distance. The general process for the optimization procedure for the $ \textrm{Corr}(p) $ is proposed by \cite{david2019riemannian}. Considering a set of $ n $ correlation matrices $ \textbf{C}_1,\dots,\textbf{C}_n$, the general optimization procedure will formally take the following steps:

\begin{enumerate}
	\item Given a  current iterate $ \textbf{C}_t \in \textrm{Corr}(p) $ for the mean of correlation matrices, find all appropriate distances to initial observations $ \textbf{C}_i $ utilizing the fiber structure of $\textrm{Corr}(p) $, recalling that such a structure is defined as quotient by $\textrm{Sym}^+(p)/\textmd{Diag}^{+}(p) $. This translates in displacing each observations $ \textbf{C}_i $ individually from $\textrm{Corr}(p) $ into a point $ \widetilde{\textbf{C}}_i \in \textrm{Sym}^+(p)$, by the action of a diagonal matrix $ \textbf{D} $.
	\item  Interpret the current iterate $ \textbf{C}_t \in \textrm{Sym}^+(p) $ for the mean of correlation matrices as well, and perform the update to a point $\textbf{P}_{t+1}\in \textrm{Sym}^+(p) $, which corresponds to the geometric mean on $\textrm{Sym}^+(p) $ of the points $ \widetilde{\textbf{C}}_i $.
	\item Obtain the next iterate in the algorithm by projecting back to $ \textrm{Corr}(p) $, that is $ \textbf{C}_t = \pi(\textbf{P}_{t+1}) $.
\end{enumerate}

The steps are described in detail later. We begin by summarizing the optimization method first on $ \textrm{Sym}^+(p) $ followed by the optimization method on $ \textrm{Corr}(p) $.

\subsubsection{Optimizing on $ \textrm{Sym}^+(p) $}
Given the observations  $ \textbf{P}_1,\dots,\textbf{P}_n \in \textrm{Sym}^+(p) $, one could consider the \textit{arithmetic mean} of the $ n $ labeled covariance matrices $ \{\textbf{P}_i\}^{n}_{i=1} $: \[ \widehat{\varvec{\upSigma}} = \frac{1}{n}\sum_{i=1}^{n}\textbf{P}_i \]which does not account for any intrinsic geometric property of $ \textrm{Sym}^+(p) $.

We consider, instead, to use the \textit{\textbf{geometric}} or \textbf{\textit{Fr\'echet mean}}, introduced in the $ \textrm{Sym}^+(p) $ context by \cite{moakher2005differential}. Such a matrix is defined as follows:

\begin{equation}\widehat{\varvec{\upSigma}} = \textmd{arg} \inf_{\substack{\varvec{\upSigma}}} \sum_{i=1}^n d^2_{\textrm{Sym}^+}(\textbf{P}_i,\varvec{\upSigma}).\label{minim}
\end{equation}
Recall from Eq. \ref{dist} that the Riemannian distance between two SPD matrices is defined as:
$$ d_{\textrm{Sym}^+}^2(\textbf{P}_i, \varvec{\upSigma})=  \text{tr}\big(\text{Log}^2(\textbf{P}^{1/2}_i\varvec{\upSigma}\textbf{P}^{1/2}_i)\big) $$
and, therefore, minimizing Eq. \ref{minim} needs to be solved numerically. \cite{moakher2006averaging}  describes a numerical fixed-point algorithm to solve the geometric mean of a set of symmetric positive-definite matrices. Other methods such as Newton’s method on Riemannian manifolds \citep{david2019riemannian} could also be used for the numerical
computation of the geometric mean. However, the fixed-point
algorithm described below is simple, does not require a
sophisticated implementation, and converges rapidly.

\begin{figure}[h!]
	\begin{center}
		\includegraphics[width=1\textwidth]{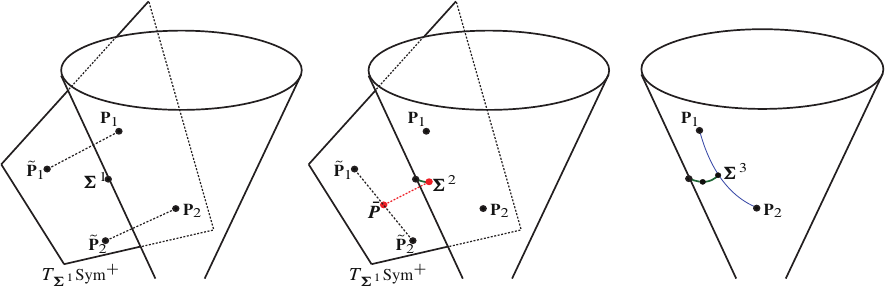}
	\end{center}
	\caption{Mean of $ 2 $ SPD matrices: Start with an initial guess point ${\varvec{\upSigma}}^{1} \in \textrm{Sym}^+(2)$. Project the two points $ \textbf{P}_1, \textbf{P}_2 \in \textrm{Sym}^+(2) $, into the tangent space of the initial guess ${\varvec{\upSigma}}^{1} $ through logarithmic maps to get $ \tilde{\textbf{P}}_1=\textmd{log}_{{\varvec{\upSigma}}^{1}}(\textbf{P}_1), \tilde{\textbf{P}}_2=\textmd{log}_{{\varvec{\upSigma}}^{1}}(\textbf{P}_2) \in T_{{\varvec{\upSigma}}^{1}}\textrm{Sym}^+(2) $ (left); get the mean of $ \textbf{P}_1$ and $\textbf{P}_2$, $ {\varvec{\bar P}}= 0.5\textbf{P}_1+ 0.5\textbf{P}_1$, and project back the resulting point into the manifold through exponential map to get the next iterative ${\varvec{\upSigma}}^{2}=\textmd{exp}_{{\varvec{\upSigma}}^{1}}({\varvec{\bar P}})$ (middle). The iterative process ends up with the geometric mean of the symmetric positive-definite matrices, which is a point lying in the geodesic joining $ \textbf{P}_1$ and $\textbf{P}_2$ (right). }
	\label{meanexa}
\end{figure}

The geometric mean $ \widehat{\varvec{\upSigma}} $ can be computed efficiently by an iterative procedure consisting in: projecting the covariance matrices in the tangent space, estimating the arithmetic mean in the tangent space and projecting the arithmetic mean back in the manifold. Then iterate the three steps until convergence. This process is illustrated in Fig. \ref{meanexa}.

The full algorithm, taken from \cite{moakher2006averaging}, is given in Algorithm 1.

\begin{algorithm}
	\caption{Mean of $ n $ SPD matrices }\label{alg:cap}
	\begin{algorithmic}[1]
		\Require a set of $ n $ SPD matrices  $ \textbf{P}_1,\dots,\textbf{P}_n \in \textrm{Sym}^+(p) $  and $ \epsilon> 0$.
		\State Initialize ${\varvec{\upSigma}}^{(1)} = \textbf{I}_p$
		\Repeat
		\State ${\varvec{\bar P}} = \frac{1}{n}\sum_{i=1}^{n}\textmd{log}_{{\varvec{\upSigma}}^{(t)}}(\textbf{P}_i)$ \Comment{Arithmetic mean in the tangent space, using \ref{logmap}}
		\State ${\varvec{\upSigma}}^{(t+1)}=\textmd{exp}_{{\varvec{\upSigma}}^{(t)}}({\varvec{\bar P}})$
		\Comment{Projecting back to SPD manifold, using \ref{exp}}
		\Until{$ \norm{{\varvec{\bar P}}} < \epsilon$}\\
		\Return $\widehat{\varvec{\upSigma}}:={\varvec{\upSigma}}^{(t+1)}$	
	\end{algorithmic}
\end{algorithm}

If we want to account for the spatial configuration of the data, we could consider the use of the \textbf{\textit{weighted Fr\'echet mean}}: \[ \widehat{\varvec{\upSigma}} = \textmd{arg} \inf_{\substack{\varvec{\upSigma}}} \sum_{i=1}^n \lambda_i d^2_{\textrm{Sym}^+}(\textbf{P}_i,\varvec{\upSigma}), \quad \sum_{i=1}^n \lambda_i = 1,\]
with $ \lambda_i $ the weights obtained, for instance, from the kriging interpolation. Algorithm 2 in this case is given by slightly modifying Algorithm 1.

\begin{algorithm}
	\caption{Weighted mean of $ n $ SPD matrices }\label{alg:cap2}
	\begin{algorithmic}[1]
		\Require a set of $ n $ SPD matrices  $ \textbf{P}_1,\dots,\textbf{P}_n \in \textrm{Sym}^+(p) $  and $ \epsilon> 0$.
		\State Initialize ${\varvec{\upSigma}}^{(1)} = \textbf{I}_p$
		\Repeat
		\State ${\varvec{\bar P}} = \sum_{i=1}^{n}\lambda_i\textmd{log}_{{\varvec{\upSigma}}^{(t)}}(\textbf{P}_i)$ \Comment{Weighted mean in the tangent space}
		\State ${\varvec{\upSigma}}^{(t+1)}=\textmd{exp}_{{\varvec{\upSigma}}^{(t)}}({\varvec{\bar P}})$  
		\Until{$ \norm{{\varvec{\bar P}}} < \epsilon$}\\
		\Return $\widehat{\varvec{\upSigma}}:={\varvec{\upSigma}}^{(t+1)}$	
	\end{algorithmic}
\end{algorithm}

\subsubsection{Optimizing Along Fibers}
\label{Optimizing}
In the same fashion, given the observations  $ \textbf{C}_1,\dots,\textbf{C}_n \in \textrm{Corr}(p) $, we are  interested in finding

\begin{equation}\widehat{\textbf{C}} = \textmd{arg} \inf_{\substack{\textbf{C}}} \sum_{i=1}^n d^2_{\textrm{Corr}}(\textbf{C}_i,\textbf{C}).\label{minim2}
\end{equation}

\begin{figure}[h!]
	\begin{center}
		\includegraphics[width=1\textwidth]{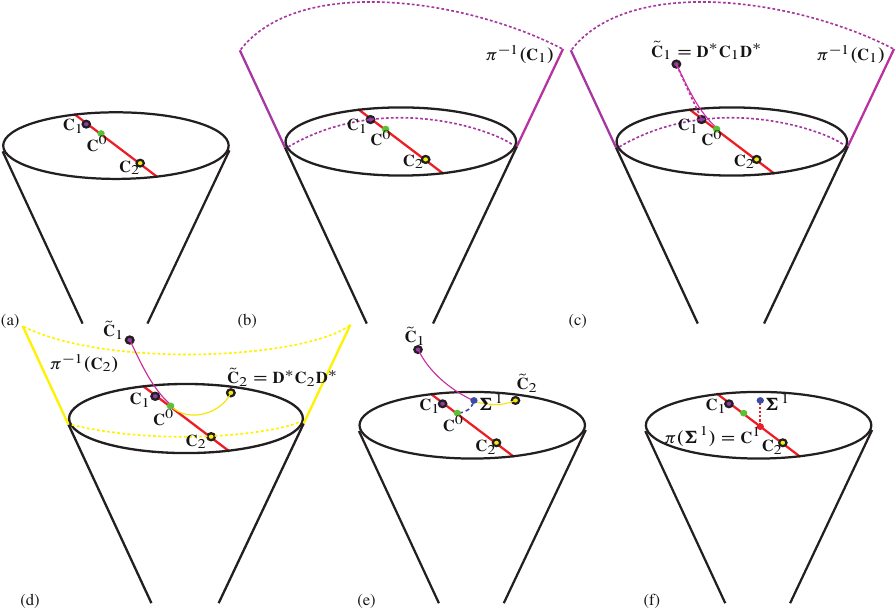}
	\end{center}
	\caption{Mean of $ 2 $ correlation matrices in $\textrm{Corr}(2) $ (red line). (a) Start with two points $ \textbf{C}_1, \textbf{C}_2 \in \textrm{Corr}(2) $, and an initial guess $ \textbf{C}^0  $. (b) Minimize the $ {\textrm{Sym}^+} $-distance between $ \textbf{C}^0 $ and $ \textbf{C}_1 $, restricted to the fiber $ \pi^{-1}(\textbf{C}_1) $ and keeping fixed the iterate $ \textbf{C}^0  $. (c) The minimization process along the $\textbf{C}_1 $-fiber ends up with an optimal point ${\tilde{\textbf{C}}}_1 = \textbf{D}^*\textbf{C}_1\textbf{D}^*$. (d) Repeat the previous step for all the given correlation matrices  (for $\textbf{C}_2 $ in this case). (e) Obtain Fr\'echet mean on $ \textrm{Sym}^+(2) $ considering ${\tilde{\textbf{C}}}_1 $ and ${\tilde{\textbf{C}}}_2 $ and taking as initial guess $ \textbf{C}^0 $. (f) The optimization process ends with a point ${\varvec{\upSigma}}^{1}$ out of $\textrm{Corr}(2) $, which has to be projected again into $\textrm{Corr}(2) $. }
	\label{meancorr}
\end{figure}

Recall again that the distance between $ \textbf{C}_i, \textbf{C} \in \textmd{Corr}(p) $ is given
by

\begin{eqnarray}
	d^2_{\textmd{Corr}}(\textbf{C}_i,\textbf{C})&= &\inf_{\substack{\textbf{D} \in \textmd{Diag}^{+}(p)}}d^2_{\textrm{Sym}^+}(\textbf{C}_i,\textbf{DC}\textbf{D}) \nonumber\\
	&= &\inf_{\substack{\textbf{D} \in \textmd{Diag}^{+}(p)}}\textmd{tr}\big[\textmd{Log}^2\big(\textbf{C}_i^{-1/2}\textbf{DC}\textbf{D}\textbf{C}_i^{-1/2}\big)\big], \nonumber
\end{eqnarray}
where we note that by symmetry we can just as well fix $ \textbf{C} $ and then optimize over the fiber of $ \textbf{C}_i $. For our purposes, we intend to minimize the distance between an iterate $ \textbf{C}_t $
of our algorithm between all of the observations $\textbf{C}_1,\dots,\textbf{C}_n $, hence we want to arrange our algorithm so that we are always keeping our iterate fixed and then optimizing along the fibers of our observations. In this way, we guarantee that we are updating our iterated point appropriately. The overall process of finding the mean of correlation matrices is illustrated in Fig. \ref{meancorr}, focusing on this updating process.

To find the optimal point, we employ a simple gradient descent in the Lie group $ \textmd{Diag}^+(p) $ with respect to the objective function \citep{david2019riemannian}
\[ g_i(\textbf{D})= d^2_{\textrm{Sym}^+}(\textbf{C},\textbf{DC}_i\textbf{D})\]
We minimize the above expression by using a gradient descent algorithm in order to find the optimal $ \textbf{D} $. We refer to \cite{david2019riemannian} for the details on the algorithm's derivation, as it requires several intermediate steps. However, one ends up with a brief two-steps iterative algorithm, by using a stepsize $ \delta > 0  $, initializing $ \textbf{D}_0 = \textbf{I}_p $ and the following iterative steps:
\begin{eqnarray}
	\varvec\upDelta_t & = &\textbf{I} \circ 2\textmd{Sym}[\textbf{D}_t\textmd{Log}(\textbf{C}_i\textbf{D}_t\textbf{C}^{-1}\textbf{D}_t)],\nonumber\\
	\textbf{D}_{t+1} & = & \textbf{D}_t\textmd{Exp}(-\delta\textbf{D}^{-1}_t\varvec\upDelta_t),\nonumber
\end{eqnarray}
with $ \textmd{Sym}(\textbf{A})=\frac{1}{2}(\textbf{A}+\textbf{A}^{T}) $, until a desired stopping criterion is reached. Once we find an optimal Lie group element $\textbf{D}^* \in \textmd{Diag}^{+}(p)$ as a result of minimizing $ g_i(\textbf{D}) $,  we define as $ \widetilde{\textbf{C}}_i $ this element over the fiber $ \pi^{-1}(\textbf{C}_i) $ which minimizes the $ {\textrm{Sym}^+} $-distance between $ \textbf{C} $ and $ \textbf{C}_i $,  ${\widetilde{\textbf{C}}}_i = \textbf{D}^*\textbf{C}_i\textbf{D}^*$.

We summarize the proposed algorithm which finds the Fr\'echet mean on $ \textmd{Corr}(p) $ in Algorithm 3.

\begin{algorithm}
	\caption{Weighted mean of $ n $ correlation matrices }\label{alg:cap3}
	\begin{algorithmic}[1]
		\Require a set of $ n $ correlation matrices  $ \textbf{C}_1,\dots,\textbf{C}_n \in \textrm{Corr}(p) $, $ \epsilon> 0$, a set of $ n $ kriging weights $ \lambda_1,\dots,\lambda_n $, initial point $ \textbf{C}^{(0)}= \textbf{I}_{p}$, and stepsize $ \delta>0 $.
		\State $ t=0 $
		\While {Stopping criterion not met}
		\For{$ i= 1,\dots,n$}
		\Require Initial point $ \textbf{D}_0 $
		\State $ k= 1$
		\While {Stopping criterion not met}
		\State $ \varvec\upDelta_t =\textbf{I} \circ 2\textmd{Sym}[\textbf{D}_k\textmd{Log}(\textbf{C}_i\textbf{D}_n\textbf{C}^{-1}_t\textbf{D}_n)] $
		\State $ \textbf{D}_{k+1} = \textbf{D}_{n}\textmd{Exp}(-\delta\textbf{D}^{-1}_k\varvec\upDelta_k) $
		\State $ k = k+1 $
		\EndWhile
		\State ${\widetilde{\textbf{C}}}_i = \textbf{D}_{k}\textbf{C}_i\textbf{D}_{k}$
		\EndFor
		\State ${\varvec{\bar P}} = \sum_{i=1}^{k}\lambda_i\textmd{log}_{\textbf{C}^{(t)}}({\widetilde{\textbf{C}}}_i)$ \Comment{Mean in the tangent space of $ \textmd{Sym}^{+}(p) $}
		\State ${\varvec{\upSigma}}^{t+1}=\textmd{exp}_{{\textbf{C}^t}}({\varvec{\bar P}})$
		\State $ \textbf{C}^{(t+1)} =\pi({\varvec{\upSigma}}^{t+1})= (\textbf{I}_p \circ \varvec{\upSigma}^{t+1})^{-1/2}\varvec{\upSigma}^{t+1} (\textbf{I}_p \circ \varvec{\upSigma}^{t+1})^{-1/2} $
		\Comment{Project back to $ \textmd{Corr}(p) $}
		\State $ t=t+1 $
		\EndWhile\\
		\Return $\textbf{C}^{(t)}$	
	\end{algorithmic}
\end{algorithm}

\end{document}